\documentstyle[]{article}
\begin{document}
\overfullrule 0 mm
\language 0
\centerline { \bf{ ONE-DIMENSIONAL CENTRAL-FORCE PROBLEM }}
\centerline { \bf{FOR SOMMERFELD SPHERE }}
\centerline { \bf {IN CLASSICAL
ELECTRODYNAMICS:}}
\centerline { \bf{ SOME NUMERICAL RESULTS }}
\vskip 0.5 cm

\centerline {\bf{ Alexander A.  Vlasov}}
\vskip 0.3 cm
\centerline {{  High Energy and Quantum Theory}}
\centerline {{Department of Physics}}
\centerline {{ Moscow State University}}
\centerline {{  Moscow, 119899}}
\centerline {{ Russia}}
\vskip 0.3cm
{\it Equation of motion of Sommerfeld sphere in the field of
Coulomb center is numerically investigated. It is shown that
contrary to Lorentz-Dirac equation in the attractive case there are
physical solutions. In the repulsive case sphere gains less energy
then that should be according to relativistic equation of motion of
point charge without radiation force.}

03.50.De
\vskip 0.3 cm
Numerical calculations of head-on collisions of two point charged
particles in  classical electrodynamics with retardation and
radiation reaction show many interesting properties of Lorentz-Dirac
equation [1-6].

Among them are:

{\bf (P.1)}  absence of physical trajectory in the attractive case -
for finite initial values of position, energy and acceleration point
charge stops before it reaches the Coulomb
center of the opposite sign and then turns back and moves away to
infinity  with velocity growing up to that of light [1,4,6];

{\bf (P.2)} in the repulsive case point charge can gain velocity,
after the turning point, much more greater then that follows from
the relativistic equations of motion without radiation force [4].

These and other effects (among them is  the effect of
preacceleration) cause much doubt in validity of standard approach to
radiation reaction.

In literature one can find the opinion that only
consistent quantum theory can solve all problems of radiation
reaction [7];

but also  one can find the point of view that the
problems lie in the first principles of classical theory, for
example, in the notion of "point" particle  (quantum theory only
rewrites classical problems in another language), and for "extended"
(in some sense) particles the situation will be different [8-11].

From the latter point of view it is interesting to consider how the
above results of numerical calculations change  for "extended", not
"point-like" charges. For this sake lets consider the famous
Sommerfeld model of extended charge with self-action.

Long time ago in Sommerfeld works [12, see also 9,13]
was derived the expression of self-force acting on
"nonrelativistically rigid charged sphere", i.e sphere with radius
$a$, its center moving along trajectory $\vec R(t)$, with total
charge $Q$ and charge density (in laboratory reference frame)
$$\rho(t,\vec r)={Q\over 4\pi a^2}\delta(|\vec r- \vec R|-a).$$
(One can treat this model in the following way: one  builds
the uniformly charged sphere in laboratory reference frame and then
begins it to accelerate in the  way that the charge density
in laboratory frame is described by the above equation while in sphere
self-frame charge density can be calculated by standard
tensor coordinate transformations.)

 In the
case of shell rectilinear motion this force has the form [9,13]
$$F_{self}={Q^2 \over 4 a^2}\left[ -c \int\limits_{T^{-}}^{T^{+}} dT
{cT-2a \over L^2} + \ln {{L^{+}\over L^{-}}} + ({1\over
\beta^2}-1)\ln { {1+\beta \over 1-\beta}} -{2\over \beta} \right]
\eqno(1)$$ here $cT^{\pm}=2a \pm L^{\pm},\ \
L^{\pm}=R(t)-R(t-T^{\pm}),\ \ L=R(t)-R(t-T),\ \ \beta=v/c,\ \ v=dR/dt
$.

The total shell equation of motion then will be $$m{d \over
dt}(\gamma v)=F_{self}  \eqno(2)$$ Here $m$ - is the "mechanical"
shell mass.

This equation has one trivial solution - the uniform motion without
radiation:
$R(t)=R_0+vt. $

Introducing dimensionless variables $y= R/2a,\ \ x=ct/2a$ one can
rewrite the shell equation of motion (2) in the form
$${d^2 y \over dx^2} =\left(1-({dy\over dx})^2\right)^{3/2} k \cdot$$
$$\cdot \left[ - \int\limits_{x^{-}}^{x^{+}} dz
{z-1 \over L^2} + \ln {{L^{+}\over L^{-}}} + ({1\over
\beta^2}-1)\ln { {1+\beta \over 1-\beta}} -{2\over \beta} \right]
\eqno(3)$$ here $$x^{\pm}=1 \pm L^{\pm},\ \
L^{\pm}=y(x)-y(x-x^{\pm}),\ \ L=y(x)-y(x-z),$$
$$ \beta=dy/dx, \
\ \ \ k= {Q^2 \over 2 m c^2 a}.$$

Lets take the charged sphere of diameter $2a$ equal to the 
"particle radius" ${Q^2 \over  m c^2 }$: $$k=1.$$

Lets place this sphere into the Coulomb field  of
charge $q$. Then the equation of motion of such
central-force problem reads
$${d^2 y \over dx^2} =\left(1-({dy\over dx})^2\right)^{3/2}  \cdot$$
$$\cdot \left[ - \int\limits_{x^{-}}^{x^{+}} dz
{z-1 \over L^2} + \ln {{L^{+}\over L^{-}}} + ({1\over
\beta^2}-1)\ln { {1+\beta \over 1-\beta}} -{2\over \beta}  +
{M\over (y-d)^2} \right] \eqno(4)$$ here   $d$ - is coordinate of
Coulomb center, $M=q/Q$ .

It is useful to compare solutions of (4) with  point
charge motion in the same field, governed by the following
relativistic equation without radiation force:
$${d^2 y \over dx^2} =\left(1-({dy\over dx})^2\right)^{3/2}  \cdot
\left[    {M\over (y-d)^2} \right] \eqno(5)$$
\vskip 0.5 cm {\bf A.}

We integrated eq.(4,5) in the repulsive case numerically with the
following initial data:

(i) Coulomb center is placed at $d=5.0$;

(ii) initial value of coordinates of the point particle and of sphere
center of mass is $y=0.0$;

(iii) initial sphere and point particle velocities ${dy\over dx}$ are
zero (and ${dy\over dx} =0.0 $ for $x<0.0$);

(iv) $M$ is taken equal to $1.0$ and to $0.1$.

Numerical results are shown on figs. (A.1-A.3):

curves $vz,\ \ vq$  correspond to velocities of
Sommerfeld sphere and of point particle (Fig. A.1 for $M=1.0$ and
Fig. A.2 for $M=0.1$);

curves $wz,\ \ wq$  correspond to accelerations of  Sommerfeld
sphere and of point charge (Fig. A.3 for  $M=1.0$);

 horizontal axis is $x$.

One can see that there is the following  main property of motion
of Sommerfeld sphere:

 sphere  gains velocity less then that should be
according to relativistic equation of motion of point charge without
radiation reaction.

This result  one can explain as simple consequence of effect of
retardation.

\vskip 0.5 cm
{\bf B.}

In the attractive case we numerically intergated eq.(4,5)
with the following initial data:

(i) Coulomb center is placed at $d=5.0$;

(ii) initial value of coordinates of the point particle and of sphere
center of mass is $y=0.0$;

(iii) initial sphere and point particle velocities ${dy\over dx}$ are
zero (and ${dy\over dx} =0.0 $ for $x<0.0$);

(iv) $M$ is taken equal to $-1.0$;

curve $vz$  corresponds to velocity of
Sommerfeld sphere;

curve $vq$  corresponds to velocity  of  point
charge;

 horizontal axis is $x$.

Numerical results are shown on fig. (B.1).

One can see that Sommerfeld sphere indeed falls on the Coulomb
center, so there is physical trajectory contrary to the
motion of point charge governed by Lorentz-Dirac equation.

Thus we conclude that extended  radiating object can solve
problems  of  Lorentz-Dirac approach.  This happens thanks to
the fact that equations of motion of extended objects are not
analytic near the zero value of their size ($a=0$) and thus
equations with $a=0$ and $a\to 0$ are essentially different equations
with different physical solutions.

I am glad to thank my colleagues:

P.A.Polyakov - for theoretical discussions;

V.A.Iljina and P.K.Silaev - for assistance rendered during numerical
calculations.

  \vskip 0.5 cm
 \centerline {\bf{REFERENCES}}

  \begin{enumerate}
\item J.Huschilt, W.E.Baylis,  Phys.Rev., D13, n
12, 3257 (1976).
\item W.E.Baylis, J.Huschilt, Phys.Rev., D13, n
12, 3262 (1976).
\item J.Huschilt, W.E.Baylis,  Phys.Rev., D17, n
4, 985 (1978).
  \item J.C.Kasher, Phys.Rev., D14, n 4, 393 (1976).
  \item E.Comay, J.Phys.A, 29, 2111 (1996).
\item S.Parrott, {\it
Relativistic Electrodynamics and Differential Geometry},
 Springer-Verlag, NY, 1987.

\item L.Landau, E.Lifshitz, {\it The Classical Theory of Fields},
Addison-Wesley, 1961.
\item I.Prigogine, F.Henin, Acad. Roy. Belgique, Sciences, T.35,
fasc. 7, 1965.
  \item P.Pearle, in {\it Electromagnetism}, ed.
D.Tepliz, Plenum, NY, 1982, p.211.  \item A.D.Yaghjian, {\it
Relativistic Dynamics of a Charged Sphere}, Lecture Notes in Physics,
11, Springer, Berlin, 1992.  \item F.Rohrlich, Am.J.Physics, 65(11),
1051 (1997).

\item A.Sommerfeld, Gottingen Nachrichten, 29 (1904), 363 (1904), 201
  (1905).
\item Alexander A.Vlasov, physics/9811019.

\end{enumerate}

\eject
\newcount\numpoint
\newcount\numpointo
\numpoint=1 \numpointo=1
\def\emmoveto#1#2{\offinterlineskip
\hbox to 0 true cm{\vbox to 0
true cm{\vskip - #2 true mm
\hskip #1 true mm \special{em:point
\the\numpoint}\vss}\hss}
\numpointo=\numpoint
\global\advance \numpoint by 1}
\def\emlineto#1#2{\offinterlineskip
\hbox to 0 true cm{\vbox to 0
true cm{\vskip - #2 true mm
\hskip #1 true mm \special{em:point
\the\numpoint}\vss}\hss}
\special{em:line
\the\numpointo,\the\numpoint}
\numpointo=\numpoint
\global\advance \numpoint by 1}
\def\emshow#1#2#3{\offinterlineskip
\hbox to 0 true cm{\vbox to 0
true cm{\vskip - #2 true mm
\hskip #1 true mm \vbox to 0
true cm{\vss\hbox{#3\hss
}}\vss}\hss}}
\special{em:linewidth 0.8pt}

\vrule width 0 mm height                0 mm depth 90.000 true mm

\special{em:linewidth 0.8pt}
\emmoveto{130.000}{10.000}
\emlineto{12.000}{10.000}
\emlineto{12.000}{80.000}
\emmoveto{71.000}{10.000}
\emlineto{71.000}{80.000}
\emmoveto{12.000}{45.000}
\emlineto{130.000}{45.000}
\emmoveto{130.000}{10.000}
\emlineto{130.000}{80.000}
\emlineto{12.000}{80.000}
\emlineto{12.000}{10.000}
\emlineto{130.000}{10.000}
\special{em:linewidth 0.4pt}
\emmoveto{12.000}{17.000}
\emlineto{130.000}{17.000}
\emmoveto{12.000}{24.000}
\emlineto{130.000}{24.000}
\emmoveto{12.000}{31.000}
\emlineto{130.000}{31.000}
\emmoveto{12.000}{38.000}
\emlineto{130.000}{38.000}
\emmoveto{12.000}{45.000}
\emlineto{130.000}{45.000}
\emmoveto{12.000}{52.000}
\emlineto{130.000}{52.000}
\emmoveto{12.000}{59.000}
\emlineto{130.000}{59.000}
\emmoveto{12.000}{66.000}
\emlineto{130.000}{66.000}
\emmoveto{12.000}{73.000}
\emlineto{130.000}{73.000}
\emmoveto{23.800}{10.000}
\emlineto{23.800}{80.000}
\emmoveto{35.600}{10.000}
\emlineto{35.600}{80.000}
\emmoveto{47.400}{10.000}
\emlineto{47.400}{80.000}
\emmoveto{59.200}{10.000}
\emlineto{59.200}{80.000}
\emmoveto{71.000}{10.000}
\emlineto{71.000}{80.000}
\emmoveto{82.800}{10.000}
\emlineto{82.800}{80.000}
\emmoveto{94.600}{10.000}
\emlineto{94.600}{80.000}
\emmoveto{106.400}{10.000}
\emlineto{106.400}{80.000}
\emmoveto{118.200}{10.000}
\emlineto{118.200}{80.000}
\special{em:linewidth 0.8pt}
\emmoveto{12.000}{80.000}
\emlineto{12.295}{79.775}
\emmoveto{12.295}{79.765}
\emlineto{12.590}{79.549}
\emmoveto{12.590}{79.539}
\emlineto{12.885}{79.332}
\emmoveto{12.885}{79.322}
\emlineto{13.180}{79.123}
\emmoveto{13.180}{79.113}
\emlineto{13.475}{78.922}
\emmoveto{13.475}{78.912}
\emlineto{13.770}{78.730}
\emmoveto{13.770}{78.720}
\emlineto{14.065}{78.545}
\emmoveto{14.065}{78.535}
\emlineto{14.360}{78.367}
\emmoveto{14.360}{78.357}
\emlineto{14.655}{78.196}
\emmoveto{14.655}{78.186}
\emlineto{14.950}{78.032}
\emmoveto{14.950}{78.022}
\emlineto{15.245}{77.875}
\emmoveto{15.245}{77.865}
\emlineto{15.540}{77.724}
\emmoveto{15.540}{77.714}
\emlineto{15.835}{77.578}
\emmoveto{15.835}{77.568}
\emlineto{16.130}{77.439}
\emmoveto{16.130}{77.429}
\emlineto{16.425}{77.305}
\emmoveto{16.425}{77.295}
\emlineto{16.720}{77.176}
\emmoveto{16.720}{77.166}
\emlineto{17.015}{77.052}
\emmoveto{17.015}{77.042}
\emlineto{17.310}{76.933}
\emmoveto{17.310}{76.923}
\emlineto{17.605}{76.819}
\emmoveto{17.605}{76.809}
\emlineto{17.900}{76.710}
\emmoveto{17.900}{76.700}
\emlineto{18.195}{76.604}
\emmoveto{18.195}{76.594}
\emlineto{18.490}{76.503}
\emmoveto{18.490}{76.493}
\emlineto{18.785}{76.406}
\emmoveto{18.785}{76.396}
\emlineto{19.080}{76.313}
\emmoveto{19.080}{76.303}
\emlineto{19.375}{76.223}
\emmoveto{19.375}{76.213}
\emlineto{19.670}{76.138}
\emmoveto{19.670}{76.128}
\emlineto{19.965}{76.055}
\emmoveto{19.965}{76.045}
\emlineto{20.260}{75.976}
\emmoveto{20.260}{75.966}
\emlineto{20.555}{75.900}
\emmoveto{20.555}{75.890}
\emlineto{20.850}{75.827}
\emmoveto{20.850}{75.817}
\emlineto{21.145}{75.757}
\emmoveto{21.145}{75.747}
\emlineto{21.440}{75.689}
\emmoveto{21.440}{75.679}
\emlineto{21.735}{75.625}
\emmoveto{21.735}{75.615}
\emlineto{22.030}{75.557}
\emmoveto{22.030}{75.547}
\emlineto{22.325}{75.487}
\emmoveto{22.325}{75.477}
\emlineto{22.620}{75.411}
\emmoveto{22.620}{75.401}
\emlineto{22.915}{75.329}
\emmoveto{22.915}{75.319}
\emlineto{23.210}{75.243}
\emmoveto{23.210}{75.233}
\emlineto{23.505}{75.153}
\emmoveto{23.505}{75.143}
\emlineto{23.800}{75.058}
\emmoveto{23.800}{75.048}
\emlineto{24.095}{74.960}
\emmoveto{24.095}{74.950}
\emlineto{24.390}{74.859}
\emmoveto{24.390}{74.849}
\emlineto{24.685}{74.756}
\emmoveto{24.685}{74.746}
\emlineto{24.980}{74.650}
\emmoveto{24.980}{74.640}
\emlineto{25.275}{74.543}
\emmoveto{25.275}{74.533}
\emlineto{25.570}{74.433}
\emmoveto{25.570}{74.423}
\emlineto{25.865}{74.323}
\emmoveto{25.865}{74.313}
\emlineto{26.160}{74.211}
\emmoveto{26.160}{74.201}
\emlineto{26.455}{74.099}
\emmoveto{26.455}{74.089}
\emlineto{26.750}{73.986}
\emmoveto{26.750}{73.976}
\emlineto{27.045}{73.873}
\emmoveto{27.045}{73.863}
\emlineto{27.340}{73.760}
\emmoveto{27.340}{73.750}
\emlineto{27.635}{73.647}
\emmoveto{27.635}{73.637}
\emlineto{27.930}{73.534}
\emmoveto{27.930}{73.524}
\emlineto{28.225}{73.421}
\emmoveto{28.225}{73.411}
\emlineto{28.520}{73.310}
\emmoveto{28.520}{73.300}
\emlineto{28.815}{73.198}
\emmoveto{28.815}{73.188}
\emlineto{29.110}{73.088}
\emmoveto{29.110}{73.078}
\emlineto{29.405}{72.979}
\emmoveto{29.405}{72.969}
\emlineto{29.700}{72.870}
\emmoveto{29.700}{72.860}
\emlineto{29.995}{72.763}
\emmoveto{29.995}{72.753}
\emlineto{30.290}{72.657}
\emmoveto{30.290}{72.647}
\emlineto{30.585}{72.553}
\emmoveto{30.585}{72.543}
\emlineto{30.880}{72.449}
\emmoveto{30.880}{72.439}
\emlineto{31.175}{72.347}
\emmoveto{31.175}{72.337}
\emlineto{31.470}{72.247}
\emmoveto{31.470}{72.237}
\emlineto{31.765}{72.148}
\emmoveto{31.765}{72.138}
\emlineto{32.060}{72.050}
\emmoveto{32.060}{72.040}
\emlineto{32.355}{71.954}
\emmoveto{32.355}{71.944}
\emlineto{32.650}{71.858}
\emmoveto{32.650}{71.848}
\emlineto{32.945}{71.764}
\emmoveto{32.945}{71.754}
\emlineto{33.240}{71.669}
\emmoveto{33.240}{71.659}
\emlineto{33.535}{71.575}
\emmoveto{33.535}{71.565}
\emlineto{33.830}{71.481}
\emmoveto{33.830}{71.471}
\emlineto{34.125}{71.387}
\emmoveto{34.125}{71.377}
\emlineto{34.420}{71.292}
\emmoveto{34.420}{71.282}
\emlineto{34.715}{71.198}
\emmoveto{34.715}{71.188}
\emlineto{35.010}{71.103}
\emmoveto{35.010}{71.093}
\emlineto{35.305}{71.008}
\emmoveto{35.305}{70.998}
\emlineto{35.600}{70.912}
\emmoveto{35.600}{70.902}
\emlineto{35.895}{70.816}
\emmoveto{35.895}{70.806}
\emlineto{36.190}{70.719}
\emmoveto{36.190}{70.709}
\emlineto{36.485}{70.622}
\emmoveto{36.485}{70.612}
\emlineto{36.780}{70.525}
\emmoveto{36.780}{70.515}
\emlineto{37.075}{70.427}
\emmoveto{37.075}{70.417}
\emlineto{37.370}{70.328}
\emmoveto{37.370}{70.318}
\emlineto{37.665}{70.229}
\emmoveto{37.665}{70.219}
\emlineto{37.960}{70.130}
\emmoveto{37.960}{70.120}
\emlineto{38.255}{70.030}
\emmoveto{38.255}{70.020}
\emlineto{38.550}{69.931}
\emmoveto{38.550}{69.921}
\emlineto{38.845}{69.830}
\emmoveto{38.845}{69.820}
\emlineto{39.140}{69.730}
\emmoveto{39.140}{69.720}
\emlineto{39.435}{69.629}
\emmoveto{39.435}{69.619}
\emlineto{39.730}{69.528}
\emmoveto{39.730}{69.518}
\emlineto{40.025}{69.427}
\emmoveto{40.025}{69.417}
\emlineto{40.320}{69.327}
\emmoveto{40.320}{69.317}
\emlineto{40.615}{69.226}
\emmoveto{40.615}{69.216}
\emlineto{40.910}{69.125}
\emmoveto{40.910}{69.115}
\emlineto{41.205}{69.024}
\emmoveto{41.205}{69.014}
\emlineto{41.500}{68.924}
\emmoveto{41.500}{68.914}
\emlineto{41.795}{68.823}
\emmoveto{41.795}{68.813}
\emlineto{42.090}{68.723}
\emmoveto{42.090}{68.713}
\emlineto{42.385}{68.624}
\emmoveto{42.385}{68.614}
\emlineto{42.680}{68.524}
\emmoveto{42.680}{68.514}
\emlineto{42.975}{68.425}
\emmoveto{42.975}{68.415}
\emlineto{43.270}{68.326}
\emmoveto{43.270}{68.316}
\emlineto{43.565}{68.228}
\emmoveto{43.565}{68.218}
\emlineto{43.860}{68.130}
\emmoveto{43.860}{68.120}
\emlineto{44.155}{68.032}
\emmoveto{44.155}{68.022}
\emlineto{44.450}{67.935}
\emmoveto{44.450}{67.925}
\emlineto{44.745}{67.838}
\emmoveto{44.745}{67.828}
\emlineto{45.040}{67.741}
\emmoveto{45.040}{67.731}
\emlineto{45.335}{67.644}
\emmoveto{45.335}{67.634}
\emlineto{45.630}{67.548}
\emmoveto{45.630}{67.538}
\emlineto{45.925}{67.452}
\emmoveto{45.925}{67.442}
\emlineto{46.220}{67.356}
\emmoveto{46.220}{67.346}
\emlineto{46.515}{67.260}
\emmoveto{46.515}{67.250}
\emlineto{46.810}{67.164}
\emmoveto{46.810}{67.154}
\emlineto{47.105}{67.068}
\emmoveto{47.105}{67.058}
\emlineto{47.400}{66.973}
\emmoveto{47.400}{66.963}
\emlineto{47.695}{66.877}
\emmoveto{47.695}{66.867}
\emlineto{47.990}{66.782}
\emmoveto{47.990}{66.772}
\emlineto{48.285}{66.686}
\emmoveto{48.285}{66.676}
\emlineto{48.580}{66.591}
\emmoveto{48.580}{66.581}
\emlineto{48.875}{66.495}
\emmoveto{48.875}{66.485}
\emlineto{49.170}{66.400}
\emmoveto{49.170}{66.390}
\emlineto{49.465}{66.304}
\emmoveto{49.465}{66.294}
\emlineto{49.760}{66.209}
\emmoveto{49.760}{66.199}
\emlineto{50.055}{66.113}
\emmoveto{50.055}{66.103}
\emlineto{50.350}{66.018}
\emmoveto{50.350}{66.008}
\emlineto{50.645}{65.922}
\emmoveto{50.645}{65.912}
\emlineto{50.940}{65.827}
\emmoveto{50.940}{65.817}
\emlineto{51.235}{65.731}
\emmoveto{51.235}{65.721}
\emlineto{51.530}{65.636}
\emmoveto{51.530}{65.626}
\emlineto{51.825}{65.540}
\emmoveto{51.825}{65.530}
\emlineto{52.120}{65.445}
\emmoveto{52.120}{65.435}
\emlineto{52.415}{65.349}
\emmoveto{52.415}{65.339}
\emlineto{52.710}{65.254}
\emmoveto{52.710}{65.244}
\emlineto{53.005}{65.159}
\emmoveto{53.005}{65.149}
\emlineto{53.300}{65.063}
\emmoveto{53.300}{65.053}
\emlineto{53.595}{64.968}
\emmoveto{53.595}{64.958}
\emlineto{53.890}{64.873}
\emmoveto{53.890}{64.863}
\emlineto{54.185}{64.778}
\emmoveto{54.185}{64.768}
\emlineto{54.480}{64.683}
\emmoveto{54.480}{64.673}
\emlineto{54.775}{64.589}
\emmoveto{54.775}{64.579}
\emlineto{55.070}{64.494}
\emmoveto{55.070}{64.484}
\emlineto{55.365}{64.400}
\emmoveto{55.365}{64.390}
\emlineto{55.660}{64.305}
\emmoveto{55.660}{64.295}
\emlineto{55.955}{64.211}
\emmoveto{55.955}{64.201}
\emlineto{56.250}{64.117}
\emmoveto{56.250}{64.107}
\emlineto{56.545}{64.023}
\emmoveto{56.545}{64.013}
\emlineto{56.840}{63.930}
\emmoveto{56.840}{63.920}
\emlineto{57.135}{63.836}
\emmoveto{57.135}{63.826}
\emlineto{57.430}{63.743}
\emmoveto{57.430}{63.733}
\emlineto{57.725}{63.649}
\emmoveto{57.725}{63.639}
\emlineto{58.020}{63.556}
\emmoveto{58.020}{63.546}
\emlineto{58.315}{63.463}
\emmoveto{58.315}{63.453}
\emlineto{58.610}{63.370}
\emmoveto{58.610}{63.360}
\emlineto{58.905}{63.277}
\emmoveto{58.905}{63.267}
\emlineto{59.200}{63.185}
\emmoveto{59.200}{63.175}
\emlineto{59.495}{63.092}
\emmoveto{59.495}{63.082}
\emlineto{59.790}{63.000}
\emmoveto{59.790}{62.990}
\emlineto{60.085}{62.908}
\emmoveto{60.085}{62.898}
\emlineto{60.380}{62.815}
\emmoveto{60.380}{62.805}
\emlineto{60.675}{62.723}
\emmoveto{60.675}{62.713}
\emlineto{60.970}{62.631}
\emmoveto{60.970}{62.621}
\emlineto{61.265}{62.539}
\emmoveto{61.265}{62.529}
\emlineto{61.560}{62.448}
\emmoveto{61.560}{62.438}
\emlineto{61.855}{62.356}
\emmoveto{61.855}{62.346}
\emlineto{62.150}{62.264}
\emmoveto{62.150}{62.254}
\emlineto{62.445}{62.173}
\emmoveto{62.445}{62.163}
\emlineto{62.740}{62.081}
\emmoveto{62.740}{62.071}
\emlineto{63.035}{61.990}
\emmoveto{63.035}{61.980}
\emlineto{63.330}{61.899}
\emmoveto{63.330}{61.889}
\emlineto{63.625}{61.807}
\emmoveto{63.625}{61.797}
\emlineto{63.920}{61.716}
\emmoveto{63.920}{61.706}
\emlineto{64.215}{61.625}
\emmoveto{64.215}{61.615}
\emlineto{64.510}{61.534}
\emmoveto{64.510}{61.524}
\emlineto{64.805}{61.444}
\emmoveto{64.805}{61.434}
\emlineto{65.100}{61.353}
\emmoveto{65.100}{61.343}
\emlineto{65.395}{61.262}
\emmoveto{65.395}{61.252}
\emlineto{65.690}{61.172}
\emmoveto{65.690}{61.162}
\emlineto{65.985}{61.081}
\emmoveto{65.985}{61.071}
\emlineto{66.280}{60.991}
\emmoveto{66.280}{60.981}
\emlineto{66.575}{60.901}
\emmoveto{66.575}{60.891}
\emlineto{66.870}{60.811}
\emmoveto{66.870}{60.801}
\emlineto{67.165}{60.721}
\emmoveto{67.165}{60.711}
\emlineto{67.460}{60.631}
\emmoveto{67.460}{60.621}
\emlineto{67.755}{60.541}
\emmoveto{67.755}{60.531}
\emlineto{68.050}{60.452}
\emmoveto{68.050}{60.442}
\emlineto{68.345}{60.362}
\emmoveto{68.345}{60.352}
\emlineto{68.640}{60.273}
\emmoveto{68.640}{60.263}
\emlineto{68.935}{60.184}
\emmoveto{68.935}{60.174}
\emlineto{69.230}{60.094}
\emmoveto{69.230}{60.084}
\emlineto{69.525}{60.005}
\emmoveto{69.525}{59.995}
\emlineto{69.820}{59.917}
\emmoveto{69.820}{59.907}
\emlineto{70.115}{59.828}
\emmoveto{70.115}{59.818}
\emlineto{70.410}{59.739}
\emmoveto{70.410}{59.729}
\emlineto{70.705}{59.651}
\emmoveto{70.705}{59.641}
\emlineto{71.000}{59.562}
\emmoveto{71.000}{59.552}
\emlineto{71.295}{59.474}
\emmoveto{71.295}{59.464}
\emlineto{71.590}{59.386}
\emmoveto{71.590}{59.376}
\emlineto{71.885}{59.298}
\emmoveto{71.885}{59.288}
\emlineto{72.180}{59.210}
\emmoveto{72.180}{59.200}
\emlineto{72.475}{59.122}
\emmoveto{72.475}{59.112}
\emlineto{72.770}{59.034}
\emmoveto{72.770}{59.024}
\emlineto{73.065}{58.947}
\emmoveto{73.065}{58.937}
\emlineto{73.360}{58.859}
\emmoveto{73.360}{58.849}
\emlineto{73.655}{58.772}
\emmoveto{73.655}{58.762}
\emlineto{73.950}{58.685}
\emmoveto{73.950}{58.675}
\emlineto{74.245}{58.598}
\emmoveto{74.245}{58.588}
\emlineto{74.540}{58.511}
\emmoveto{74.540}{58.501}
\emlineto{74.835}{58.424}
\emmoveto{74.835}{58.414}
\emlineto{75.130}{58.337}
\emmoveto{75.130}{58.327}
\emlineto{75.425}{58.251}
\emmoveto{75.425}{58.241}
\emlineto{75.720}{58.164}
\emmoveto{75.720}{58.154}
\emlineto{76.015}{58.078}
\emmoveto{76.015}{58.068}
\emlineto{76.310}{57.992}
\emmoveto{76.310}{57.982}
\emlineto{76.605}{57.905}
\emmoveto{76.605}{57.895}
\emlineto{76.900}{57.819}
\emmoveto{76.900}{57.809}
\emlineto{77.195}{57.734}
\emmoveto{77.195}{57.724}
\emlineto{77.490}{57.648}
\emmoveto{77.490}{57.638}
\emlineto{77.785}{57.562}
\emmoveto{77.785}{57.552}
\emlineto{78.080}{57.477}
\emmoveto{78.080}{57.467}
\emlineto{78.375}{57.391}
\emmoveto{78.375}{57.381}
\emlineto{78.670}{57.306}
\emmoveto{78.670}{57.296}
\emlineto{78.965}{57.221}
\emmoveto{78.965}{57.211}
\emlineto{79.260}{57.136}
\emmoveto{79.260}{57.126}
\emlineto{79.555}{57.051}
\emmoveto{79.555}{57.041}
\emlineto{79.850}{56.966}
\emmoveto{79.850}{56.956}
\emlineto{80.145}{56.881}
\emmoveto{80.145}{56.871}
\emlineto{80.440}{56.796}
\emmoveto{80.440}{56.786}
\emlineto{80.735}{56.712}
\emmoveto{80.735}{56.702}
\emlineto{81.030}{56.628}
\emmoveto{81.030}{56.618}
\emlineto{81.325}{56.543}
\emmoveto{81.325}{56.533}
\emlineto{81.620}{56.459}
\emmoveto{81.620}{56.449}
\emlineto{81.915}{56.375}
\emmoveto{81.915}{56.365}
\emlineto{82.210}{56.292}
\emmoveto{82.210}{56.282}
\emlineto{82.505}{56.208}
\emmoveto{82.505}{56.198}
\emlineto{82.800}{56.124}
\emmoveto{82.800}{56.114}
\emlineto{83.095}{56.041}
\emmoveto{83.095}{56.031}
\emlineto{83.390}{55.957}
\emmoveto{83.390}{55.947}
\emlineto{83.685}{55.874}
\emmoveto{83.685}{55.864}
\emlineto{83.980}{55.791}
\emmoveto{83.980}{55.781}
\emlineto{84.275}{55.708}
\emmoveto{84.275}{55.698}
\emlineto{84.570}{55.625}
\emmoveto{84.570}{55.615}
\emlineto{84.865}{55.543}
\emmoveto{84.865}{55.533}
\emlineto{85.160}{55.460}
\emmoveto{85.160}{55.450}
\emlineto{85.455}{55.378}
\emmoveto{85.455}{55.368}
\emlineto{85.750}{55.295}
\emmoveto{85.750}{55.285}
\emlineto{86.045}{55.213}
\emmoveto{86.045}{55.203}
\emlineto{86.340}{55.131}
\emmoveto{86.340}{55.121}
\emlineto{86.635}{55.049}
\emmoveto{86.635}{55.039}
\emlineto{86.930}{54.967}
\emmoveto{86.930}{54.957}
\emlineto{87.225}{54.885}
\emmoveto{87.225}{54.875}
\emlineto{87.520}{54.804}
\emmoveto{87.520}{54.794}
\emlineto{87.815}{54.722}
\emmoveto{87.815}{54.712}
\emlineto{88.110}{54.641}
\emmoveto{88.110}{54.631}
\emlineto{88.405}{54.560}
\emmoveto{88.405}{54.550}
\emlineto{88.700}{54.479}
\emmoveto{88.700}{54.469}
\emlineto{88.995}{54.398}
\emmoveto{88.995}{54.388}
\emlineto{89.290}{54.317}
\emmoveto{89.290}{54.307}
\emlineto{89.585}{54.236}
\emmoveto{89.585}{54.226}
\emlineto{89.880}{54.156}
\emmoveto{89.880}{54.146}
\emlineto{90.175}{54.075}
\emmoveto{90.175}{54.065}
\emlineto{90.470}{53.995}
\emmoveto{90.470}{53.985}
\emlineto{90.765}{53.915}
\emmoveto{90.765}{53.905}
\emlineto{91.060}{53.835}
\emmoveto{91.060}{53.825}
\emlineto{91.355}{53.755}
\emmoveto{91.355}{53.745}
\emlineto{91.650}{53.675}
\emmoveto{91.650}{53.665}
\emlineto{91.945}{53.595}
\emmoveto{91.945}{53.585}
\emlineto{92.240}{53.516}
\emmoveto{92.240}{53.506}
\emlineto{92.535}{53.436}
\emmoveto{92.535}{53.426}
\emlineto{92.830}{53.357}
\emmoveto{92.830}{53.347}
\emlineto{93.125}{53.278}
\emmoveto{93.125}{53.268}
\emlineto{93.420}{53.199}
\emmoveto{93.420}{53.189}
\emlineto{93.715}{53.120}
\emmoveto{93.715}{53.110}
\emlineto{94.010}{53.041}
\emmoveto{94.010}{53.031}
\emlineto{94.305}{52.962}
\emmoveto{94.305}{52.952}
\emlineto{94.600}{52.884}
\emmoveto{94.600}{52.874}
\emlineto{94.895}{52.805}
\emmoveto{94.895}{52.795}
\emlineto{95.190}{52.727}
\emmoveto{95.190}{52.717}
\emlineto{95.485}{52.649}
\emmoveto{95.485}{52.639}
\emlineto{95.780}{52.571}
\emmoveto{95.780}{52.561}
\emlineto{96.075}{52.493}
\emmoveto{96.075}{52.483}
\emlineto{96.370}{52.415}
\emmoveto{96.370}{52.405}
\emlineto{96.665}{52.337}
\emmoveto{96.665}{52.327}
\emlineto{96.960}{52.260}
\emmoveto{96.960}{52.250}
\emlineto{97.255}{52.183}
\emmoveto{97.255}{52.173}
\emlineto{97.550}{52.105}
\emmoveto{97.550}{52.095}
\emlineto{97.845}{52.028}
\emmoveto{97.845}{52.018}
\emlineto{98.140}{51.951}
\emmoveto{98.140}{51.941}
\emlineto{98.435}{51.874}
\emmoveto{98.435}{51.864}
\emlineto{98.730}{51.797}
\emmoveto{98.730}{51.787}
\emlineto{99.025}{51.721}
\emmoveto{99.025}{51.711}
\emlineto{99.320}{51.644}
\emmoveto{99.320}{51.634}
\emlineto{99.615}{51.568}
\emmoveto{99.615}{51.558}
\emlineto{99.910}{51.492}
\emmoveto{99.910}{51.482}
\emlineto{100.205}{51.416}
\emmoveto{100.205}{51.406}
\emlineto{100.500}{51.340}
\emmoveto{100.500}{51.330}
\emlineto{100.795}{51.264}
\emmoveto{100.795}{51.254}
\emlineto{101.090}{51.188}
\emmoveto{101.090}{51.178}
\emlineto{101.385}{51.112}
\emmoveto{101.385}{51.102}
\emlineto{101.680}{51.037}
\emmoveto{101.680}{51.027}
\emlineto{101.975}{50.962}
\emmoveto{101.975}{50.952}
\emlineto{102.270}{50.886}
\emmoveto{102.270}{50.876}
\emlineto{102.565}{50.811}
\emmoveto{102.565}{50.801}
\emlineto{102.860}{50.736}
\emmoveto{102.860}{50.726}
\emlineto{103.155}{50.661}
\emmoveto{103.155}{50.651}
\emlineto{103.450}{50.587}
\emmoveto{103.450}{50.577}
\emlineto{103.745}{50.512}
\emmoveto{103.745}{50.502}
\emlineto{104.040}{50.438}
\emmoveto{104.040}{50.428}
\emlineto{104.335}{50.363}
\emmoveto{104.335}{50.353}
\emlineto{104.630}{50.289}
\emmoveto{104.630}{50.279}
\emlineto{104.925}{50.215}
\emmoveto{104.925}{50.205}
\emlineto{105.220}{50.141}
\emmoveto{105.220}{50.131}
\emlineto{105.515}{50.067}
\emmoveto{105.515}{50.057}
\emlineto{105.810}{49.994}
\emmoveto{105.810}{49.984}
\emlineto{106.105}{49.920}
\emmoveto{106.105}{49.910}
\emlineto{106.400}{49.847}
\emmoveto{106.400}{49.837}
\emlineto{106.695}{49.773}
\emmoveto{106.695}{49.763}
\emlineto{106.990}{49.700}
\emmoveto{106.990}{49.690}
\emlineto{107.285}{49.627}
\emmoveto{107.285}{49.617}
\emlineto{107.580}{49.554}
\emmoveto{107.580}{49.544}
\emlineto{107.875}{49.482}
\emmoveto{107.875}{49.472}
\emlineto{108.170}{49.409}
\emmoveto{108.170}{49.399}
\emlineto{108.465}{49.336}
\emmoveto{108.465}{49.326}
\emlineto{108.760}{49.264}
\emmoveto{108.760}{49.254}
\emlineto{109.055}{49.192}
\emmoveto{109.055}{49.182}
\emlineto{109.350}{49.119}
\emmoveto{109.350}{49.109}
\emlineto{109.645}{49.047}
\emmoveto{109.645}{49.037}
\emlineto{109.940}{48.976}
\emmoveto{109.940}{48.966}
\emlineto{110.235}{48.904}
\emmoveto{110.235}{48.894}
\emlineto{110.530}{48.832}
\emmoveto{110.530}{48.822}
\emlineto{110.825}{48.761}
\emmoveto{110.825}{48.751}
\emlineto{111.120}{48.689}
\emmoveto{111.120}{48.679}
\emlineto{111.415}{48.618}
\emmoveto{111.415}{48.608}
\emlineto{111.710}{48.547}
\emmoveto{111.710}{48.537}
\emlineto{112.005}{48.476}
\emmoveto{112.005}{48.466}
\emlineto{112.300}{48.405}
\emmoveto{112.300}{48.395}
\emlineto{112.595}{48.334}
\emmoveto{112.595}{48.324}
\emlineto{112.890}{48.264}
\emmoveto{112.890}{48.254}
\emlineto{113.185}{48.193}
\emmoveto{113.185}{48.183}
\emlineto{113.480}{48.123}
\emmoveto{113.480}{48.113}
\emlineto{113.775}{48.052}
\emmoveto{113.775}{48.042}
\emlineto{114.070}{47.982}
\emmoveto{114.070}{47.972}
\emlineto{114.365}{47.912}
\emmoveto{114.365}{47.902}
\emlineto{114.660}{47.842}
\emmoveto{114.660}{47.832}
\emlineto{114.955}{47.773}
\emmoveto{114.955}{47.763}
\emlineto{115.250}{47.703}
\emmoveto{115.250}{47.693}
\emlineto{115.545}{47.634}
\emmoveto{115.545}{47.624}
\emlineto{115.840}{47.564}
\emmoveto{115.840}{47.554}
\emlineto{116.135}{47.495}
\emmoveto{116.135}{47.485}
\emlineto{116.430}{47.426}
\emmoveto{116.430}{47.416}
\emlineto{116.725}{47.357}
\emmoveto{116.725}{47.347}
\emlineto{117.020}{47.288}
\emmoveto{117.020}{47.278}
\emlineto{117.315}{47.219}
\emmoveto{117.315}{47.209}
\emlineto{117.610}{47.151}
\emmoveto{117.610}{47.141}
\emlineto{117.905}{47.082}
\emmoveto{117.905}{47.072}
\emlineto{118.200}{47.014}
\emmoveto{118.200}{47.004}
\emlineto{118.495}{46.945}
\emmoveto{118.495}{46.935}
\emlineto{118.790}{46.877}
\emmoveto{118.790}{46.867}
\emlineto{119.085}{46.809}
\emmoveto{119.085}{46.799}
\emlineto{119.380}{46.742}
\emmoveto{119.380}{46.732}
\emlineto{119.675}{46.674}
\emmoveto{119.675}{46.664}
\emlineto{119.970}{46.606}
\emmoveto{119.970}{46.596}
\emlineto{120.265}{46.539}
\emmoveto{120.265}{46.529}
\emlineto{120.560}{46.471}
\emmoveto{120.560}{46.461}
\emlineto{120.855}{46.404}
\emmoveto{120.855}{46.394}
\emlineto{121.150}{46.337}
\emmoveto{121.150}{46.327}
\emlineto{121.445}{46.270}
\emmoveto{121.445}{46.260}
\emlineto{121.740}{46.203}
\emmoveto{121.740}{46.193}
\emlineto{122.035}{46.136}
\emmoveto{122.035}{46.126}
\emlineto{122.330}{46.070}
\emmoveto{122.330}{46.060}
\emlineto{122.625}{46.003}
\emmoveto{122.625}{45.993}
\emlineto{122.920}{45.937}
\emmoveto{122.920}{45.927}
\emlineto{123.215}{45.871}
\emmoveto{123.215}{45.861}
\emlineto{123.510}{45.804}
\emmoveto{123.510}{45.794}
\emlineto{123.805}{45.738}
\emmoveto{123.805}{45.728}
\emlineto{124.100}{45.673}
\emmoveto{124.100}{45.663}
\emlineto{124.395}{45.607}
\emmoveto{124.395}{45.597}
\emlineto{124.690}{45.541}
\emmoveto{124.690}{45.531}
\emlineto{124.985}{45.476}
\emmoveto{124.985}{45.466}
\emlineto{125.280}{45.410}
\emmoveto{125.280}{45.400}
\emlineto{125.575}{45.345}
\emmoveto{125.575}{45.335}
\emlineto{125.870}{45.280}
\emmoveto{125.870}{45.270}
\emlineto{126.165}{45.215}
\emmoveto{126.165}{45.205}
\emlineto{126.460}{45.150}
\emmoveto{126.460}{45.140}
\emlineto{126.755}{45.085}
\emmoveto{126.755}{45.075}
\emlineto{127.050}{45.021}
\emmoveto{127.050}{45.011}
\emlineto{127.345}{44.956}
\emmoveto{127.345}{44.946}
\emlineto{127.640}{44.892}
\emmoveto{127.640}{44.882}
\emlineto{127.935}{44.827}
\emmoveto{127.935}{44.817}
\emlineto{128.230}{44.763}
\emmoveto{128.230}{44.753}
\emlineto{128.525}{44.699}
\emmoveto{128.525}{44.689}
\emlineto{128.820}{44.635}
\emmoveto{128.820}{44.625}
\emlineto{129.115}{44.571}
\emmoveto{129.115}{44.561}
\emlineto{129.410}{44.508}
\emmoveto{129.410}{44.498}
\emlineto{129.705}{44.444}
\emshow{83.980}{45.700}{vz}
\emmoveto{12.000}{80.000}
\emlineto{12.295}{79.770}
\emmoveto{12.295}{79.760}
\emlineto{12.590}{79.530}
\emmoveto{12.590}{79.520}
\emlineto{12.885}{79.290}
\emmoveto{12.885}{79.280}
\emlineto{13.180}{79.050}
\emmoveto{13.180}{79.040}
\emlineto{13.475}{78.810}
\emmoveto{13.475}{78.800}
\emlineto{13.770}{78.570}
\emmoveto{13.770}{78.560}
\emlineto{14.065}{78.330}
\emmoveto{14.065}{78.320}
\emlineto{14.360}{78.090}
\emmoveto{14.360}{78.080}
\emlineto{14.655}{77.851}
\emmoveto{14.655}{77.841}
\emlineto{14.950}{77.611}
\emmoveto{14.950}{77.601}
\emlineto{15.245}{77.371}
\emmoveto{15.245}{77.361}
\emlineto{15.540}{77.131}
\emmoveto{15.540}{77.121}
\emlineto{15.835}{76.892}
\emmoveto{15.835}{76.882}
\emlineto{16.130}{76.652}
\emmoveto{16.130}{76.642}
\emlineto{16.425}{76.413}
\emmoveto{16.425}{76.403}
\emlineto{16.720}{76.173}
\emmoveto{16.720}{76.163}
\emlineto{17.015}{75.934}
\emmoveto{17.015}{75.924}
\emlineto{17.310}{75.694}
\emmoveto{17.310}{75.684}
\emlineto{17.605}{75.455}
\emmoveto{17.605}{75.445}
\emlineto{17.900}{75.216}
\emmoveto{17.900}{75.206}
\emlineto{18.195}{74.977}
\emmoveto{18.195}{74.967}
\emlineto{18.490}{74.738}
\emmoveto{18.490}{74.728}
\emlineto{18.785}{74.499}
\emmoveto{18.785}{74.489}
\emlineto{19.080}{74.260}
\emmoveto{19.080}{74.250}
\emlineto{19.375}{74.022}
\emmoveto{19.375}{74.012}
\emlineto{19.670}{73.783}
\emmoveto{19.670}{73.773}
\emlineto{19.965}{73.545}
\emmoveto{19.965}{73.535}
\emlineto{20.260}{73.306}
\emmoveto{20.260}{73.296}
\emlineto{20.555}{73.068}
\emmoveto{20.555}{73.058}
\emlineto{20.850}{72.830}
\emmoveto{20.850}{72.820}
\emlineto{21.145}{72.592}
\emmoveto{21.145}{72.582}
\emlineto{21.440}{72.354}
\emmoveto{21.440}{72.344}
\emlineto{21.735}{72.117}
\emmoveto{21.735}{72.107}
\emlineto{22.030}{71.879}
\emmoveto{22.030}{71.869}
\emlineto{22.325}{71.642}
\emmoveto{22.325}{71.632}
\emlineto{22.620}{71.405}
\emmoveto{22.620}{71.395}
\emlineto{22.915}{71.168}
\emmoveto{22.915}{71.158}
\emlineto{23.210}{70.931}
\emmoveto{23.210}{70.921}
\emlineto{23.505}{70.694}
\emmoveto{23.505}{70.684}
\emlineto{23.800}{70.458}
\emmoveto{23.800}{70.448}
\emlineto{24.095}{70.221}
\emmoveto{24.095}{70.211}
\emlineto{24.390}{69.985}
\emmoveto{24.390}{69.975}
\emlineto{24.685}{69.749}
\emmoveto{24.685}{69.739}
\emlineto{24.980}{69.513}
\emmoveto{24.980}{69.503}
\emlineto{25.275}{69.278}
\emmoveto{25.275}{69.268}
\emlineto{25.570}{69.042}
\emmoveto{25.570}{69.032}
\emlineto{25.865}{68.807}
\emmoveto{25.865}{68.797}
\emlineto{26.160}{68.572}
\emmoveto{26.160}{68.562}
\emlineto{26.455}{68.337}
\emmoveto{26.455}{68.327}
\emlineto{26.750}{68.102}
\emmoveto{26.750}{68.092}
\emlineto{27.045}{67.868}
\emmoveto{27.045}{67.858}
\emlineto{27.340}{67.634}
\emmoveto{27.340}{67.624}
\emlineto{27.635}{67.400}
\emmoveto{27.635}{67.390}
\emlineto{27.930}{67.166}
\emmoveto{27.930}{67.156}
\emlineto{28.225}{66.933}
\emmoveto{28.225}{66.923}
\emlineto{28.520}{66.699}
\emmoveto{28.520}{66.689}
\emlineto{28.815}{66.466}
\emmoveto{28.815}{66.456}
\emlineto{29.110}{66.234}
\emmoveto{29.110}{66.224}
\emlineto{29.405}{66.001}
\emmoveto{29.405}{65.991}
\emlineto{29.700}{65.769}
\emmoveto{29.700}{65.759}
\emlineto{29.995}{65.537}
\emmoveto{29.995}{65.527}
\emlineto{30.290}{65.305}
\emmoveto{30.290}{65.295}
\emlineto{30.585}{65.074}
\emmoveto{30.585}{65.064}
\emlineto{30.880}{64.842}
\emmoveto{30.880}{64.832}
\emlineto{31.175}{64.611}
\emmoveto{31.175}{64.601}
\emlineto{31.470}{64.381}
\emmoveto{31.470}{64.371}
\emlineto{31.765}{64.150}
\emmoveto{31.765}{64.140}
\emlineto{32.060}{63.920}
\emmoveto{32.060}{63.910}
\emlineto{32.355}{63.690}
\emmoveto{32.355}{63.680}
\emlineto{32.650}{63.461}
\emmoveto{32.650}{63.451}
\emlineto{32.945}{63.231}
\emmoveto{32.945}{63.221}
\emlineto{33.240}{63.002}
\emmoveto{33.240}{62.992}
\emlineto{33.535}{62.774}
\emmoveto{33.535}{62.764}
\emlineto{33.830}{62.545}
\emmoveto{33.830}{62.535}
\emlineto{34.125}{62.317}
\emmoveto{34.125}{62.307}
\emlineto{34.420}{62.089}
\emmoveto{34.420}{62.079}
\emlineto{34.715}{61.862}
\emmoveto{34.715}{61.852}
\emlineto{35.010}{61.635}
\emmoveto{35.010}{61.625}
\emlineto{35.305}{61.408}
\emmoveto{35.305}{61.398}
\emlineto{35.600}{61.181}
\emmoveto{35.600}{61.171}
\emlineto{35.895}{60.955}
\emmoveto{35.895}{60.945}
\emlineto{36.190}{60.729}
\emmoveto{36.190}{60.719}
\emlineto{36.485}{60.504}
\emmoveto{36.485}{60.494}
\emlineto{36.780}{60.278}
\emmoveto{36.780}{60.268}
\emlineto{37.075}{60.054}
\emmoveto{37.075}{60.044}
\emlineto{37.370}{59.829}
\emmoveto{37.370}{59.819}
\emlineto{37.665}{59.605}
\emmoveto{37.665}{59.595}
\emlineto{37.960}{59.381}
\emmoveto{37.960}{59.371}
\emlineto{38.255}{59.157}
\emmoveto{38.255}{59.147}
\emlineto{38.550}{58.934}
\emmoveto{38.550}{58.924}
\emlineto{38.845}{58.711}
\emmoveto{38.845}{58.701}
\emlineto{39.140}{58.489}
\emmoveto{39.140}{58.479}
\emlineto{39.435}{58.267}
\emmoveto{39.435}{58.257}
\emlineto{39.730}{58.045}
\emmoveto{39.730}{58.035}
\emlineto{40.025}{57.824}
\emmoveto{40.025}{57.814}
\emlineto{40.320}{57.603}
\emmoveto{40.320}{57.593}
\emlineto{40.615}{57.382}
\emmoveto{40.615}{57.372}
\emlineto{40.910}{57.162}
\emmoveto{40.910}{57.152}
\emlineto{41.205}{56.942}
\emmoveto{41.205}{56.932}
\emlineto{41.500}{56.723}
\emmoveto{41.500}{56.713}
\emlineto{41.795}{56.503}
\emmoveto{41.795}{56.493}
\emlineto{42.090}{56.285}
\emmoveto{42.090}{56.275}
\emlineto{42.385}{56.066}
\emmoveto{42.385}{56.056}
\emlineto{42.680}{55.848}
\emmoveto{42.680}{55.838}
\emlineto{42.975}{55.631}
\emmoveto{42.975}{55.621}
\emlineto{43.270}{55.414}
\emmoveto{43.270}{55.404}
\emlineto{43.565}{55.197}
\emmoveto{43.565}{55.187}
\emlineto{43.860}{54.981}
\emmoveto{43.860}{54.971}
\emlineto{44.155}{54.765}
\emmoveto{44.155}{54.755}
\emlineto{44.450}{54.549}
\emmoveto{44.450}{54.539}
\emlineto{44.745}{54.334}
\emmoveto{44.745}{54.324}
\emlineto{45.040}{54.119}
\emmoveto{45.040}{54.109}
\emlineto{45.335}{53.905}
\emmoveto{45.335}{53.895}
\emlineto{45.630}{53.691}
\emmoveto{45.630}{53.681}
\emlineto{45.925}{53.477}
\emmoveto{45.925}{53.467}
\emlineto{46.220}{53.264}
\emmoveto{46.220}{53.254}
\emlineto{46.515}{53.051}
\emmoveto{46.515}{53.041}
\emlineto{46.810}{52.839}
\emmoveto{46.810}{52.829}
\emlineto{47.105}{52.627}
\emmoveto{47.105}{52.617}
\emlineto{47.400}{52.416}
\emmoveto{47.400}{52.406}
\emlineto{47.695}{52.205}
\emmoveto{47.695}{52.195}
\emlineto{47.990}{51.994}
\emmoveto{47.990}{51.984}
\emlineto{48.285}{51.784}
\emmoveto{48.285}{51.774}
\emlineto{48.580}{51.574}
\emmoveto{48.580}{51.564}
\emlineto{48.875}{51.365}
\emmoveto{48.875}{51.355}
\emlineto{49.170}{51.156}
\emmoveto{49.170}{51.146}
\emlineto{49.465}{50.948}
\emmoveto{49.465}{50.938}
\emlineto{49.760}{50.740}
\emmoveto{49.760}{50.730}
\emlineto{50.055}{50.533}
\emmoveto{50.055}{50.523}
\emlineto{50.350}{50.325}
\emmoveto{50.350}{50.315}
\emlineto{50.645}{50.119}
\emmoveto{50.645}{50.109}
\emlineto{50.940}{49.913}
\emmoveto{50.940}{49.903}
\emlineto{51.235}{49.707}
\emmoveto{51.235}{49.697}
\emlineto{51.530}{49.502}
\emmoveto{51.530}{49.492}
\emlineto{51.825}{49.297}
\emmoveto{51.825}{49.287}
\emlineto{52.120}{49.092}
\emmoveto{52.120}{49.082}
\emlineto{52.415}{48.889}
\emmoveto{52.415}{48.879}
\emlineto{52.710}{48.685}
\emmoveto{52.710}{48.675}
\emlineto{53.005}{48.482}
\emmoveto{53.005}{48.472}
\emlineto{53.300}{48.280}
\emmoveto{53.300}{48.270}
\emlineto{53.595}{48.077}
\emmoveto{53.595}{48.067}
\emlineto{53.890}{47.876}
\emmoveto{53.890}{47.866}
\emlineto{54.185}{47.675}
\emmoveto{54.185}{47.665}
\emlineto{54.480}{47.474}
\emmoveto{54.480}{47.464}
\emlineto{54.775}{47.274}
\emmoveto{54.775}{47.264}
\emlineto{55.070}{47.074}
\emmoveto{55.070}{47.064}
\emlineto{55.365}{46.875}
\emmoveto{55.365}{46.865}
\emlineto{55.660}{46.676}
\emmoveto{55.660}{46.666}
\emlineto{55.955}{46.478}
\emmoveto{55.955}{46.468}
\emlineto{56.250}{46.280}
\emmoveto{56.250}{46.270}
\emlineto{56.545}{46.082}
\emmoveto{56.545}{46.072}
\emlineto{56.840}{45.885}
\emmoveto{56.840}{45.875}
\emlineto{57.135}{45.689}
\emmoveto{57.135}{45.679}
\emlineto{57.430}{45.493}
\emmoveto{57.430}{45.483}
\emlineto{57.725}{45.297}
\emmoveto{57.725}{45.287}
\emlineto{58.020}{45.102}
\emmoveto{58.020}{45.092}
\emlineto{58.315}{44.908}
\emmoveto{58.315}{44.898}
\emlineto{58.610}{44.714}
\emmoveto{58.610}{44.704}
\emlineto{58.905}{44.520}
\emmoveto{58.905}{44.510}
\emlineto{59.200}{44.327}
\emmoveto{59.200}{44.317}
\emlineto{59.495}{44.134}
\emmoveto{59.495}{44.124}
\emlineto{59.790}{43.942}
\emmoveto{59.790}{43.932}
\emlineto{60.085}{43.751}
\emmoveto{60.085}{43.741}
\emlineto{60.380}{43.559}
\emmoveto{60.380}{43.549}
\emlineto{60.675}{43.369}
\emmoveto{60.675}{43.359}
\emlineto{60.970}{43.178}
\emmoveto{60.970}{43.168}
\emlineto{61.265}{42.989}
\emmoveto{61.265}{42.979}
\emlineto{61.560}{42.800}
\emmoveto{61.560}{42.790}
\emlineto{61.855}{42.611}
\emmoveto{61.855}{42.601}
\emlineto{62.150}{42.422}
\emmoveto{62.150}{42.412}
\emlineto{62.445}{42.235}
\emmoveto{62.445}{42.225}
\emlineto{62.740}{42.047}
\emmoveto{62.740}{42.037}
\emlineto{63.035}{41.861}
\emmoveto{63.035}{41.851}
\emlineto{63.330}{41.674}
\emmoveto{63.330}{41.664}
\emlineto{63.625}{41.489}
\emmoveto{63.625}{41.479}
\emlineto{63.920}{41.303}
\emmoveto{63.920}{41.293}
\emlineto{64.215}{41.118}
\emmoveto{64.215}{41.108}
\emlineto{64.510}{40.934}
\emmoveto{64.510}{40.924}
\emlineto{64.805}{40.750}
\emmoveto{64.805}{40.740}
\emlineto{65.100}{40.567}
\emmoveto{65.100}{40.557}
\emlineto{65.395}{40.384}
\emmoveto{65.395}{40.374}
\emlineto{65.690}{40.202}
\emmoveto{65.690}{40.192}
\emlineto{65.985}{40.020}
\emmoveto{65.985}{40.010}
\emlineto{66.280}{39.838}
\emmoveto{66.280}{39.828}
\emlineto{66.575}{39.658}
\emmoveto{66.575}{39.648}
\emlineto{66.870}{39.477}
\emmoveto{66.870}{39.467}
\emlineto{67.165}{39.297}
\emmoveto{67.165}{39.287}
\emlineto{67.460}{39.118}
\emmoveto{67.460}{39.108}
\emlineto{67.755}{38.939}
\emmoveto{67.755}{38.929}
\emlineto{68.050}{38.761}
\emmoveto{68.050}{38.751}
\emlineto{68.345}{38.583}
\emmoveto{68.345}{38.573}
\emlineto{68.640}{38.405}
\emmoveto{68.640}{38.395}
\emlineto{68.935}{38.228}
\emmoveto{68.935}{38.218}
\emlineto{69.230}{38.052}
\emmoveto{69.230}{38.042}
\emlineto{69.525}{37.876}
\emmoveto{69.525}{37.866}
\emlineto{69.820}{37.701}
\emmoveto{69.820}{37.691}
\emlineto{70.115}{37.526}
\emmoveto{70.115}{37.516}
\emlineto{70.410}{37.351}
\emmoveto{70.410}{37.341}
\emlineto{70.705}{37.177}
\emmoveto{70.705}{37.167}
\emlineto{71.000}{37.004}
\emmoveto{71.000}{36.994}
\emlineto{71.295}{36.831}
\emmoveto{71.295}{36.821}
\emlineto{71.590}{36.658}
\emmoveto{71.590}{36.648}
\emlineto{71.885}{36.486}
\emmoveto{71.885}{36.476}
\emlineto{72.180}{36.315}
\emmoveto{72.180}{36.305}
\emlineto{72.475}{36.144}
\emmoveto{72.475}{36.134}
\emlineto{72.770}{35.973}
\emmoveto{72.770}{35.963}
\emlineto{73.065}{35.803}
\emmoveto{73.065}{35.793}
\emlineto{73.360}{35.634}
\emmoveto{73.360}{35.624}
\emlineto{73.655}{35.465}
\emmoveto{73.655}{35.455}
\emlineto{73.950}{35.296}
\emmoveto{73.950}{35.286}
\emlineto{74.245}{35.128}
\emmoveto{74.245}{35.118}
\emlineto{74.540}{34.961}
\emmoveto{74.540}{34.951}
\emlineto{74.835}{34.794}
\emmoveto{74.835}{34.784}
\emlineto{75.130}{34.627}
\emmoveto{75.130}{34.617}
\emlineto{75.425}{34.461}
\emmoveto{75.425}{34.451}
\emlineto{75.720}{34.296}
\emmoveto{75.720}{34.286}
\emlineto{76.015}{34.131}
\emmoveto{76.015}{34.121}
\emlineto{76.310}{33.966}
\emmoveto{76.310}{33.956}
\emlineto{76.605}{33.802}
\emmoveto{76.605}{33.792}
\emlineto{76.900}{33.638}
\emmoveto{76.900}{33.628}
\emlineto{77.195}{33.475}
\emmoveto{77.195}{33.465}
\emlineto{77.490}{33.313}
\emmoveto{77.490}{33.303}
\emlineto{77.785}{33.150}
\emmoveto{77.785}{33.140}
\emlineto{78.080}{32.989}
\emmoveto{78.080}{32.979}
\emlineto{78.375}{32.828}
\emmoveto{78.375}{32.818}
\emlineto{78.670}{32.667}
\emmoveto{78.670}{32.657}
\emlineto{78.965}{32.507}
\emmoveto{78.965}{32.497}
\emlineto{79.260}{32.347}
\emmoveto{79.260}{32.337}
\emlineto{79.555}{32.188}
\emmoveto{79.555}{32.178}
\emlineto{79.850}{32.029}
\emmoveto{79.850}{32.019}
\emlineto{80.145}{31.871}
\emmoveto{80.145}{31.861}
\emlineto{80.440}{31.713}
\emmoveto{80.440}{31.703}
\emlineto{80.735}{31.556}
\emmoveto{80.735}{31.546}
\emlineto{81.030}{31.399}
\emmoveto{81.030}{31.389}
\emlineto{81.325}{31.243}
\emmoveto{81.325}{31.233}
\emlineto{81.620}{31.087}
\emmoveto{81.620}{31.077}
\emlineto{81.915}{30.931}
\emmoveto{81.915}{30.921}
\emlineto{82.210}{30.777}
\emmoveto{82.210}{30.767}
\emlineto{82.505}{30.622}
\emmoveto{82.505}{30.612}
\emlineto{82.800}{30.468}
\emmoveto{82.800}{30.458}
\emlineto{83.095}{30.315}
\emmoveto{83.095}{30.305}
\emlineto{83.390}{30.162}
\emmoveto{83.390}{30.152}
\emlineto{83.685}{30.009}
\emmoveto{83.685}{29.999}
\emlineto{83.980}{29.857}
\emmoveto{83.980}{29.847}
\emlineto{84.275}{29.706}
\emmoveto{84.275}{29.696}
\emlineto{84.570}{29.555}
\emmoveto{84.570}{29.545}
\emlineto{84.865}{29.404}
\emmoveto{84.865}{29.394}
\emlineto{85.160}{29.254}
\emmoveto{85.160}{29.244}
\emlineto{85.455}{29.104}
\emmoveto{85.455}{29.094}
\emlineto{85.750}{28.955}
\emmoveto{85.750}{28.945}
\emlineto{86.045}{28.806}
\emmoveto{86.045}{28.796}
\emlineto{86.340}{28.658}
\emmoveto{86.340}{28.648}
\emlineto{86.635}{28.510}
\emmoveto{86.635}{28.500}
\emlineto{86.930}{28.363}
\emmoveto{86.930}{28.353}
\emlineto{87.225}{28.216}
\emmoveto{87.225}{28.206}
\emlineto{87.520}{28.070}
\emmoveto{87.520}{28.060}
\emlineto{87.815}{27.924}
\emmoveto{87.815}{27.914}
\emlineto{88.110}{27.778}
\emmoveto{88.110}{27.768}
\emlineto{88.405}{27.633}
\emmoveto{88.405}{27.623}
\emlineto{88.700}{27.489}
\emmoveto{88.700}{27.479}
\emlineto{88.995}{27.345}
\emmoveto{88.995}{27.335}
\emlineto{89.290}{27.201}
\emmoveto{89.290}{27.191}
\emlineto{89.585}{27.058}
\emmoveto{89.585}{27.048}
\emlineto{89.880}{26.915}
\emmoveto{89.880}{26.905}
\emlineto{90.175}{26.773}
\emmoveto{90.175}{26.763}
\emlineto{90.470}{26.631}
\emmoveto{90.470}{26.621}
\emlineto{90.765}{26.490}
\emmoveto{90.765}{26.480}
\emlineto{91.060}{26.349}
\emmoveto{91.060}{26.339}
\emlineto{91.355}{26.208}
\emmoveto{91.355}{26.198}
\emlineto{91.650}{26.068}
\emmoveto{91.650}{26.058}
\emlineto{91.945}{25.929}
\emmoveto{91.945}{25.919}
\emlineto{92.240}{25.790}
\emmoveto{92.240}{25.780}
\emlineto{92.535}{25.651}
\emmoveto{92.535}{25.641}
\emlineto{92.830}{25.513}
\emmoveto{92.830}{25.503}
\emlineto{93.125}{25.375}
\emmoveto{93.125}{25.365}
\emlineto{93.420}{25.238}
\emmoveto{93.420}{25.228}
\emlineto{93.715}{25.101}
\emmoveto{93.715}{25.091}
\emlineto{94.010}{24.965}
\emmoveto{94.010}{24.955}
\emlineto{94.305}{24.829}
\emmoveto{94.305}{24.819}
\emlineto{94.600}{24.693}
\emmoveto{94.600}{24.683}
\emlineto{94.895}{24.558}
\emmoveto{94.895}{24.548}
\emlineto{95.190}{24.423}
\emmoveto{95.190}{24.413}
\emlineto{95.485}{24.289}
\emmoveto{95.485}{24.279}
\emlineto{95.780}{24.155}
\emmoveto{95.780}{24.145}
\emlineto{96.075}{24.022}
\emmoveto{96.075}{24.012}
\emlineto{96.370}{23.889}
\emmoveto{96.370}{23.879}
\emlineto{96.665}{23.757}
\emmoveto{96.665}{23.747}
\emlineto{96.960}{23.625}
\emmoveto{96.960}{23.615}
\emlineto{97.255}{23.493}
\emmoveto{97.255}{23.483}
\emlineto{97.550}{23.362}
\emmoveto{97.550}{23.352}
\emlineto{97.845}{23.231}
\emmoveto{97.845}{23.221}
\emlineto{98.140}{23.101}
\emmoveto{98.140}{23.091}
\emlineto{98.435}{22.971}
\emmoveto{98.435}{22.961}
\emlineto{98.730}{22.842}
\emmoveto{98.730}{22.832}
\emlineto{99.025}{22.712}
\emmoveto{99.025}{22.702}
\emlineto{99.320}{22.584}
\emmoveto{99.320}{22.574}
\emlineto{99.615}{22.456}
\emmoveto{99.615}{22.446}
\emlineto{99.910}{22.328}
\emmoveto{99.910}{22.318}
\emlineto{100.205}{22.201}
\emmoveto{100.205}{22.191}
\emlineto{100.500}{22.074}
\emmoveto{100.500}{22.064}
\emlineto{100.795}{21.947}
\emmoveto{100.795}{21.937}
\emlineto{101.090}{21.821}
\emmoveto{101.090}{21.811}
\emlineto{101.385}{21.695}
\emmoveto{101.385}{21.685}
\emlineto{101.680}{21.570}
\emmoveto{101.680}{21.560}
\emlineto{101.975}{21.445}
\emmoveto{101.975}{21.435}
\emlineto{102.270}{21.321}
\emmoveto{102.270}{21.311}
\emlineto{102.565}{21.197}
\emmoveto{102.565}{21.187}
\emlineto{102.860}{21.073}
\emmoveto{102.860}{21.063}
\emlineto{103.155}{20.950}
\emmoveto{103.155}{20.940}
\emlineto{103.450}{20.827}
\emmoveto{103.450}{20.817}
\emlineto{103.745}{20.705}
\emmoveto{103.745}{20.695}
\emlineto{104.040}{20.583}
\emmoveto{104.040}{20.573}
\emlineto{104.335}{20.461}
\emmoveto{104.335}{20.451}
\emlineto{104.630}{20.340}
\emmoveto{104.630}{20.330}
\emlineto{104.925}{20.219}
\emmoveto{104.925}{20.209}
\emlineto{105.220}{20.099}
\emmoveto{105.220}{20.089}
\emlineto{105.515}{19.979}
\emmoveto{105.515}{19.969}
\emlineto{105.810}{19.859}
\emmoveto{105.810}{19.849}
\emlineto{106.105}{19.740}
\emmoveto{106.105}{19.730}
\emlineto{106.400}{19.621}
\emmoveto{106.400}{19.611}
\emlineto{106.695}{19.503}
\emmoveto{106.695}{19.493}
\emlineto{106.990}{19.385}
\emmoveto{106.990}{19.375}
\emlineto{107.285}{19.267}
\emmoveto{107.285}{19.257}
\emlineto{107.580}{19.150}
\emmoveto{107.580}{19.140}
\emlineto{107.875}{19.033}
\emmoveto{107.875}{19.023}
\emlineto{108.170}{18.917}
\emmoveto{108.170}{18.907}
\emlineto{108.465}{18.801}
\emmoveto{108.465}{18.791}
\emlineto{108.760}{18.685}
\emmoveto{108.760}{18.675}
\emlineto{109.055}{18.570}
\emmoveto{109.055}{18.560}
\emlineto{109.350}{18.455}
\emmoveto{109.350}{18.445}
\emlineto{109.645}{18.340}
\emmoveto{109.645}{18.330}
\emlineto{109.940}{18.226}
\emmoveto{109.940}{18.216}
\emlineto{110.235}{18.112}
\emmoveto{110.235}{18.102}
\emlineto{110.530}{17.999}
\emmoveto{110.530}{17.989}
\emlineto{110.825}{17.886}
\emmoveto{110.825}{17.876}
\emlineto{111.120}{17.773}
\emmoveto{111.120}{17.763}
\emlineto{111.415}{17.661}
\emmoveto{111.415}{17.651}
\emlineto{111.710}{17.549}
\emmoveto{111.710}{17.539}
\emlineto{112.005}{17.438}
\emmoveto{112.005}{17.428}
\emlineto{112.300}{17.327}
\emmoveto{112.300}{17.317}
\emlineto{112.595}{17.216}
\emmoveto{112.595}{17.206}
\emlineto{112.890}{17.106}
\emmoveto{112.890}{17.096}
\emlineto{113.185}{16.996}
\emmoveto{113.185}{16.986}
\emlineto{113.480}{16.886}
\emmoveto{113.480}{16.876}
\emlineto{113.775}{16.777}
\emmoveto{113.775}{16.767}
\emlineto{114.070}{16.668}
\emmoveto{114.070}{16.658}
\emlineto{114.365}{16.559}
\emmoveto{114.365}{16.549}
\emlineto{114.660}{16.451}
\emmoveto{114.660}{16.441}
\emlineto{114.955}{16.343}
\emmoveto{114.955}{16.333}
\emlineto{115.250}{16.236}
\emmoveto{115.250}{16.226}
\emlineto{115.545}{16.129}
\emmoveto{115.545}{16.119}
\emlineto{115.840}{16.022}
\emmoveto{115.840}{16.012}
\emlineto{116.135}{15.915}
\emmoveto{116.135}{15.905}
\emlineto{116.430}{15.809}
\emmoveto{116.430}{15.799}
\emlineto{116.725}{15.704}
\emmoveto{116.725}{15.694}
\emlineto{117.020}{15.598}
\emmoveto{117.020}{15.588}
\emlineto{117.315}{15.493}
\emmoveto{117.315}{15.483}
\emlineto{117.610}{15.389}
\emmoveto{117.610}{15.379}
\emlineto{117.905}{15.284}
\emmoveto{117.905}{15.274}
\emlineto{118.200}{15.180}
\emmoveto{118.200}{15.170}
\emlineto{118.495}{15.077}
\emmoveto{118.495}{15.067}
\emlineto{118.790}{14.974}
\emmoveto{118.790}{14.964}
\emlineto{119.085}{14.871}
\emmoveto{119.085}{14.861}
\emlineto{119.380}{14.768}
\emmoveto{119.380}{14.758}
\emlineto{119.675}{14.666}
\emmoveto{119.675}{14.656}
\emlineto{119.970}{14.564}
\emmoveto{119.970}{14.554}
\emlineto{120.265}{14.463}
\emmoveto{120.265}{14.453}
\emlineto{120.560}{14.361}
\emmoveto{120.560}{14.351}
\emlineto{120.855}{14.260}
\emmoveto{120.855}{14.250}
\emlineto{121.150}{14.160}
\emmoveto{121.150}{14.150}
\emlineto{121.445}{14.060}
\emmoveto{121.445}{14.050}
\emlineto{121.740}{13.960}
\emmoveto{121.740}{13.950}
\emlineto{122.035}{13.860}
\emmoveto{122.035}{13.850}
\emlineto{122.330}{13.761}
\emmoveto{122.330}{13.751}
\emlineto{122.625}{13.662}
\emmoveto{122.625}{13.652}
\emlineto{122.920}{13.564}
\emmoveto{122.920}{13.554}
\emlineto{123.215}{13.466}
\emmoveto{123.215}{13.456}
\emlineto{123.510}{13.368}
\emmoveto{123.510}{13.358}
\emlineto{123.805}{13.270}
\emmoveto{123.805}{13.260}
\emlineto{124.100}{13.173}
\emmoveto{124.100}{13.163}
\emlineto{124.395}{13.076}
\emmoveto{124.395}{13.066}
\emlineto{124.690}{12.980}
\emmoveto{124.690}{12.970}
\emlineto{124.985}{12.883}
\emmoveto{124.985}{12.873}
\emlineto{125.280}{12.787}
\emmoveto{125.280}{12.777}
\emlineto{125.575}{12.692}
\emmoveto{125.575}{12.682}
\emlineto{125.870}{12.596}
\emmoveto{125.870}{12.586}
\emlineto{126.165}{12.501}
\emmoveto{126.165}{12.491}
\emlineto{126.460}{12.407}
\emmoveto{126.460}{12.397}
\emlineto{126.755}{12.312}
\emmoveto{126.755}{12.302}
\emlineto{127.050}{12.218}
\emmoveto{127.050}{12.208}
\emlineto{127.345}{12.125}
\emmoveto{127.345}{12.115}
\emlineto{127.640}{12.031}
\emmoveto{127.640}{12.021}
\emlineto{127.935}{11.938}
\emmoveto{127.935}{11.928}
\emlineto{128.230}{11.845}
\emmoveto{128.230}{11.835}
\emlineto{128.525}{11.753}
\emmoveto{128.525}{11.743}
\emlineto{128.820}{11.661}
\emmoveto{128.820}{11.651}
\emlineto{129.115}{11.569}
\emmoveto{129.115}{11.559}
\emlineto{129.410}{11.477}
\emmoveto{129.410}{11.467}
\emlineto{129.705}{11.386}
\emshow{48.580}{38.700}{vq}
\emshow{1.000}{10.000}{-3.50e-1}
\emshow{1.000}{17.000}{-3.15e-1}
\emshow{1.000}{24.000}{-2.80e-1}
\emshow{1.000}{31.000}{-2.45e-1}
\emshow{1.000}{38.000}{-2.10e-1}
\emshow{1.000}{45.000}{-1.75e-1}
\emshow{1.000}{52.000}{-1.40e-1}
\emshow{1.000}{59.000}{-1.05e-1}
\emshow{1.000}{66.000}{-7.00e-2}
\emshow{1.000}{73.000}{-3.50e-2}
\emshow{1.000}{80.000}{0.00e0}
\emshow{12.000}{5.000}{0.00e0}
\emshow{23.800}{5.000}{1.20e0}
\emshow{35.600}{5.000}{2.40e0}
\emshow{47.400}{5.000}{3.60e0}
\emshow{59.200}{5.000}{4.80e0}
\emshow{71.000}{5.000}{6.00e0}
\emshow{82.800}{5.000}{7.20e0}
\emshow{94.600}{5.000}{8.40e0}
\emshow{106.400}{5.000}{9.60e0}
\emshow{118.200}{5.000}{1.08e1}
\emshow{130.000}{5.000}{1.20e1}

\centerline {\bf {Fig. A. 1}}
\eject
\newcount\numpoint
\newcount\numpointo
\numpoint=1 \numpointo=1
\def\emmoveto#1#2{\offinterlineskip
\hbox to 0 true cm{\vbox to 0
true cm{\vskip - #2 true mm
\hskip #1 true mm \special{em:point
\the\numpoint}\vss}\hss}
\numpointo=\numpoint
\global\advance \numpoint by 1}
\def\emlineto#1#2{\offinterlineskip
\hbox to 0 true cm{\vbox to 0
true cm{\vskip - #2 true mm
\hskip #1 true mm \special{em:point
\the\numpoint}\vss}\hss}
\special{em:line
\the\numpointo,\the\numpoint}
\numpointo=\numpoint
\global\advance \numpoint by 1}
\def\emshow#1#2#3{\offinterlineskip
\hbox to 0 true cm{\vbox to 0
true cm{\vskip - #2 true mm
\hskip #1 true mm \vbox to 0
true cm{\vss\hbox{#3\hss
}}\vss}\hss}}
\special{em:linewidth 0.8pt}

\vrule width 0 mm height                0 mm depth 90.000 true mm

\special{em:linewidth 0.8pt}
\emmoveto{130.000}{10.000}
\emlineto{12.000}{10.000}
\emlineto{12.000}{80.000}
\emmoveto{71.000}{10.000}
\emlineto{71.000}{80.000}
\emmoveto{12.000}{45.000}
\emlineto{130.000}{45.000}
\emmoveto{130.000}{10.000}
\emlineto{130.000}{80.000}
\emlineto{12.000}{80.000}
\emlineto{12.000}{10.000}
\emlineto{130.000}{10.000}
\special{em:linewidth 0.4pt}
\emmoveto{12.000}{17.000}
\emlineto{130.000}{17.000}
\emmoveto{12.000}{24.000}
\emlineto{130.000}{24.000}
\emmoveto{12.000}{31.000}
\emlineto{130.000}{31.000}
\emmoveto{12.000}{38.000}
\emlineto{130.000}{38.000}
\emmoveto{12.000}{45.000}
\emlineto{130.000}{45.000}
\emmoveto{12.000}{52.000}
\emlineto{130.000}{52.000}
\emmoveto{12.000}{59.000}
\emlineto{130.000}{59.000}
\emmoveto{12.000}{66.000}
\emlineto{130.000}{66.000}
\emmoveto{12.000}{73.000}
\emlineto{130.000}{73.000}
\emmoveto{23.800}{10.000}
\emlineto{23.800}{80.000}
\emmoveto{35.600}{10.000}
\emlineto{35.600}{80.000}
\emmoveto{47.400}{10.000}
\emlineto{47.400}{80.000}
\emmoveto{59.200}{10.000}
\emlineto{59.200}{80.000}
\emmoveto{71.000}{10.000}
\emlineto{71.000}{80.000}
\emmoveto{82.800}{10.000}
\emlineto{82.800}{80.000}
\emmoveto{94.600}{10.000}
\emlineto{94.600}{80.000}
\emmoveto{106.400}{10.000}
\emlineto{106.400}{80.000}
\emmoveto{118.200}{10.000}
\emlineto{118.200}{80.000}
\special{em:linewidth 0.8pt}
\emmoveto{12.000}{80.000}
\emlineto{12.197}{79.890}
\emmoveto{12.197}{79.880}
\emlineto{12.393}{79.770}
\emmoveto{12.393}{79.760}
\emlineto{12.590}{79.650}
\emmoveto{12.590}{79.640}
\emlineto{12.787}{79.530}
\emmoveto{12.787}{79.520}
\emlineto{12.983}{79.410}
\emmoveto{12.983}{79.400}
\emlineto{13.180}{79.290}
\emmoveto{13.180}{79.280}
\emlineto{13.377}{79.170}
\emmoveto{13.377}{79.160}
\emlineto{13.573}{79.050}
\emmoveto{13.573}{79.040}
\emlineto{13.770}{78.930}
\emmoveto{13.770}{78.920}
\emlineto{13.967}{78.810}
\emmoveto{13.967}{78.800}
\emlineto{14.163}{78.690}
\emmoveto{14.163}{78.680}
\emlineto{14.360}{78.570}
\emmoveto{14.360}{78.560}
\emlineto{14.557}{78.450}
\emmoveto{14.557}{78.440}
\emlineto{14.753}{78.330}
\emmoveto{14.753}{78.320}
\emlineto{14.950}{78.210}
\emmoveto{14.950}{78.200}
\emlineto{15.147}{78.090}
\emmoveto{15.147}{78.080}
\emlineto{15.343}{77.970}
\emmoveto{15.343}{77.960}
\emlineto{15.540}{77.850}
\emmoveto{15.540}{77.840}
\emlineto{15.737}{77.730}
\emmoveto{15.737}{77.720}
\emlineto{15.933}{77.610}
\emmoveto{15.933}{77.600}
\emlineto{16.130}{77.490}
\emmoveto{16.130}{77.480}
\emlineto{16.327}{77.370}
\emmoveto{16.327}{77.360}
\emlineto{16.523}{77.250}
\emmoveto{16.523}{77.240}
\emlineto{16.720}{77.130}
\emmoveto{16.720}{77.120}
\emlineto{16.917}{77.010}
\emmoveto{16.917}{77.000}
\emlineto{17.113}{76.891}
\emmoveto{17.113}{76.881}
\emlineto{17.310}{76.771}
\emmoveto{17.310}{76.761}
\emlineto{17.507}{76.651}
\emmoveto{17.507}{76.641}
\emlineto{17.703}{76.531}
\emmoveto{17.703}{76.521}
\emlineto{17.900}{76.411}
\emmoveto{17.900}{76.401}
\emlineto{18.097}{76.291}
\emmoveto{18.097}{76.281}
\emlineto{18.293}{76.171}
\emmoveto{18.293}{76.161}
\emlineto{18.490}{76.051}
\emmoveto{18.490}{76.041}
\emlineto{18.687}{75.931}
\emmoveto{18.687}{75.921}
\emlineto{18.883}{75.811}
\emmoveto{18.883}{75.801}
\emlineto{19.080}{75.691}
\emmoveto{19.080}{75.681}
\emlineto{19.277}{75.572}
\emmoveto{19.277}{75.562}
\emlineto{19.473}{75.452}
\emmoveto{19.473}{75.442}
\emlineto{19.670}{75.332}
\emmoveto{19.670}{75.322}
\emlineto{19.867}{75.212}
\emmoveto{19.867}{75.202}
\emlineto{20.063}{75.092}
\emmoveto{20.063}{75.082}
\emlineto{20.260}{74.972}
\emmoveto{20.260}{74.962}
\emlineto{20.457}{74.852}
\emmoveto{20.457}{74.842}
\emlineto{20.653}{74.733}
\emmoveto{20.653}{74.723}
\emlineto{20.850}{74.613}
\emmoveto{20.850}{74.603}
\emlineto{21.047}{74.493}
\emmoveto{21.047}{74.483}
\emlineto{21.243}{74.373}
\emmoveto{21.243}{74.363}
\emlineto{21.440}{74.253}
\emmoveto{21.440}{74.243}
\emlineto{21.637}{74.133}
\emmoveto{21.637}{74.123}
\emlineto{21.833}{74.014}
\emmoveto{21.833}{74.004}
\emlineto{22.030}{73.894}
\emmoveto{22.030}{73.884}
\emlineto{22.227}{73.774}
\emmoveto{22.227}{73.764}
\emlineto{22.423}{73.654}
\emmoveto{22.423}{73.644}
\emlineto{22.620}{73.535}
\emmoveto{22.620}{73.525}
\emlineto{22.817}{73.415}
\emmoveto{22.817}{73.405}
\emlineto{23.013}{73.295}
\emmoveto{23.013}{73.285}
\emlineto{23.210}{73.175}
\emmoveto{23.210}{73.165}
\emlineto{23.407}{73.056}
\emmoveto{23.407}{73.046}
\emlineto{23.603}{72.936}
\emmoveto{23.603}{72.926}
\emlineto{23.800}{72.816}
\emmoveto{23.800}{72.806}
\emlineto{23.997}{72.697}
\emmoveto{23.997}{72.687}
\emlineto{24.193}{72.577}
\emmoveto{24.193}{72.567}
\emlineto{24.390}{72.457}
\emmoveto{24.390}{72.447}
\emlineto{24.587}{72.338}
\emmoveto{24.587}{72.328}
\emlineto{24.783}{72.218}
\emmoveto{24.783}{72.208}
\emlineto{24.980}{72.099}
\emmoveto{24.980}{72.089}
\emlineto{25.177}{71.979}
\emmoveto{25.177}{71.969}
\emlineto{25.373}{71.859}
\emmoveto{25.373}{71.849}
\emlineto{25.570}{71.740}
\emmoveto{25.570}{71.730}
\emlineto{25.767}{71.620}
\emmoveto{25.767}{71.610}
\emlineto{25.963}{71.501}
\emmoveto{25.963}{71.491}
\emlineto{26.160}{71.381}
\emmoveto{26.160}{71.371}
\emlineto{26.357}{71.262}
\emmoveto{26.357}{71.252}
\emlineto{26.553}{71.142}
\emmoveto{26.553}{71.132}
\emlineto{26.750}{71.022}
\emmoveto{26.750}{71.012}
\emlineto{26.947}{70.903}
\emmoveto{26.947}{70.893}
\emlineto{27.143}{70.784}
\emmoveto{27.143}{70.774}
\emlineto{27.340}{70.664}
\emmoveto{27.340}{70.654}
\emlineto{27.537}{70.545}
\emmoveto{27.537}{70.535}
\emlineto{27.733}{70.425}
\emmoveto{27.733}{70.415}
\emlineto{27.930}{70.306}
\emmoveto{27.930}{70.296}
\emlineto{28.127}{70.186}
\emmoveto{28.127}{70.176}
\emlineto{28.323}{70.067}
\emmoveto{28.323}{70.057}
\emlineto{28.520}{69.948}
\emmoveto{28.520}{69.938}
\emlineto{28.717}{69.828}
\emmoveto{28.717}{69.818}
\emlineto{28.913}{69.709}
\emmoveto{28.913}{69.699}
\emlineto{29.110}{69.589}
\emmoveto{29.110}{69.579}
\emlineto{29.307}{69.470}
\emmoveto{29.307}{69.460}
\emlineto{29.503}{69.351}
\emmoveto{29.503}{69.341}
\emlineto{29.700}{69.232}
\emmoveto{29.700}{69.222}
\emlineto{29.897}{69.112}
\emmoveto{29.897}{69.102}
\emlineto{30.093}{68.993}
\emmoveto{30.093}{68.983}
\emlineto{30.290}{68.874}
\emmoveto{30.290}{68.864}
\emlineto{30.487}{68.755}
\emmoveto{30.487}{68.745}
\emlineto{30.683}{68.635}
\emmoveto{30.683}{68.625}
\emlineto{30.880}{68.516}
\emmoveto{30.880}{68.506}
\emlineto{31.077}{68.397}
\emmoveto{31.077}{68.387}
\emlineto{31.273}{68.278}
\emmoveto{31.273}{68.268}
\emlineto{31.470}{68.159}
\emmoveto{31.470}{68.149}
\emlineto{31.667}{68.040}
\emmoveto{31.667}{68.030}
\emlineto{31.863}{67.920}
\emmoveto{31.863}{67.910}
\emlineto{32.060}{67.801}
\emmoveto{32.060}{67.791}
\emlineto{32.257}{67.682}
\emmoveto{32.257}{67.672}
\emlineto{32.453}{67.563}
\emmoveto{32.453}{67.553}
\emlineto{32.650}{67.444}
\emmoveto{32.650}{67.434}
\emlineto{32.847}{67.325}
\emmoveto{32.847}{67.315}
\emlineto{33.043}{67.206}
\emmoveto{33.043}{67.196}
\emlineto{33.240}{67.087}
\emmoveto{33.240}{67.077}
\emlineto{33.437}{66.968}
\emmoveto{33.437}{66.958}
\emlineto{33.633}{66.849}
\emmoveto{33.633}{66.839}
\emlineto{33.830}{66.730}
\emmoveto{33.830}{66.720}
\emlineto{34.027}{66.611}
\emmoveto{34.027}{66.601}
\emlineto{34.223}{66.493}
\emmoveto{34.223}{66.483}
\emlineto{34.420}{66.374}
\emmoveto{34.420}{66.364}
\emlineto{34.617}{66.255}
\emmoveto{34.617}{66.245}
\emlineto{34.813}{66.136}
\emmoveto{34.813}{66.126}
\emlineto{35.010}{66.017}
\emmoveto{35.010}{66.007}
\emlineto{35.207}{65.898}
\emmoveto{35.207}{65.888}
\emlineto{35.403}{65.780}
\emmoveto{35.403}{65.770}
\emlineto{35.600}{65.661}
\emmoveto{35.600}{65.651}
\emlineto{35.797}{65.542}
\emmoveto{35.797}{65.532}
\emlineto{35.993}{65.424}
\emmoveto{35.993}{65.414}
\emlineto{36.190}{65.305}
\emmoveto{36.190}{65.295}
\emlineto{36.387}{65.186}
\emmoveto{36.387}{65.176}
\emlineto{36.583}{65.068}
\emmoveto{36.583}{65.058}
\emlineto{36.780}{64.949}
\emmoveto{36.780}{64.939}
\emlineto{36.977}{64.830}
\emmoveto{36.977}{64.820}
\emlineto{37.173}{64.712}
\emmoveto{37.173}{64.702}
\emlineto{37.370}{64.593}
\emmoveto{37.370}{64.583}
\emlineto{37.567}{64.475}
\emmoveto{37.567}{64.465}
\emlineto{37.763}{64.356}
\emmoveto{37.763}{64.346}
\emlineto{37.960}{64.238}
\emmoveto{37.960}{64.228}
\emlineto{38.157}{64.119}
\emmoveto{38.157}{64.109}
\emlineto{38.353}{64.001}
\emmoveto{38.353}{63.991}
\emlineto{38.550}{63.882}
\emmoveto{38.550}{63.872}
\emlineto{38.747}{63.764}
\emmoveto{38.747}{63.754}
\emlineto{38.943}{63.646}
\emmoveto{38.943}{63.636}
\emlineto{39.140}{63.527}
\emmoveto{39.140}{63.517}
\emlineto{39.337}{63.409}
\emmoveto{39.337}{63.399}
\emlineto{39.533}{63.291}
\emmoveto{39.533}{63.281}
\emlineto{39.730}{63.172}
\emmoveto{39.730}{63.162}
\emlineto{39.927}{63.054}
\emmoveto{39.927}{63.044}
\emlineto{40.123}{62.936}
\emmoveto{40.123}{62.926}
\emlineto{40.320}{62.818}
\emmoveto{40.320}{62.808}
\emlineto{40.517}{62.700}
\emmoveto{40.517}{62.690}
\emlineto{40.713}{62.582}
\emmoveto{40.713}{62.572}
\emlineto{40.910}{62.463}
\emmoveto{40.910}{62.453}
\emlineto{41.107}{62.345}
\emmoveto{41.107}{62.335}
\emlineto{41.303}{62.227}
\emmoveto{41.303}{62.217}
\emlineto{41.500}{62.109}
\emmoveto{41.500}{62.099}
\emlineto{41.697}{61.991}
\emmoveto{41.697}{61.981}
\emlineto{41.893}{61.873}
\emmoveto{41.893}{61.863}
\emlineto{42.090}{61.755}
\emmoveto{42.090}{61.745}
\emlineto{42.287}{61.637}
\emmoveto{42.287}{61.627}
\emlineto{42.483}{61.519}
\emmoveto{42.483}{61.509}
\emlineto{42.680}{61.402}
\emmoveto{42.680}{61.392}
\emlineto{42.877}{61.284}
\emmoveto{42.877}{61.274}
\emlineto{43.073}{61.166}
\emmoveto{43.073}{61.156}
\emlineto{43.270}{61.048}
\emmoveto{43.270}{61.038}
\emlineto{43.467}{60.930}
\emmoveto{43.467}{60.920}
\emlineto{43.663}{60.813}
\emmoveto{43.663}{60.803}
\emlineto{43.860}{60.695}
\emmoveto{43.860}{60.685}
\emlineto{44.057}{60.577}
\emmoveto{44.057}{60.567}
\emlineto{44.253}{60.459}
\emmoveto{44.253}{60.449}
\emlineto{44.450}{60.342}
\emmoveto{44.450}{60.332}
\emlineto{44.647}{60.224}
\emmoveto{44.647}{60.214}
\emlineto{44.843}{60.107}
\emmoveto{44.843}{60.097}
\emlineto{45.040}{59.989}
\emmoveto{45.040}{59.979}
\emlineto{45.237}{59.872}
\emmoveto{45.237}{59.862}
\emlineto{45.433}{59.754}
\emmoveto{45.433}{59.744}
\emlineto{45.630}{59.637}
\emmoveto{45.630}{59.627}
\emlineto{45.827}{59.519}
\emmoveto{45.827}{59.509}
\emlineto{46.023}{59.402}
\emmoveto{46.023}{59.392}
\emlineto{46.220}{59.284}
\emmoveto{46.220}{59.274}
\emlineto{46.417}{59.167}
\emmoveto{46.417}{59.157}
\emlineto{46.613}{59.050}
\emmoveto{46.613}{59.040}
\emlineto{46.810}{58.932}
\emmoveto{46.810}{58.922}
\emlineto{47.007}{58.815}
\emmoveto{47.007}{58.805}
\emlineto{47.203}{58.698}
\emmoveto{47.203}{58.688}
\emlineto{47.400}{58.581}
\emmoveto{47.400}{58.571}
\emlineto{47.597}{58.464}
\emmoveto{47.597}{58.454}
\emlineto{47.793}{58.346}
\emmoveto{47.793}{58.336}
\emlineto{47.990}{58.229}
\emmoveto{47.990}{58.219}
\emlineto{48.187}{58.112}
\emmoveto{48.187}{58.102}
\emlineto{48.383}{57.995}
\emmoveto{48.383}{57.985}
\emlineto{48.580}{57.878}
\emmoveto{48.580}{57.868}
\emlineto{48.777}{57.761}
\emmoveto{48.777}{57.751}
\emlineto{48.973}{57.644}
\emmoveto{48.973}{57.634}
\emlineto{49.170}{57.527}
\emmoveto{49.170}{57.517}
\emlineto{49.367}{57.411}
\emmoveto{49.367}{57.401}
\emlineto{49.563}{57.294}
\emmoveto{49.563}{57.284}
\emlineto{49.760}{57.177}
\emmoveto{49.760}{57.167}
\emlineto{49.957}{57.060}
\emmoveto{49.957}{57.050}
\emlineto{50.153}{56.943}
\emmoveto{50.153}{56.933}
\emlineto{50.350}{56.827}
\emmoveto{50.350}{56.817}
\emlineto{50.547}{56.710}
\emmoveto{50.547}{56.700}
\emlineto{50.743}{56.593}
\emmoveto{50.743}{56.583}
\emlineto{50.940}{56.477}
\emmoveto{50.940}{56.467}
\emlineto{51.137}{56.360}
\emmoveto{51.137}{56.350}
\emlineto{51.333}{56.244}
\emmoveto{51.333}{56.234}
\emlineto{51.530}{56.127}
\emmoveto{51.530}{56.117}
\emlineto{51.727}{56.011}
\emmoveto{51.727}{56.001}
\emlineto{51.923}{55.894}
\emmoveto{51.923}{55.884}
\emlineto{52.120}{55.778}
\emmoveto{52.120}{55.768}
\emlineto{52.317}{55.661}
\emmoveto{52.317}{55.651}
\emlineto{52.513}{55.545}
\emmoveto{52.513}{55.535}
\emlineto{52.710}{55.429}
\emmoveto{52.710}{55.419}
\emlineto{52.907}{55.312}
\emmoveto{52.907}{55.302}
\emlineto{53.103}{55.196}
\emmoveto{53.103}{55.186}
\emlineto{53.300}{55.080}
\emmoveto{53.300}{55.070}
\emlineto{53.497}{54.964}
\emmoveto{53.497}{54.954}
\emlineto{53.693}{54.848}
\emmoveto{53.693}{54.838}
\emlineto{53.890}{54.731}
\emmoveto{53.890}{54.721}
\emlineto{54.087}{54.615}
\emmoveto{54.087}{54.605}
\emlineto{54.283}{54.499}
\emmoveto{54.283}{54.489}
\emlineto{54.480}{54.383}
\emmoveto{54.480}{54.373}
\emlineto{54.677}{54.267}
\emmoveto{54.677}{54.257}
\emlineto{54.873}{54.152}
\emmoveto{54.873}{54.142}
\emlineto{55.070}{54.036}
\emmoveto{55.070}{54.026}
\emlineto{55.267}{53.920}
\emmoveto{55.267}{53.910}
\emlineto{55.463}{53.804}
\emmoveto{55.463}{53.794}
\emlineto{55.660}{53.688}
\emmoveto{55.660}{53.678}
\emlineto{55.857}{53.572}
\emmoveto{55.857}{53.562}
\emlineto{56.053}{53.457}
\emmoveto{56.053}{53.447}
\emlineto{56.250}{53.341}
\emmoveto{56.250}{53.331}
\emlineto{56.447}{53.225}
\emmoveto{56.447}{53.215}
\emlineto{56.643}{53.110}
\emmoveto{56.643}{53.100}
\emlineto{56.840}{52.994}
\emmoveto{56.840}{52.984}
\emlineto{57.037}{52.879}
\emmoveto{57.037}{52.869}
\emlineto{57.233}{52.763}
\emmoveto{57.233}{52.753}
\emlineto{57.430}{52.648}
\emmoveto{57.430}{52.638}
\emlineto{57.627}{52.532}
\emmoveto{57.627}{52.522}
\emlineto{57.823}{52.417}
\emmoveto{57.823}{52.407}
\emlineto{58.020}{52.302}
\emmoveto{58.020}{52.292}
\emlineto{58.217}{52.187}
\emmoveto{58.217}{52.177}
\emlineto{58.413}{52.071}
\emmoveto{58.413}{52.061}
\emlineto{58.610}{51.956}
\emmoveto{58.610}{51.946}
\emlineto{58.807}{51.841}
\emmoveto{58.807}{51.831}
\emlineto{59.003}{51.726}
\emmoveto{59.003}{51.716}
\emlineto{59.200}{51.611}
\emmoveto{59.200}{51.601}
\emlineto{59.397}{51.496}
\emmoveto{59.397}{51.486}
\emlineto{59.593}{51.381}
\emmoveto{59.593}{51.371}
\emlineto{59.790}{51.266}
\emmoveto{59.790}{51.256}
\emlineto{59.987}{51.151}
\emmoveto{59.987}{51.141}
\emlineto{60.183}{51.036}
\emmoveto{60.183}{51.026}
\emlineto{60.380}{50.921}
\emmoveto{60.380}{50.911}
\emlineto{60.577}{50.806}
\emmoveto{60.577}{50.796}
\emlineto{60.773}{50.691}
\emmoveto{60.773}{50.681}
\emlineto{60.970}{50.577}
\emmoveto{60.970}{50.567}
\emlineto{61.167}{50.462}
\emmoveto{61.167}{50.452}
\emlineto{61.363}{50.347}
\emmoveto{61.363}{50.337}
\emlineto{61.560}{50.233}
\emmoveto{61.560}{50.223}
\emlineto{61.757}{50.118}
\emmoveto{61.757}{50.108}
\emlineto{61.953}{50.004}
\emmoveto{61.953}{49.994}
\emlineto{62.150}{49.889}
\emmoveto{62.150}{49.879}
\emlineto{62.347}{49.775}
\emmoveto{62.347}{49.765}
\emlineto{62.543}{49.660}
\emmoveto{62.543}{49.650}
\emlineto{62.740}{49.546}
\emmoveto{62.740}{49.536}
\emlineto{62.937}{49.432}
\emmoveto{62.937}{49.422}
\emlineto{63.133}{49.317}
\emmoveto{63.133}{49.307}
\emlineto{63.330}{49.203}
\emmoveto{63.330}{49.193}
\emlineto{63.527}{49.089}
\emmoveto{63.527}{49.079}
\emlineto{63.723}{48.975}
\emmoveto{63.723}{48.965}
\emlineto{63.920}{48.861}
\emmoveto{63.920}{48.851}
\emlineto{64.117}{48.747}
\emmoveto{64.117}{48.737}
\emlineto{64.313}{48.633}
\emmoveto{64.313}{48.623}
\emlineto{64.510}{48.519}
\emmoveto{64.510}{48.509}
\emlineto{64.707}{48.405}
\emmoveto{64.707}{48.395}
\emlineto{64.903}{48.291}
\emmoveto{64.903}{48.281}
\emlineto{65.100}{48.177}
\emmoveto{65.100}{48.167}
\emlineto{65.297}{48.063}
\emmoveto{65.297}{48.053}
\emlineto{65.493}{47.950}
\emmoveto{65.493}{47.940}
\emlineto{65.690}{47.836}
\emmoveto{65.690}{47.826}
\emlineto{65.887}{47.722}
\emmoveto{65.887}{47.712}
\emlineto{66.083}{47.609}
\emmoveto{66.083}{47.599}
\emlineto{66.280}{47.495}
\emmoveto{66.280}{47.485}
\emlineto{66.477}{47.381}
\emmoveto{66.477}{47.371}
\emlineto{66.673}{47.268}
\emmoveto{66.673}{47.258}
\emlineto{66.870}{47.155}
\emmoveto{66.870}{47.145}
\emlineto{67.067}{47.041}
\emmoveto{67.067}{47.031}
\emlineto{67.263}{46.928}
\emmoveto{67.263}{46.918}
\emlineto{67.460}{46.814}
\emmoveto{67.460}{46.804}
\emlineto{67.657}{46.701}
\emmoveto{67.657}{46.691}
\emlineto{67.853}{46.588}
\emmoveto{67.853}{46.578}
\emlineto{68.050}{46.475}
\emmoveto{68.050}{46.465}
\emlineto{68.247}{46.362}
\emmoveto{68.247}{46.352}
\emlineto{68.443}{46.249}
\emmoveto{68.443}{46.239}
\emlineto{68.640}{46.136}
\emmoveto{68.640}{46.126}
\emlineto{68.837}{46.023}
\emmoveto{68.837}{46.013}
\emlineto{69.033}{45.910}
\emmoveto{69.033}{45.900}
\emlineto{69.230}{45.797}
\emmoveto{69.230}{45.787}
\emlineto{69.427}{45.684}
\emmoveto{69.427}{45.674}
\emlineto{69.623}{45.571}
\emmoveto{69.623}{45.561}
\emlineto{69.820}{45.458}
\emmoveto{69.820}{45.448}
\emlineto{70.017}{45.346}
\emmoveto{70.017}{45.336}
\emlineto{70.213}{45.233}
\emmoveto{70.213}{45.223}
\emlineto{70.410}{45.120}
\emmoveto{70.410}{45.110}
\emlineto{70.607}{45.008}
\emmoveto{70.607}{44.998}
\emlineto{70.803}{44.895}
\emmoveto{70.803}{44.885}
\emlineto{71.000}{44.783}
\emmoveto{71.000}{44.773}
\emlineto{71.197}{44.670}
\emmoveto{71.197}{44.660}
\emlineto{71.393}{44.558}
\emmoveto{71.393}{44.548}
\emlineto{71.590}{44.446}
\emmoveto{71.590}{44.436}
\emlineto{71.787}{44.333}
\emmoveto{71.787}{44.323}
\emlineto{71.983}{44.221}
\emmoveto{71.983}{44.211}
\emlineto{72.180}{44.109}
\emmoveto{72.180}{44.099}
\emlineto{72.377}{43.997}
\emmoveto{72.377}{43.987}
\emlineto{72.573}{43.885}
\emmoveto{72.573}{43.875}
\emlineto{72.770}{43.773}
\emmoveto{72.770}{43.763}
\emlineto{72.967}{43.661}
\emmoveto{72.967}{43.651}
\emlineto{73.163}{43.549}
\emmoveto{73.163}{43.539}
\emlineto{73.360}{43.437}
\emmoveto{73.360}{43.427}
\emlineto{73.557}{43.325}
\emmoveto{73.557}{43.315}
\emlineto{73.753}{43.213}
\emmoveto{73.753}{43.203}
\emlineto{73.950}{43.101}
\emmoveto{73.950}{43.091}
\emlineto{74.147}{42.990}
\emmoveto{74.147}{42.980}
\emlineto{74.343}{42.878}
\emmoveto{74.343}{42.868}
\emlineto{74.540}{42.766}
\emmoveto{74.540}{42.756}
\emlineto{74.737}{42.655}
\emmoveto{74.737}{42.645}
\emlineto{74.933}{42.543}
\emmoveto{74.933}{42.533}
\emlineto{75.130}{42.432}
\emmoveto{75.130}{42.422}
\emlineto{75.327}{42.320}
\emmoveto{75.327}{42.310}
\emlineto{75.523}{42.209}
\emmoveto{75.523}{42.199}
\emlineto{75.720}{42.098}
\emmoveto{75.720}{42.088}
\emlineto{75.917}{41.987}
\emmoveto{75.917}{41.977}
\emlineto{76.113}{41.875}
\emmoveto{76.113}{41.865}
\emlineto{76.310}{41.764}
\emmoveto{76.310}{41.754}
\emlineto{76.507}{41.653}
\emmoveto{76.507}{41.643}
\emlineto{76.703}{41.542}
\emmoveto{76.703}{41.532}
\emlineto{76.900}{41.431}
\emmoveto{76.900}{41.421}
\emlineto{77.097}{41.320}
\emmoveto{77.097}{41.310}
\emlineto{77.293}{41.209}
\emmoveto{77.293}{41.199}
\emlineto{77.490}{41.098}
\emmoveto{77.490}{41.088}
\emlineto{77.687}{40.987}
\emmoveto{77.687}{40.977}
\emlineto{77.883}{40.877}
\emmoveto{77.883}{40.867}
\emlineto{78.080}{40.766}
\emmoveto{78.080}{40.756}
\emlineto{78.277}{40.655}
\emmoveto{78.277}{40.645}
\emlineto{78.473}{40.545}
\emmoveto{78.473}{40.535}
\emlineto{78.670}{40.434}
\emmoveto{78.670}{40.424}
\emlineto{78.867}{40.324}
\emmoveto{78.867}{40.314}
\emlineto{79.063}{40.213}
\emmoveto{79.063}{40.203}
\emlineto{79.260}{40.103}
\emmoveto{79.260}{40.093}
\emlineto{79.457}{39.993}
\emmoveto{79.457}{39.983}
\emlineto{79.653}{39.882}
\emmoveto{79.653}{39.872}
\emlineto{79.850}{39.772}
\emmoveto{79.850}{39.762}
\emlineto{80.047}{39.662}
\emmoveto{80.047}{39.652}
\emlineto{80.243}{39.552}
\emmoveto{80.243}{39.542}
\emlineto{80.440}{39.442}
\emmoveto{80.440}{39.432}
\emlineto{80.637}{39.332}
\emmoveto{80.637}{39.322}
\emlineto{80.833}{39.222}
\emmoveto{80.833}{39.212}
\emlineto{81.030}{39.112}
\emmoveto{81.030}{39.102}
\emlineto{81.227}{39.002}
\emmoveto{81.227}{38.992}
\emlineto{81.423}{38.892}
\emmoveto{81.423}{38.882}
\emlineto{81.620}{38.782}
\emmoveto{81.620}{38.772}
\emlineto{81.817}{38.673}
\emmoveto{81.817}{38.663}
\emlineto{82.013}{38.563}
\emmoveto{82.013}{38.553}
\emlineto{82.210}{38.453}
\emmoveto{82.210}{38.443}
\emlineto{82.407}{38.344}
\emmoveto{82.407}{38.334}
\emlineto{82.603}{38.234}
\emmoveto{82.603}{38.224}
\emlineto{82.800}{38.125}
\emmoveto{82.800}{38.115}
\emlineto{82.997}{38.016}
\emmoveto{82.997}{38.006}
\emlineto{83.193}{37.906}
\emmoveto{83.193}{37.896}
\emlineto{83.390}{37.797}
\emmoveto{83.390}{37.787}
\emlineto{83.587}{37.688}
\emmoveto{83.587}{37.678}
\emlineto{83.783}{37.579}
\emmoveto{83.783}{37.569}
\emlineto{83.980}{37.470}
\emmoveto{83.980}{37.460}
\emlineto{84.177}{37.361}
\emmoveto{84.177}{37.351}
\emlineto{84.373}{37.252}
\emmoveto{84.373}{37.242}
\emlineto{84.570}{37.143}
\emmoveto{84.570}{37.133}
\emlineto{84.767}{37.034}
\emmoveto{84.767}{37.024}
\emlineto{84.963}{36.925}
\emmoveto{84.963}{36.915}
\emlineto{85.160}{36.816}
\emmoveto{85.160}{36.806}
\emlineto{85.357}{36.707}
\emmoveto{85.357}{36.697}
\emlineto{85.553}{36.599}
\emmoveto{85.553}{36.589}
\emlineto{85.750}{36.490}
\emmoveto{85.750}{36.480}
\emlineto{85.947}{36.382}
\emmoveto{85.947}{36.372}
\emlineto{86.143}{36.273}
\emmoveto{86.143}{36.263}
\emlineto{86.340}{36.165}
\emmoveto{86.340}{36.155}
\emlineto{86.537}{36.056}
\emmoveto{86.537}{36.046}
\emlineto{86.733}{35.948}
\emmoveto{86.733}{35.938}
\emlineto{86.930}{35.840}
\emmoveto{86.930}{35.830}
\emlineto{87.127}{35.731}
\emmoveto{87.127}{35.721}
\emlineto{87.323}{35.623}
\emmoveto{87.323}{35.613}
\emlineto{87.520}{35.515}
\emmoveto{87.520}{35.505}
\emlineto{87.717}{35.407}
\emmoveto{87.717}{35.397}
\emlineto{87.913}{35.299}
\emmoveto{87.913}{35.289}
\emlineto{88.110}{35.191}
\emmoveto{88.110}{35.181}
\emlineto{88.307}{35.083}
\emmoveto{88.307}{35.073}
\emlineto{88.503}{34.975}
\emmoveto{88.503}{34.965}
\emlineto{88.700}{34.868}
\emmoveto{88.700}{34.858}
\emlineto{88.897}{34.760}
\emmoveto{88.897}{34.750}
\emlineto{89.093}{34.652}
\emmoveto{89.093}{34.642}
\emlineto{89.290}{34.545}
\emmoveto{89.290}{34.535}
\emlineto{89.487}{34.437}
\emmoveto{89.487}{34.427}
\emlineto{89.683}{34.330}
\emmoveto{89.683}{34.320}
\emlineto{89.880}{34.222}
\emmoveto{89.880}{34.212}
\emlineto{90.077}{34.115}
\emmoveto{90.077}{34.105}
\emlineto{90.273}{34.008}
\emmoveto{90.273}{33.998}
\emlineto{90.470}{33.900}
\emmoveto{90.470}{33.890}
\emlineto{90.667}{33.793}
\emmoveto{90.667}{33.783}
\emlineto{90.863}{33.686}
\emmoveto{90.863}{33.676}
\emlineto{91.060}{33.579}
\emmoveto{91.060}{33.569}
\emlineto{91.257}{33.472}
\emmoveto{91.257}{33.462}
\emlineto{91.453}{33.365}
\emmoveto{91.453}{33.355}
\emlineto{91.650}{33.258}
\emmoveto{91.650}{33.248}
\emlineto{91.847}{33.151}
\emmoveto{91.847}{33.141}
\emlineto{92.043}{33.045}
\emmoveto{92.043}{33.035}
\emlineto{92.240}{32.938}
\emmoveto{92.240}{32.928}
\emlineto{92.437}{32.831}
\emmoveto{92.437}{32.821}
\emlineto{92.633}{32.724}
\emmoveto{92.633}{32.714}
\emlineto{92.830}{32.618}
\emmoveto{92.830}{32.608}
\emlineto{93.027}{32.511}
\emmoveto{93.027}{32.501}
\emlineto{93.223}{32.405}
\emmoveto{93.223}{32.395}
\emlineto{93.420}{32.299}
\emmoveto{93.420}{32.289}
\emlineto{93.617}{32.192}
\emmoveto{93.617}{32.182}
\emlineto{93.813}{32.086}
\emmoveto{93.813}{32.076}
\emlineto{94.010}{31.980}
\emmoveto{94.010}{31.970}
\emlineto{94.207}{31.874}
\emmoveto{94.207}{31.864}
\emlineto{94.403}{31.768}
\emmoveto{94.403}{31.758}
\emlineto{94.600}{31.662}
\emmoveto{94.600}{31.652}
\emlineto{94.797}{31.556}
\emmoveto{94.797}{31.546}
\emlineto{94.993}{31.450}
\emmoveto{94.993}{31.440}
\emlineto{95.190}{31.344}
\emmoveto{95.190}{31.334}
\emlineto{95.387}{31.238}
\emmoveto{95.387}{31.228}
\emlineto{95.583}{31.132}
\emmoveto{95.583}{31.122}
\emlineto{95.780}{31.027}
\emmoveto{95.780}{31.017}
\emlineto{95.977}{30.921}
\emmoveto{95.977}{30.911}
\emlineto{96.173}{30.816}
\emmoveto{96.173}{30.806}
\emlineto{96.370}{30.710}
\emmoveto{96.370}{30.700}
\emlineto{96.567}{30.605}
\emmoveto{96.567}{30.595}
\emlineto{96.763}{30.499}
\emmoveto{96.763}{30.489}
\emlineto{96.960}{30.394}
\emmoveto{96.960}{30.384}
\emlineto{97.157}{30.289}
\emmoveto{97.157}{30.279}
\emlineto{97.353}{30.184}
\emmoveto{97.353}{30.174}
\emlineto{97.550}{30.079}
\emmoveto{97.550}{30.069}
\emlineto{97.747}{29.974}
\emmoveto{97.747}{29.964}
\emlineto{97.943}{29.869}
\emmoveto{97.943}{29.859}
\emlineto{98.140}{29.764}
\emmoveto{98.140}{29.754}
\emlineto{98.337}{29.659}
\emmoveto{98.337}{29.649}
\emlineto{98.533}{29.554}
\emmoveto{98.533}{29.544}
\emlineto{98.730}{29.449}
\emmoveto{98.730}{29.439}
\emlineto{98.927}{29.344}
\emmoveto{98.927}{29.334}
\emlineto{99.123}{29.240}
\emmoveto{99.123}{29.230}
\emlineto{99.320}{29.135}
\emmoveto{99.320}{29.125}
\emlineto{99.517}{29.031}
\emmoveto{99.517}{29.021}
\emlineto{99.713}{28.926}
\emmoveto{99.713}{28.916}
\emlineto{99.910}{28.822}
\emmoveto{99.910}{28.812}
\emlineto{100.107}{28.718}
\emmoveto{100.107}{28.708}
\emlineto{100.303}{28.613}
\emmoveto{100.303}{28.603}
\emlineto{100.500}{28.509}
\emmoveto{100.500}{28.499}
\emlineto{100.697}{28.405}
\emmoveto{100.697}{28.395}
\emlineto{100.893}{28.301}
\emmoveto{100.893}{28.291}
\emlineto{101.090}{28.197}
\emmoveto{101.090}{28.187}
\emlineto{101.287}{28.093}
\emmoveto{101.287}{28.083}
\emlineto{101.483}{27.989}
\emmoveto{101.483}{27.979}
\emlineto{101.680}{27.885}
\emmoveto{101.680}{27.875}
\emlineto{101.877}{27.781}
\emmoveto{101.877}{27.771}
\emlineto{102.073}{27.678}
\emmoveto{102.073}{27.668}
\emlineto{102.270}{27.574}
\emmoveto{102.270}{27.564}
\emlineto{102.467}{27.471}
\emmoveto{102.467}{27.461}
\emlineto{102.663}{27.367}
\emmoveto{102.663}{27.357}
\emlineto{102.860}{27.264}
\emmoveto{102.860}{27.254}
\emlineto{103.057}{27.160}
\emmoveto{103.057}{27.150}
\emlineto{103.253}{27.057}
\emmoveto{103.253}{27.047}
\emlineto{103.450}{26.954}
\emmoveto{103.450}{26.944}
\emlineto{103.647}{26.850}
\emmoveto{103.647}{26.840}
\emlineto{103.843}{26.747}
\emmoveto{103.843}{26.737}
\emlineto{104.040}{26.644}
\emmoveto{104.040}{26.634}
\emlineto{104.237}{26.541}
\emmoveto{104.237}{26.531}
\emlineto{104.433}{26.438}
\emmoveto{104.433}{26.428}
\emlineto{104.630}{26.335}
\emmoveto{104.630}{26.325}
\emlineto{104.827}{26.233}
\emmoveto{104.827}{26.223}
\emlineto{105.023}{26.130}
\emmoveto{105.023}{26.120}
\emlineto{105.220}{26.027}
\emmoveto{105.220}{26.017}
\emlineto{105.417}{25.924}
\emmoveto{105.417}{25.914}
\emlineto{105.613}{25.822}
\emmoveto{105.613}{25.812}
\emlineto{105.810}{25.719}
\emmoveto{105.810}{25.709}
\emlineto{106.007}{25.617}
\emmoveto{106.007}{25.607}
\emlineto{106.203}{25.515}
\emmoveto{106.203}{25.505}
\emlineto{106.400}{25.412}
\emmoveto{106.400}{25.402}
\emlineto{106.597}{25.310}
\emmoveto{106.597}{25.300}
\emlineto{106.793}{25.208}
\emmoveto{106.793}{25.198}
\emlineto{106.990}{25.106}
\emmoveto{106.990}{25.096}
\emlineto{107.187}{25.004}
\emmoveto{107.187}{24.994}
\emlineto{107.383}{24.902}
\emmoveto{107.383}{24.892}
\emlineto{107.580}{24.800}
\emmoveto{107.580}{24.790}
\emlineto{107.777}{24.698}
\emmoveto{107.777}{24.688}
\emlineto{107.973}{24.596}
\emmoveto{107.973}{24.586}
\emlineto{108.170}{24.494}
\emmoveto{108.170}{24.484}
\emlineto{108.367}{24.393}
\emmoveto{108.367}{24.383}
\emlineto{108.563}{24.291}
\emmoveto{108.563}{24.281}
\emlineto{108.760}{24.189}
\emmoveto{108.760}{24.179}
\emlineto{108.957}{24.088}
\emmoveto{108.957}{24.078}
\emlineto{109.153}{23.986}
\emmoveto{109.153}{23.976}
\emlineto{109.350}{23.885}
\emmoveto{109.350}{23.875}
\emlineto{109.547}{23.784}
\emmoveto{109.547}{23.774}
\emlineto{109.743}{23.683}
\emmoveto{109.743}{23.673}
\emlineto{109.940}{23.581}
\emmoveto{109.940}{23.571}
\emlineto{110.137}{23.480}
\emmoveto{110.137}{23.470}
\emlineto{110.333}{23.379}
\emmoveto{110.333}{23.369}
\emlineto{110.530}{23.278}
\emmoveto{110.530}{23.268}
\emlineto{110.727}{23.177}
\emmoveto{110.727}{23.167}
\emlineto{110.923}{23.077}
\emmoveto{110.923}{23.067}
\emlineto{111.120}{22.976}
\emmoveto{111.120}{22.966}
\emlineto{111.317}{22.875}
\emmoveto{111.317}{22.865}
\emlineto{111.513}{22.774}
\emmoveto{111.513}{22.764}
\emlineto{111.710}{22.674}
\emmoveto{111.710}{22.664}
\emlineto{111.907}{22.573}
\emmoveto{111.907}{22.563}
\emlineto{112.103}{22.473}
\emmoveto{112.103}{22.463}
\emlineto{112.300}{22.373}
\emmoveto{112.300}{22.363}
\emlineto{112.497}{22.272}
\emmoveto{112.497}{22.262}
\emlineto{112.693}{22.172}
\emmoveto{112.693}{22.162}
\emlineto{112.890}{22.072}
\emmoveto{112.890}{22.062}
\emlineto{113.087}{21.972}
\emmoveto{113.087}{21.962}
\emlineto{113.283}{21.872}
\emmoveto{113.283}{21.862}
\emlineto{113.480}{21.772}
\emmoveto{113.480}{21.762}
\emlineto{113.677}{21.672}
\emmoveto{113.677}{21.662}
\emlineto{113.873}{21.572}
\emmoveto{113.873}{21.562}
\emlineto{114.070}{21.472}
\emmoveto{114.070}{21.462}
\emlineto{114.267}{21.372}
\emmoveto{114.267}{21.362}
\emlineto{114.463}{21.273}
\emmoveto{114.463}{21.263}
\emlineto{114.660}{21.173}
\emmoveto{114.660}{21.163}
\emlineto{114.857}{21.074}
\emmoveto{114.857}{21.064}
\emlineto{115.053}{20.974}
\emmoveto{115.053}{20.964}
\emlineto{115.250}{20.875}
\emmoveto{115.250}{20.865}
\emlineto{115.447}{20.775}
\emmoveto{115.447}{20.765}
\emlineto{115.643}{20.676}
\emmoveto{115.643}{20.666}
\emlineto{115.840}{20.577}
\emmoveto{115.840}{20.567}
\emlineto{116.037}{20.478}
\emmoveto{116.037}{20.468}
\emlineto{116.233}{20.379}
\emmoveto{116.233}{20.369}
\emlineto{116.430}{20.280}
\emmoveto{116.430}{20.270}
\emlineto{116.627}{20.181}
\emmoveto{116.627}{20.171}
\emlineto{116.823}{20.082}
\emmoveto{116.823}{20.072}
\emlineto{117.020}{19.983}
\emmoveto{117.020}{19.973}
\emlineto{117.217}{19.884}
\emmoveto{117.217}{19.874}
\emlineto{117.413}{19.786}
\emmoveto{117.413}{19.776}
\emlineto{117.610}{19.687}
\emmoveto{117.610}{19.677}
\emlineto{117.807}{19.589}
\emmoveto{117.807}{19.579}
\emlineto{118.003}{19.490}
\emmoveto{118.003}{19.480}
\emlineto{118.200}{19.392}
\emmoveto{118.200}{19.382}
\emlineto{118.397}{19.293}
\emmoveto{118.397}{19.283}
\emlineto{118.593}{19.195}
\emmoveto{118.593}{19.185}
\emlineto{118.790}{19.097}
\emmoveto{118.790}{19.087}
\emlineto{118.987}{18.999}
\emmoveto{118.987}{18.989}
\emlineto{119.183}{18.901}
\emmoveto{119.183}{18.891}
\emlineto{119.380}{18.803}
\emmoveto{119.380}{18.793}
\emlineto{119.577}{18.705}
\emmoveto{119.577}{18.695}
\emlineto{119.773}{18.607}
\emmoveto{119.773}{18.597}
\emlineto{119.970}{18.509}
\emmoveto{119.970}{18.499}
\emlineto{120.167}{18.411}
\emmoveto{120.167}{18.401}
\emlineto{120.363}{18.314}
\emmoveto{120.363}{18.304}
\emlineto{120.560}{18.216}
\emmoveto{120.560}{18.206}
\emlineto{120.757}{18.119}
\emmoveto{120.757}{18.109}
\emlineto{120.953}{18.021}
\emmoveto{120.953}{18.011}
\emlineto{121.150}{17.924}
\emmoveto{121.150}{17.914}
\emlineto{121.347}{17.827}
\emmoveto{121.347}{17.817}
\emlineto{121.543}{17.729}
\emmoveto{121.543}{17.719}
\emlineto{121.740}{17.632}
\emmoveto{121.740}{17.622}
\emlineto{121.937}{17.535}
\emmoveto{121.937}{17.525}
\emlineto{122.133}{17.438}
\emmoveto{122.133}{17.428}
\emlineto{122.330}{17.341}
\emmoveto{122.330}{17.331}
\emlineto{122.527}{17.244}
\emmoveto{122.527}{17.234}
\emlineto{122.723}{17.147}
\emmoveto{122.723}{17.137}
\emlineto{122.920}{17.050}
\emmoveto{122.920}{17.040}
\emlineto{123.117}{16.954}
\emmoveto{123.117}{16.944}
\emlineto{123.313}{16.857}
\emmoveto{123.313}{16.847}
\emlineto{123.510}{16.760}
\emmoveto{123.510}{16.750}
\emlineto{123.707}{16.664}
\emmoveto{123.707}{16.654}
\emlineto{123.903}{16.567}
\emmoveto{123.903}{16.557}
\emlineto{124.100}{16.471}
\emmoveto{124.100}{16.461}
\emlineto{124.297}{16.375}
\emmoveto{124.297}{16.365}
\emlineto{124.493}{16.278}
\emmoveto{124.493}{16.268}
\emlineto{124.690}{16.182}
\emmoveto{124.690}{16.172}
\emlineto{124.887}{16.086}
\emmoveto{124.887}{16.076}
\emlineto{125.083}{15.990}
\emmoveto{125.083}{15.980}
\emlineto{125.280}{15.894}
\emmoveto{125.280}{15.884}
\emlineto{125.477}{15.798}
\emmoveto{125.477}{15.788}
\emlineto{125.673}{15.702}
\emmoveto{125.673}{15.692}
\emlineto{125.870}{15.607}
\emmoveto{125.870}{15.597}
\emlineto{126.067}{15.511}
\emmoveto{126.067}{15.501}
\emlineto{126.263}{15.415}
\emmoveto{126.263}{15.405}
\emlineto{126.460}{15.320}
\emmoveto{126.460}{15.310}
\emlineto{126.657}{15.224}
\emmoveto{126.657}{15.214}
\emlineto{126.853}{15.129}
\emmoveto{126.853}{15.119}
\emlineto{127.050}{15.033}
\emmoveto{127.050}{15.023}
\emlineto{127.247}{14.938}
\emmoveto{127.247}{14.928}
\emlineto{127.443}{14.843}
\emmoveto{127.443}{14.833}
\emlineto{127.640}{14.748}
\emmoveto{127.640}{14.738}
\emlineto{127.837}{14.653}
\emmoveto{127.837}{14.643}
\emlineto{128.033}{14.558}
\emmoveto{128.033}{14.548}
\emlineto{128.230}{14.463}
\emmoveto{128.230}{14.453}
\emlineto{128.427}{14.368}
\emmoveto{128.427}{14.358}
\emlineto{128.623}{14.273}
\emmoveto{128.623}{14.263}
\emlineto{128.820}{14.178}
\emmoveto{128.820}{14.168}
\emlineto{129.017}{14.084}
\emmoveto{129.017}{14.074}
\emlineto{129.213}{13.989}
\emmoveto{129.213}{13.979}
\emlineto{129.410}{13.894}
\emmoveto{129.410}{13.884}
\emlineto{129.607}{13.800}
\emmoveto{129.607}{13.790}
\emlineto{129.803}{13.706}
\emshow{48.580}{48.500}{vq}
\emmoveto{12.000}{80.000}
\emlineto{12.197}{79.892}
\emmoveto{12.197}{79.882}
\emlineto{12.393}{79.779}
\emmoveto{12.393}{79.769}
\emlineto{12.590}{79.671}
\emmoveto{12.590}{79.661}
\emlineto{12.787}{79.566}
\emmoveto{12.787}{79.556}
\emlineto{12.983}{79.466}
\emmoveto{12.983}{79.456}
\emlineto{13.180}{79.370}
\emmoveto{13.180}{79.360}
\emlineto{13.377}{79.277}
\emmoveto{13.377}{79.267}
\emlineto{13.573}{79.188}
\emmoveto{13.573}{79.178}
\emlineto{13.770}{79.103}
\emmoveto{13.770}{79.093}
\emlineto{13.967}{79.021}
\emmoveto{13.967}{79.011}
\emlineto{14.163}{78.942}
\emmoveto{14.163}{78.932}
\emlineto{14.360}{78.866}
\emmoveto{14.360}{78.856}
\emlineto{14.557}{78.794}
\emmoveto{14.557}{78.784}
\emlineto{14.753}{78.724}
\emmoveto{14.753}{78.714}
\emlineto{14.950}{78.656}
\emmoveto{14.950}{78.646}
\emlineto{15.147}{78.592}
\emmoveto{15.147}{78.582}
\emlineto{15.343}{78.530}
\emmoveto{15.343}{78.520}
\emlineto{15.540}{78.470}
\emmoveto{15.540}{78.460}
\emlineto{15.737}{78.413}
\emmoveto{15.737}{78.403}
\emlineto{15.933}{78.358}
\emmoveto{15.933}{78.348}
\emlineto{16.130}{78.305}
\emmoveto{16.130}{78.295}
\emlineto{16.327}{78.255}
\emmoveto{16.327}{78.245}
\emlineto{16.523}{78.206}
\emmoveto{16.523}{78.196}
\emlineto{16.720}{78.159}
\emmoveto{16.720}{78.149}
\emlineto{16.917}{78.114}
\emmoveto{16.917}{78.104}
\emlineto{17.113}{78.071}
\emmoveto{17.113}{78.061}
\emlineto{17.310}{78.029}
\emmoveto{17.310}{78.019}
\emlineto{17.507}{77.989}
\emmoveto{17.507}{77.979}
\emlineto{17.703}{77.951}
\emmoveto{17.703}{77.941}
\emlineto{17.900}{77.914}
\emmoveto{17.900}{77.904}
\emlineto{18.097}{77.879}
\emmoveto{18.097}{77.869}
\emlineto{18.293}{77.845}
\emmoveto{18.293}{77.835}
\emlineto{18.490}{77.812}
\emmoveto{18.490}{77.802}
\emlineto{18.687}{77.743}
\emmoveto{18.687}{77.733}
\emlineto{18.883}{77.720}
\emmoveto{18.883}{77.710}
\emlineto{19.080}{77.686}
\emmoveto{19.080}{77.676}
\emlineto{19.277}{77.644}
\emmoveto{19.277}{77.634}
\emlineto{19.473}{77.595}
\emmoveto{19.473}{77.585}
\emlineto{19.670}{77.539}
\emmoveto{19.670}{77.529}
\emlineto{19.867}{77.529}
\emmoveto{19.867}{77.519}
\emlineto{20.063}{77.485}
\emmoveto{20.063}{77.475}
\emlineto{20.260}{77.416}
\emmoveto{20.260}{77.406}
\emlineto{20.457}{77.347}
\emmoveto{20.457}{77.337}
\emlineto{20.653}{77.332}
\emmoveto{20.653}{77.322}
\emlineto{20.850}{77.258}
\emmoveto{20.850}{77.248}
\emlineto{21.047}{77.189}
\emmoveto{21.047}{77.179}
\emlineto{21.243}{77.158}
\emmoveto{21.243}{77.148}
\emlineto{21.440}{77.085}
\emmoveto{21.440}{77.075}
\emlineto{21.637}{77.049}
\emmoveto{21.637}{77.039}
\emlineto{21.833}{76.977}
\emmoveto{21.833}{76.967}
\emlineto{22.030}{76.924}
\emmoveto{22.030}{76.914}
\emlineto{22.227}{76.866}
\emmoveto{22.227}{76.856}
\emlineto{22.423}{76.802}
\emmoveto{22.423}{76.792}
\emlineto{22.620}{76.754}
\emmoveto{22.620}{76.744}
\emlineto{22.817}{76.691}
\emmoveto{22.817}{76.681}
\emlineto{23.013}{76.640}
\emmoveto{23.013}{76.630}
\emlineto{23.210}{76.581}
\emmoveto{23.210}{76.571}
\emlineto{23.407}{76.525}
\emmoveto{23.407}{76.515}
\emlineto{23.603}{76.470}
\emmoveto{23.603}{76.460}
\emlineto{23.800}{76.416}
\emmoveto{23.800}{76.406}
\emlineto{23.997}{76.358}
\emmoveto{23.997}{76.348}
\emlineto{24.193}{76.306}
\emmoveto{24.193}{76.296}
\emlineto{24.390}{76.257}
\emmoveto{24.390}{76.247}
\emlineto{24.587}{76.196}
\emmoveto{24.587}{76.186}
\emlineto{24.783}{76.148}
\emmoveto{24.783}{76.138}
\emlineto{24.980}{76.102}
\emmoveto{24.980}{76.092}
\emlineto{25.177}{76.054}
\emmoveto{25.177}{76.044}
\emlineto{25.373}{75.991}
\emmoveto{25.373}{75.981}
\emlineto{25.570}{75.947}
\emmoveto{25.570}{75.937}
\emlineto{25.767}{75.905}
\emmoveto{25.767}{75.895}
\emlineto{25.963}{75.861}
\emmoveto{25.963}{75.851}
\emlineto{26.160}{75.791}
\emmoveto{26.160}{75.781}
\emlineto{26.357}{75.747}
\emmoveto{26.357}{75.737}
\emlineto{26.553}{75.705}
\emmoveto{26.553}{75.695}
\emlineto{26.750}{75.665}
\emmoveto{26.750}{75.655}
\emlineto{26.947}{75.603}
\emmoveto{26.947}{75.593}
\emlineto{27.143}{75.544}
\emmoveto{27.143}{75.534}
\emlineto{27.340}{75.505}
\emmoveto{27.340}{75.495}
\emlineto{27.537}{75.468}
\emmoveto{27.537}{75.458}
\emlineto{27.733}{75.416}
\emmoveto{27.733}{75.406}
\emlineto{27.930}{75.346}
\emmoveto{27.930}{75.336}
\emlineto{28.127}{75.306}
\emmoveto{28.127}{75.296}
\emlineto{28.323}{75.271}
\emmoveto{28.323}{75.261}
\emlineto{28.520}{75.214}
\emmoveto{28.520}{75.204}
\emlineto{28.717}{75.146}
\emmoveto{28.717}{75.136}
\emlineto{28.913}{75.109}
\emmoveto{28.913}{75.099}
\emlineto{29.110}{75.070}
\emmoveto{29.110}{75.060}
\emlineto{29.307}{75.003}
\emmoveto{29.307}{74.993}
\emlineto{29.503}{74.945}
\emmoveto{29.503}{74.935}
\emlineto{29.700}{74.917}
\emmoveto{29.700}{74.907}
\emlineto{29.897}{74.852}
\emmoveto{29.897}{74.842}
\emlineto{30.093}{74.789}
\emmoveto{30.093}{74.779}
\emlineto{30.290}{74.756}
\emmoveto{30.290}{74.746}
\emlineto{30.487}{74.698}
\emmoveto{30.487}{74.688}
\emlineto{30.683}{74.637}
\emmoveto{30.683}{74.627}
\emlineto{30.880}{74.596}
\emmoveto{30.880}{74.586}
\emlineto{31.077}{74.541}
\emmoveto{31.077}{74.531}
\emlineto{31.273}{74.483}
\emmoveto{31.273}{74.473}
\emlineto{31.470}{74.435}
\emmoveto{31.470}{74.425}
\emlineto{31.667}{74.384}
\emmoveto{31.667}{74.374}
\emlineto{31.863}{74.329}
\emmoveto{31.863}{74.319}
\emlineto{32.060}{74.277}
\emmoveto{32.060}{74.267}
\emlineto{32.257}{74.226}
\emmoveto{32.257}{74.216}
\emlineto{32.453}{74.174}
\emmoveto{32.453}{74.164}
\emlineto{32.650}{74.125}
\emmoveto{32.650}{74.115}
\emlineto{32.847}{74.067}
\emmoveto{32.847}{74.057}
\emlineto{33.043}{74.018}
\emmoveto{33.043}{74.008}
\emlineto{33.240}{73.972}
\emmoveto{33.240}{73.962}
\emlineto{33.437}{73.909}
\emmoveto{33.437}{73.899}
\emlineto{33.633}{73.861}
\emmoveto{33.633}{73.851}
\emlineto{33.830}{73.817}
\emmoveto{33.830}{73.807}
\emlineto{34.027}{73.766}
\emmoveto{34.027}{73.756}
\emlineto{34.223}{73.703}
\emmoveto{34.223}{73.693}
\emlineto{34.420}{73.661}
\emmoveto{34.420}{73.651}
\emlineto{34.617}{73.621}
\emmoveto{34.617}{73.611}
\emlineto{34.813}{73.554}
\emmoveto{34.813}{73.544}
\emlineto{35.010}{73.503}
\emmoveto{35.010}{73.493}
\emlineto{35.207}{73.465}
\emmoveto{35.207}{73.455}
\emlineto{35.403}{73.410}
\emmoveto{35.403}{73.400}
\emlineto{35.600}{73.348}
\emmoveto{35.600}{73.338}
\emlineto{35.797}{73.305}
\emmoveto{35.797}{73.295}
\emlineto{35.993}{73.261}
\emmoveto{35.993}{73.251}
\emlineto{36.190}{73.202}
\emmoveto{36.190}{73.192}
\emlineto{36.387}{73.145}
\emmoveto{36.387}{73.135}
\emlineto{36.583}{73.110}
\emmoveto{36.583}{73.100}
\emlineto{36.780}{73.052}
\emmoveto{36.780}{73.042}
\emlineto{36.977}{72.997}
\emmoveto{36.977}{72.987}
\emlineto{37.173}{72.948}
\emmoveto{37.173}{72.938}
\emlineto{37.370}{72.899}
\emmoveto{37.370}{72.889}
\emlineto{37.567}{72.846}
\emmoveto{37.567}{72.836}
\emlineto{37.763}{72.796}
\emmoveto{37.763}{72.786}
\emlineto{37.960}{72.744}
\emmoveto{37.960}{72.734}
\emlineto{38.157}{72.693}
\emmoveto{38.157}{72.683}
\emlineto{38.353}{72.645}
\emmoveto{38.353}{72.635}
\emlineto{38.550}{72.590}
\emmoveto{38.550}{72.580}
\emlineto{38.747}{72.538}
\emmoveto{38.747}{72.528}
\emlineto{38.943}{72.492}
\emmoveto{38.943}{72.482}
\emlineto{39.140}{72.442}
\emmoveto{39.140}{72.432}
\emlineto{39.337}{72.380}
\emmoveto{39.337}{72.370}
\emlineto{39.533}{72.337}
\emmoveto{39.533}{72.327}
\emlineto{39.730}{72.293}
\emmoveto{39.730}{72.283}
\emlineto{39.927}{72.229}
\emmoveto{39.927}{72.219}
\emlineto{40.123}{72.180}
\emmoveto{40.123}{72.170}
\emlineto{40.320}{72.141}
\emmoveto{40.320}{72.131}
\emlineto{40.517}{72.080}
\emmoveto{40.517}{72.070}
\emlineto{40.713}{72.021}
\emmoveto{40.713}{72.011}
\emlineto{40.910}{71.985}
\emmoveto{40.910}{71.975}
\emlineto{41.107}{71.928}
\emmoveto{41.107}{71.918}
\emlineto{41.303}{71.872}
\emmoveto{41.303}{71.862}
\emlineto{41.500}{71.827}
\emmoveto{41.500}{71.817}
\emlineto{41.697}{71.775}
\emmoveto{41.697}{71.765}
\emlineto{41.893}{71.722}
\emmoveto{41.893}{71.712}
\emlineto{42.090}{71.671}
\emmoveto{42.090}{71.661}
\emlineto{42.287}{71.620}
\emmoveto{42.287}{71.610}
\emlineto{42.483}{71.570}
\emmoveto{42.483}{71.560}
\emlineto{42.680}{71.522}
\emmoveto{42.680}{71.512}
\emlineto{42.877}{71.464}
\emmoveto{42.877}{71.454}
\emlineto{43.073}{71.416}
\emmoveto{43.073}{71.406}
\emlineto{43.270}{71.370}
\emmoveto{43.270}{71.360}
\emlineto{43.467}{71.315}
\emmoveto{43.467}{71.305}
\emlineto{43.663}{71.260}
\emmoveto{43.663}{71.250}
\emlineto{43.860}{71.217}
\emmoveto{43.860}{71.207}
\emlineto{44.057}{71.166}
\emmoveto{44.057}{71.156}
\emlineto{44.253}{71.106}
\emmoveto{44.253}{71.096}
\emlineto{44.450}{71.062}
\emmoveto{44.450}{71.052}
\emlineto{44.647}{71.015}
\emmoveto{44.647}{71.005}
\emlineto{44.843}{70.957}
\emmoveto{44.843}{70.947}
\emlineto{45.040}{70.905}
\emmoveto{45.040}{70.895}
\emlineto{45.237}{70.862}
\emmoveto{45.237}{70.852}
\emlineto{45.433}{70.807}
\emmoveto{45.433}{70.797}
\emlineto{45.630}{70.754}
\emmoveto{45.630}{70.744}
\emlineto{45.827}{70.706}
\emmoveto{45.827}{70.696}
\emlineto{46.023}{70.654}
\emmoveto{46.023}{70.644}
\emlineto{46.220}{70.604}
\emmoveto{46.220}{70.594}
\emlineto{46.417}{70.552}
\emmoveto{46.417}{70.542}
\emlineto{46.613}{70.500}
\emmoveto{46.613}{70.490}
\emlineto{46.810}{70.453}
\emmoveto{46.810}{70.443}
\emlineto{47.007}{70.404}
\emmoveto{47.007}{70.394}
\emlineto{47.203}{70.344}
\emmoveto{47.203}{70.334}
\emlineto{47.400}{70.300}
\emmoveto{47.400}{70.290}
\emlineto{47.597}{70.254}
\emmoveto{47.597}{70.244}
\emlineto{47.793}{70.194}
\emmoveto{47.793}{70.184}
\emlineto{47.990}{70.145}
\emmoveto{47.990}{70.135}
\emlineto{48.187}{70.102}
\emmoveto{48.187}{70.092}
\emlineto{48.383}{70.044}
\emmoveto{48.383}{70.034}
\emlineto{48.580}{69.990}
\emmoveto{48.580}{69.980}
\emlineto{48.777}{69.947}
\emmoveto{48.777}{69.937}
\emlineto{48.973}{69.893}
\emmoveto{48.973}{69.883}
\emlineto{49.170}{69.841}
\emmoveto{49.170}{69.831}
\emlineto{49.367}{69.792}
\emmoveto{49.367}{69.782}
\emlineto{49.563}{69.740}
\emmoveto{49.563}{69.730}
\emlineto{49.760}{69.691}
\emmoveto{49.760}{69.681}
\emlineto{49.957}{69.639}
\emmoveto{49.957}{69.629}
\emlineto{50.153}{69.585}
\emmoveto{50.153}{69.575}
\emlineto{50.350}{69.539}
\emmoveto{50.350}{69.529}
\emlineto{50.547}{69.490}
\emmoveto{50.547}{69.480}
\emlineto{50.743}{69.430}
\emmoveto{50.743}{69.420}
\emlineto{50.940}{69.386}
\emmoveto{50.940}{69.376}
\emlineto{51.137}{69.339}
\emmoveto{51.137}{69.329}
\emlineto{51.333}{69.281}
\emmoveto{51.333}{69.271}
\emlineto{51.530}{69.232}
\emmoveto{51.530}{69.222}
\emlineto{51.727}{69.186}
\emmoveto{51.727}{69.176}
\emlineto{51.923}{69.131}
\emmoveto{51.923}{69.121}
\emlineto{52.120}{69.079}
\emmoveto{52.120}{69.069}
\emlineto{52.317}{69.032}
\emmoveto{52.317}{69.022}
\emlineto{52.513}{68.980}
\emmoveto{52.513}{68.970}
\emlineto{52.710}{68.931}
\emmoveto{52.710}{68.921}
\emlineto{52.907}{68.876}
\emmoveto{52.907}{68.866}
\emlineto{53.103}{68.827}
\emmoveto{53.103}{68.817}
\emlineto{53.300}{68.781}
\emmoveto{53.300}{68.771}
\emlineto{53.497}{68.726}
\emmoveto{53.497}{68.716}
\emlineto{53.693}{68.673}
\emmoveto{53.693}{68.663}
\emlineto{53.890}{68.629}
\emmoveto{53.890}{68.619}
\emlineto{54.087}{68.576}
\emmoveto{54.087}{68.566}
\emlineto{54.283}{68.520}
\emmoveto{54.283}{68.510}
\emlineto{54.480}{68.477}
\emmoveto{54.480}{68.467}
\emlineto{54.677}{68.425}
\emmoveto{54.677}{68.415}
\emlineto{54.873}{68.371}
\emmoveto{54.873}{68.361}
\emlineto{55.070}{68.323}
\emmoveto{55.070}{68.313}
\emlineto{55.267}{68.272}
\emmoveto{55.267}{68.262}
\emlineto{55.463}{68.222}
\emmoveto{55.463}{68.212}
\emlineto{55.660}{68.174}
\emmoveto{55.660}{68.164}
\emlineto{55.857}{68.118}
\emmoveto{55.857}{68.108}
\emlineto{56.053}{68.071}
\emmoveto{56.053}{68.061}
\emlineto{56.250}{68.024}
\emmoveto{56.250}{68.014}
\emlineto{56.447}{67.964}
\emmoveto{56.447}{67.954}
\emlineto{56.643}{67.918}
\emmoveto{56.643}{67.908}
\emlineto{56.840}{67.872}
\emmoveto{56.840}{67.862}
\emlineto{57.037}{67.815}
\emmoveto{57.037}{67.805}
\emlineto{57.233}{67.765}
\emmoveto{57.233}{67.755}
\emlineto{57.430}{67.719}
\emmoveto{57.430}{67.709}
\emlineto{57.627}{67.666}
\emmoveto{57.627}{67.656}
\emlineto{57.823}{67.614}
\emmoveto{57.823}{67.604}
\emlineto{58.020}{67.566}
\emmoveto{58.020}{67.556}
\emlineto{58.217}{67.515}
\emmoveto{58.217}{67.505}
\emlineto{58.413}{67.467}
\emmoveto{58.413}{67.457}
\emlineto{58.610}{67.411}
\emmoveto{58.610}{67.401}
\emlineto{58.807}{67.363}
\emmoveto{58.807}{67.353}
\emlineto{59.003}{67.318}
\emmoveto{59.003}{67.308}
\emlineto{59.200}{67.261}
\emmoveto{59.200}{67.251}
\emlineto{59.397}{67.210}
\emmoveto{59.397}{67.200}
\emlineto{59.593}{67.167}
\emmoveto{59.593}{67.157}
\emlineto{59.790}{67.111}
\emmoveto{59.790}{67.101}
\emlineto{59.987}{67.058}
\emmoveto{59.987}{67.048}
\emlineto{60.183}{67.014}
\emmoveto{60.183}{67.004}
\emlineto{60.380}{66.961}
\emmoveto{60.380}{66.951}
\emlineto{60.577}{66.911}
\emmoveto{60.577}{66.901}
\emlineto{60.773}{66.859}
\emmoveto{60.773}{66.849}
\emlineto{60.970}{66.810}
\emmoveto{60.970}{66.800}
\emlineto{61.167}{66.763}
\emmoveto{61.167}{66.753}
\emlineto{61.363}{66.707}
\emmoveto{61.363}{66.697}
\emlineto{61.560}{66.658}
\emmoveto{61.560}{66.648}
\emlineto{61.757}{66.614}
\emmoveto{61.757}{66.604}
\emlineto{61.953}{66.558}
\emmoveto{61.953}{66.548}
\emlineto{62.150}{66.505}
\emmoveto{62.150}{66.495}
\emlineto{62.347}{66.462}
\emmoveto{62.347}{66.452}
\emlineto{62.543}{66.408}
\emmoveto{62.543}{66.398}
\emlineto{62.740}{66.356}
\emmoveto{62.740}{66.346}
\emlineto{62.937}{66.308}
\emmoveto{62.937}{66.298}
\emlineto{63.133}{66.257}
\emmoveto{63.133}{66.247}
\emlineto{63.330}{66.209}
\emmoveto{63.330}{66.199}
\emlineto{63.527}{66.154}
\emmoveto{63.527}{66.144}
\emlineto{63.723}{66.106}
\emmoveto{63.723}{66.096}
\emlineto{63.920}{66.061}
\emmoveto{63.920}{66.051}
\emlineto{64.117}{66.005}
\emmoveto{64.117}{65.995}
\emlineto{64.313}{65.954}
\emmoveto{64.313}{65.944}
\emlineto{64.510}{65.910}
\emmoveto{64.510}{65.900}
\emlineto{64.707}{65.855}
\emmoveto{64.707}{65.845}
\emlineto{64.903}{65.804}
\emmoveto{64.903}{65.794}
\emlineto{65.100}{65.757}
\emmoveto{65.100}{65.747}
\emlineto{65.297}{65.705}
\emmoveto{65.297}{65.695}
\emlineto{65.493}{65.657}
\emmoveto{65.493}{65.647}
\emlineto{65.690}{65.603}
\emmoveto{65.690}{65.593}
\emlineto{65.887}{65.555}
\emmoveto{65.887}{65.545}
\emlineto{66.083}{65.510}
\emmoveto{66.083}{65.500}
\emlineto{66.280}{65.452}
\emmoveto{66.280}{65.442}
\emlineto{66.477}{65.404}
\emmoveto{66.477}{65.394}
\emlineto{66.673}{65.358}
\emmoveto{66.673}{65.348}
\emlineto{66.870}{65.304}
\emmoveto{66.870}{65.294}
\emlineto{67.067}{65.253}
\emmoveto{67.067}{65.243}
\emlineto{67.263}{65.206}
\emmoveto{67.263}{65.196}
\emlineto{67.460}{65.155}
\emmoveto{67.460}{65.145}
\emlineto{67.657}{65.106}
\emmoveto{67.657}{65.096}
\emlineto{67.853}{65.053}
\emmoveto{67.853}{65.043}
\emlineto{68.050}{65.006}
\emmoveto{68.050}{64.996}
\emlineto{68.247}{64.957}
\emmoveto{68.247}{64.947}
\emlineto{68.443}{64.901}
\emmoveto{68.443}{64.891}
\emlineto{68.640}{64.856}
\emmoveto{68.640}{64.846}
\emlineto{68.837}{64.807}
\emmoveto{68.837}{64.797}
\emlineto{69.033}{64.753}
\emmoveto{69.033}{64.743}
\emlineto{69.230}{64.707}
\emmoveto{69.230}{64.697}
\emlineto{69.427}{64.656}
\emmoveto{69.427}{64.646}
\emlineto{69.623}{64.605}
\emmoveto{69.623}{64.595}
\emlineto{69.820}{64.554}
\emmoveto{69.820}{64.544}
\emlineto{70.017}{64.505}
\emmoveto{70.017}{64.495}
\emlineto{70.213}{64.458}
\emmoveto{70.213}{64.448}
\emlineto{70.410}{64.405}
\emmoveto{70.410}{64.395}
\emlineto{70.607}{64.354}
\emmoveto{70.607}{64.344}
\emlineto{70.803}{64.310}
\emmoveto{70.803}{64.300}
\emlineto{71.000}{64.256}
\emmoveto{71.000}{64.246}
\emlineto{71.197}{64.204}
\emmoveto{71.197}{64.194}
\emlineto{71.393}{64.158}
\emmoveto{71.393}{64.148}
\emlineto{71.590}{64.107}
\emmoveto{71.590}{64.097}
\emlineto{71.787}{64.058}
\emmoveto{71.787}{64.048}
\emlineto{71.983}{64.005}
\emmoveto{71.983}{63.995}
\emlineto{72.180}{63.958}
\emmoveto{72.180}{63.948}
\emlineto{72.377}{63.911}
\emmoveto{72.377}{63.901}
\emlineto{72.573}{63.854}
\emmoveto{72.573}{63.844}
\emlineto{72.770}{63.809}
\emmoveto{72.770}{63.799}
\emlineto{72.967}{63.760}
\emmoveto{72.967}{63.750}
\emlineto{73.163}{63.707}
\emmoveto{73.163}{63.697}
\emlineto{73.360}{63.660}
\emmoveto{73.360}{63.650}
\emlineto{73.557}{63.609}
\emmoveto{73.557}{63.599}
\emlineto{73.753}{63.560}
\emmoveto{73.753}{63.550}
\emlineto{73.950}{63.509}
\emmoveto{73.950}{63.499}
\emlineto{74.147}{63.459}
\emmoveto{74.147}{63.449}
\emlineto{74.343}{63.413}
\emmoveto{74.343}{63.403}
\emlineto{74.540}{63.360}
\emmoveto{74.540}{63.350}
\emlineto{74.737}{63.309}
\emmoveto{74.737}{63.299}
\emlineto{74.933}{63.264}
\emmoveto{74.933}{63.254}
\emlineto{75.130}{63.211}
\emmoveto{75.130}{63.201}
\emlineto{75.327}{63.161}
\emmoveto{75.327}{63.151}
\emlineto{75.523}{63.113}
\emmoveto{75.523}{63.103}
\emlineto{75.720}{63.063}
\emmoveto{75.720}{63.053}
\emlineto{75.917}{63.015}
\emmoveto{75.917}{63.005}
\emlineto{76.113}{62.962}
\emmoveto{76.113}{62.952}
\emlineto{76.310}{62.916}
\emmoveto{76.310}{62.906}
\emlineto{76.507}{62.865}
\emmoveto{76.507}{62.855}
\emlineto{76.703}{62.811}
\emmoveto{76.703}{62.801}
\emlineto{76.900}{62.769}
\emmoveto{76.900}{62.759}
\emlineto{77.097}{62.716}
\emmoveto{77.097}{62.706}
\emlineto{77.293}{62.666}
\emmoveto{77.293}{62.656}
\emlineto{77.490}{62.617}
\emmoveto{77.490}{62.607}
\emlineto{77.687}{62.567}
\emmoveto{77.687}{62.557}
\emlineto{77.883}{62.521}
\emmoveto{77.883}{62.511}
\emlineto{78.080}{62.465}
\emmoveto{78.080}{62.455}
\emlineto{78.277}{62.420}
\emmoveto{78.277}{62.410}
\emlineto{78.473}{62.371}
\emmoveto{78.473}{62.361}
\emlineto{78.670}{62.318}
\emmoveto{78.670}{62.308}
\emlineto{78.867}{62.273}
\emmoveto{78.867}{62.263}
\emlineto{79.063}{62.221}
\emmoveto{79.063}{62.211}
\emlineto{79.260}{62.172}
\emmoveto{79.260}{62.162}
\emlineto{79.457}{62.121}
\emmoveto{79.457}{62.111}
\emlineto{79.653}{62.072}
\emmoveto{79.653}{62.062}
\emlineto{79.850}{62.027}
\emmoveto{79.850}{62.017}
\emlineto{80.047}{61.972}
\emmoveto{80.047}{61.962}
\emlineto{80.243}{61.925}
\emmoveto{80.243}{61.915}
\emlineto{80.440}{61.877}
\emmoveto{80.440}{61.867}
\emlineto{80.637}{61.825}
\emmoveto{80.637}{61.815}
\emlineto{80.833}{61.778}
\emmoveto{80.833}{61.768}
\emlineto{81.030}{61.727}
\emmoveto{81.030}{61.717}
\emlineto{81.227}{61.679}
\emmoveto{81.227}{61.669}
\emlineto{81.423}{61.627}
\emmoveto{81.423}{61.617}
\emlineto{81.620}{61.579}
\emmoveto{81.620}{61.569}
\emlineto{81.817}{61.534}
\emmoveto{81.817}{61.524}
\emlineto{82.013}{61.479}
\emmoveto{82.013}{61.469}
\emlineto{82.210}{61.432}
\emmoveto{82.210}{61.422}
\emlineto{82.407}{61.383}
\emmoveto{82.407}{61.373}
\emlineto{82.603}{61.333}
\emmoveto{82.603}{61.323}
\emlineto{82.800}{61.284}
\emmoveto{82.800}{61.274}
\emlineto{82.997}{61.234}
\emmoveto{82.997}{61.224}
\emlineto{83.193}{61.188}
\emmoveto{83.193}{61.178}
\emlineto{83.390}{61.135}
\emmoveto{83.390}{61.125}
\emlineto{83.587}{61.087}
\emmoveto{83.587}{61.077}
\emlineto{83.783}{61.040}
\emmoveto{83.783}{61.030}
\emlineto{83.980}{60.988}
\emmoveto{83.980}{60.978}
\emlineto{84.177}{60.941}
\emmoveto{84.177}{60.931}
\emlineto{84.373}{60.891}
\emmoveto{84.373}{60.881}
\emlineto{84.570}{60.842}
\emmoveto{84.570}{60.832}
\emlineto{84.767}{60.791}
\emmoveto{84.767}{60.781}
\emlineto{84.963}{60.743}
\emmoveto{84.963}{60.733}
\emlineto{85.160}{60.698}
\emmoveto{85.160}{60.688}
\emlineto{85.357}{60.643}
\emmoveto{85.357}{60.633}
\emlineto{85.553}{60.596}
\emmoveto{85.553}{60.586}
\emlineto{85.750}{60.548}
\emmoveto{85.750}{60.538}
\emlineto{85.947}{60.498}
\emmoveto{85.947}{60.488}
\emlineto{86.143}{60.448}
\emmoveto{86.143}{60.438}
\emlineto{86.340}{60.399}
\emmoveto{86.340}{60.389}
\emlineto{86.537}{60.354}
\emmoveto{86.537}{60.344}
\emlineto{86.733}{60.299}
\emmoveto{86.733}{60.289}
\emlineto{86.930}{60.253}
\emmoveto{86.930}{60.243}
\emlineto{87.127}{60.205}
\emmoveto{87.127}{60.195}
\emlineto{87.323}{60.153}
\emmoveto{87.323}{60.143}
\emlineto{87.520}{60.107}
\emmoveto{87.520}{60.097}
\emlineto{87.717}{60.057}
\emmoveto{87.717}{60.047}
\emlineto{87.913}{60.010}
\emmoveto{87.913}{60.000}
\emlineto{88.110}{59.957}
\emmoveto{88.110}{59.947}
\emlineto{88.307}{59.911}
\emmoveto{88.307}{59.901}
\emlineto{88.503}{59.863}
\emmoveto{88.503}{59.853}
\emlineto{88.700}{59.810}
\emmoveto{88.700}{59.800}
\emlineto{88.897}{59.766}
\emmoveto{88.897}{59.756}
\emlineto{89.093}{59.715}
\emmoveto{89.093}{59.705}
\emlineto{89.290}{59.666}
\emmoveto{89.290}{59.656}
\emlineto{89.487}{59.616}
\emmoveto{89.487}{59.606}
\emlineto{89.683}{59.569}
\emmoveto{89.683}{59.559}
\emlineto{89.880}{59.520}
\emmoveto{89.880}{59.510}
\emlineto{90.077}{59.468}
\emmoveto{90.077}{59.458}
\emlineto{90.273}{59.425}
\emmoveto{90.273}{59.415}
\emlineto{90.470}{59.373}
\emmoveto{90.470}{59.363}
\emlineto{90.667}{59.324}
\emmoveto{90.667}{59.314}
\emlineto{90.863}{59.275}
\emmoveto{90.863}{59.265}
\emlineto{91.060}{59.227}
\emmoveto{91.060}{59.217}
\emlineto{91.257}{59.178}
\emmoveto{91.257}{59.168}
\emlineto{91.453}{59.128}
\emmoveto{91.453}{59.118}
\emlineto{91.650}{59.084}
\emmoveto{91.650}{59.074}
\emlineto{91.847}{59.031}
\emmoveto{91.847}{59.021}
\emlineto{92.043}{58.983}
\emmoveto{92.043}{58.973}
\emlineto{92.240}{58.935}
\emmoveto{92.240}{58.925}
\emlineto{92.437}{58.886}
\emmoveto{92.437}{58.876}
\emlineto{92.633}{58.836}
\emmoveto{92.633}{58.826}
\emlineto{92.830}{58.788}
\emmoveto{92.830}{58.778}
\emlineto{93.027}{58.743}
\emmoveto{93.027}{58.733}
\emlineto{93.223}{58.689}
\emmoveto{93.223}{58.679}
\emlineto{93.420}{58.645}
\emmoveto{93.420}{58.635}
\emlineto{93.617}{58.595}
\emmoveto{93.617}{58.585}
\emlineto{93.813}{58.546}
\emmoveto{93.813}{58.536}
\emlineto{94.010}{58.496}
\emmoveto{94.010}{58.486}
\emlineto{94.207}{58.449}
\emmoveto{94.207}{58.439}
\emlineto{94.403}{58.401}
\emmoveto{94.403}{58.391}
\emlineto{94.600}{58.349}
\emmoveto{94.600}{58.339}
\emlineto{94.797}{58.306}
\emmoveto{94.797}{58.296}
\emlineto{94.993}{58.254}
\emmoveto{94.993}{58.244}
\emlineto{95.190}{58.206}
\emmoveto{95.190}{58.196}
\emlineto{95.387}{58.157}
\emmoveto{95.387}{58.147}
\emlineto{95.583}{58.110}
\emmoveto{95.583}{58.100}
\emlineto{95.780}{58.060}
\emmoveto{95.780}{58.050}
\emlineto{95.977}{58.012}
\emmoveto{95.977}{58.002}
\emlineto{96.173}{57.966}
\emmoveto{96.173}{57.956}
\emlineto{96.370}{57.914}
\emmoveto{96.370}{57.904}
\emlineto{96.567}{57.869}
\emmoveto{96.567}{57.859}
\emlineto{96.763}{57.819}
\emmoveto{96.763}{57.809}
\emlineto{96.960}{57.772}
\emmoveto{96.960}{57.762}
\emlineto{97.157}{57.720}
\emmoveto{97.157}{57.710}
\emlineto{97.353}{57.675}
\emmoveto{97.353}{57.665}
\emlineto{97.550}{57.626}
\emmoveto{97.550}{57.616}
\emlineto{97.747}{57.575}
\emmoveto{97.747}{57.565}
\emlineto{97.943}{57.530}
\emmoveto{97.943}{57.520}
\emlineto{98.140}{57.480}
\emmoveto{98.140}{57.470}
\emlineto{98.337}{57.433}
\emmoveto{98.337}{57.423}
\emlineto{98.533}{57.383}
\emmoveto{98.533}{57.373}
\emlineto{98.730}{57.339}
\emmoveto{98.730}{57.329}
\emlineto{98.927}{57.286}
\emmoveto{98.927}{57.276}
\emlineto{99.123}{57.240}
\emmoveto{99.123}{57.230}
\emlineto{99.320}{57.192}
\emmoveto{99.320}{57.182}
\emlineto{99.517}{57.143}
\emmoveto{99.517}{57.133}
\emlineto{99.713}{57.094}
\emmoveto{99.713}{57.084}
\emlineto{99.910}{57.047}
\emmoveto{99.910}{57.037}
\emlineto{100.107}{56.999}
\emmoveto{100.107}{56.989}
\emlineto{100.303}{56.949}
\emmoveto{100.303}{56.939}
\emlineto{100.500}{56.905}
\emmoveto{100.500}{56.895}
\emlineto{100.697}{56.854}
\emmoveto{100.697}{56.844}
\emlineto{100.893}{56.807}
\emmoveto{100.893}{56.797}
\emlineto{101.090}{56.758}
\emmoveto{101.090}{56.748}
\emlineto{101.287}{56.712}
\emmoveto{101.287}{56.702}
\emlineto{101.483}{56.659}
\emmoveto{101.483}{56.649}
\emlineto{101.680}{56.615}
\emmoveto{101.680}{56.605}
\emlineto{101.877}{56.566}
\emmoveto{101.877}{56.556}
\emlineto{102.073}{56.516}
\emmoveto{102.073}{56.506}
\emlineto{102.270}{56.470}
\emmoveto{102.270}{56.460}
\emlineto{102.467}{56.422}
\emmoveto{102.467}{56.412}
\emlineto{102.663}{56.374}
\emmoveto{102.663}{56.364}
\emlineto{102.860}{56.325}
\emmoveto{102.860}{56.315}
\emlineto{103.057}{56.281}
\emmoveto{103.057}{56.271}
\emlineto{103.253}{56.229}
\emmoveto{103.253}{56.219}
\emlineto{103.450}{56.184}
\emmoveto{103.450}{56.174}
\emlineto{103.647}{56.134}
\emmoveto{103.647}{56.124}
\emlineto{103.843}{56.087}
\emmoveto{103.843}{56.077}
\emlineto{104.040}{56.037}
\emmoveto{104.040}{56.027}
\emlineto{104.237}{55.992}
\emmoveto{104.237}{55.982}
\emlineto{104.433}{55.942}
\emmoveto{104.433}{55.932}
\emlineto{104.630}{55.894}
\emmoveto{104.630}{55.884}
\emlineto{104.827}{55.848}
\emmoveto{104.827}{55.838}
\emlineto{105.023}{55.799}
\emmoveto{105.023}{55.789}
\emlineto{105.220}{55.751}
\emmoveto{105.220}{55.741}
\emlineto{105.417}{55.704}
\emmoveto{105.417}{55.694}
\emlineto{105.613}{55.657}
\emmoveto{105.613}{55.647}
\emlineto{105.810}{55.606}
\emmoveto{105.810}{55.596}
\emlineto{106.007}{55.563}
\emmoveto{106.007}{55.553}
\emlineto{106.203}{55.512}
\emmoveto{106.203}{55.502}
\emlineto{106.400}{55.465}
\emmoveto{106.400}{55.455}
\emlineto{106.597}{55.417}
\emmoveto{106.597}{55.407}
\emlineto{106.793}{55.372}
\emmoveto{106.793}{55.362}
\emlineto{106.990}{55.319}
\emmoveto{106.990}{55.309}
\emlineto{107.187}{55.276}
\emmoveto{107.187}{55.266}
\emlineto{107.383}{55.227}
\emmoveto{107.383}{55.217}
\emlineto{107.580}{55.179}
\emmoveto{107.580}{55.169}
\emlineto{107.777}{55.131}
\emmoveto{107.777}{55.121}
\emlineto{107.973}{55.085}
\emmoveto{107.973}{55.075}
\emlineto{108.170}{55.034}
\emmoveto{108.170}{55.024}
\emlineto{108.367}{54.989}
\emmoveto{108.367}{54.979}
\emlineto{108.563}{54.941}
\emmoveto{108.563}{54.931}
\emlineto{108.760}{54.892}
\emmoveto{108.760}{54.882}
\emlineto{108.957}{54.846}
\emmoveto{108.957}{54.836}
\emlineto{109.153}{54.799}
\emmoveto{109.153}{54.789}
\emlineto{109.350}{54.750}
\emmoveto{109.350}{54.740}
\emlineto{109.547}{54.703}
\emmoveto{109.547}{54.693}
\emlineto{109.743}{54.657}
\emmoveto{109.743}{54.647}
\emlineto{109.940}{54.607}
\emmoveto{109.940}{54.597}
\emlineto{110.137}{54.562}
\emmoveto{110.137}{54.552}
\emlineto{110.333}{54.514}
\emmoveto{110.333}{54.504}
\emlineto{110.530}{54.466}
\emmoveto{110.530}{54.456}
\emlineto{110.727}{54.418}
\emmoveto{110.727}{54.408}
\emlineto{110.923}{54.373}
\emmoveto{110.923}{54.363}
\emlineto{111.120}{54.323}
\emmoveto{111.120}{54.313}
\emlineto{111.317}{54.278}
\emmoveto{111.317}{54.268}
\emlineto{111.513}{54.229}
\emmoveto{111.513}{54.219}
\emlineto{111.710}{54.183}
\emmoveto{111.710}{54.173}
\emlineto{111.907}{54.134}
\emmoveto{111.907}{54.124}
\emlineto{112.103}{54.090}
\emmoveto{112.103}{54.080}
\emlineto{112.300}{54.038}
\emmoveto{112.300}{54.028}
\emlineto{112.497}{53.995}
\emmoveto{112.497}{53.985}
\emlineto{112.693}{53.946}
\emmoveto{112.693}{53.936}
\emlineto{112.890}{53.900}
\emmoveto{112.890}{53.890}
\emlineto{113.087}{53.851}
\emmoveto{113.087}{53.841}
\emlineto{113.283}{53.807}
\emmoveto{113.283}{53.797}
\emlineto{113.480}{53.755}
\emmoveto{113.480}{53.745}
\emlineto{113.677}{53.712}
\emmoveto{113.677}{53.702}
\emlineto{113.873}{53.662}
\emmoveto{113.873}{53.652}
\emlineto{114.070}{53.616}
\emmoveto{114.070}{53.606}
\emlineto{114.267}{53.568}
\emmoveto{114.267}{53.558}
\emlineto{114.463}{53.524}
\emmoveto{114.463}{53.514}
\emlineto{114.660}{53.472}
\emmoveto{114.660}{53.462}
\emlineto{114.857}{53.429}
\emmoveto{114.857}{53.419}
\emlineto{115.053}{53.380}
\emmoveto{115.053}{53.370}
\emlineto{115.250}{53.333}
\emmoveto{115.250}{53.323}
\emlineto{115.447}{53.285}
\emmoveto{115.447}{53.275}
\emlineto{115.643}{53.241}
\emmoveto{115.643}{53.231}
\emlineto{115.840}{53.189}
\emmoveto{115.840}{53.179}
\emlineto{116.037}{53.147}
\emmoveto{116.037}{53.137}
\emlineto{116.233}{53.097}
\emmoveto{116.233}{53.087}
\emlineto{116.430}{53.051}
\emmoveto{116.430}{53.041}
\emlineto{116.627}{53.003}
\emmoveto{116.627}{52.993}
\emlineto{116.823}{52.959}
\emmoveto{116.823}{52.949}
\emlineto{117.020}{52.908}
\emmoveto{117.020}{52.898}
\emlineto{117.217}{52.865}
\emmoveto{117.217}{52.855}
\emlineto{117.413}{52.815}
\emmoveto{117.413}{52.805}
\emlineto{117.610}{52.770}
\emmoveto{117.610}{52.760}
\emlineto{117.807}{52.722}
\emmoveto{117.807}{52.712}
\emlineto{118.003}{52.677}
\emmoveto{118.003}{52.667}
\emlineto{118.200}{52.627}
\emmoveto{118.200}{52.617}
\emlineto{118.397}{52.584}
\emmoveto{118.397}{52.574}
\emlineto{118.593}{52.534}
\emmoveto{118.593}{52.524}
\emlineto{118.790}{52.489}
\emmoveto{118.790}{52.479}
\emlineto{118.987}{52.441}
\emmoveto{118.987}{52.431}
\emlineto{119.183}{52.395}
\emmoveto{119.183}{52.385}
\emlineto{119.380}{52.347}
\emmoveto{119.380}{52.337}
\emlineto{119.577}{52.302}
\emmoveto{119.577}{52.292}
\emlineto{119.773}{52.253}
\emmoveto{119.773}{52.243}
\emlineto{119.970}{52.208}
\emmoveto{119.970}{52.198}
\emlineto{120.167}{52.161}
\emmoveto{120.167}{52.151}
\emlineto{120.363}{52.114}
\emmoveto{120.363}{52.104}
\emlineto{120.560}{52.068}
\emmoveto{120.560}{52.058}
\emlineto{120.757}{52.022}
\emmoveto{120.757}{52.012}
\emlineto{120.953}{51.973}
\emmoveto{120.953}{51.963}
\emlineto{121.150}{51.928}
\emmoveto{121.150}{51.918}
\emlineto{121.347}{51.882}
\emmoveto{121.347}{51.872}
\emlineto{121.543}{51.833}
\emmoveto{121.543}{51.823}
\emlineto{121.740}{51.789}
\emmoveto{121.740}{51.779}
\emlineto{121.937}{51.741}
\emmoveto{121.937}{51.731}
\emlineto{122.133}{51.695}
\emmoveto{122.133}{51.685}
\emlineto{122.330}{51.648}
\emmoveto{122.330}{51.638}
\emlineto{122.527}{51.603}
\emmoveto{122.527}{51.593}
\emlineto{122.723}{51.554}
\emmoveto{122.723}{51.544}
\emlineto{122.920}{51.512}
\emmoveto{122.920}{51.502}
\emlineto{123.117}{51.461}
\emmoveto{123.117}{51.451}
\emlineto{123.313}{51.418}
\emmoveto{123.313}{51.408}
\emlineto{123.510}{51.370}
\emmoveto{123.510}{51.360}
\emlineto{123.707}{51.324}
\emmoveto{123.707}{51.314}
\emlineto{123.903}{51.276}
\emmoveto{123.903}{51.266}
\emlineto{124.100}{51.232}
\emmoveto{124.100}{51.222}
\emlineto{124.297}{51.182}
\emmoveto{124.297}{51.172}
\emlineto{124.493}{51.138}
\emmoveto{124.493}{51.128}
\emlineto{124.690}{51.092}
\emmoveto{124.690}{51.082}
\emlineto{124.887}{51.044}
\emmoveto{124.887}{51.034}
\emlineto{125.083}{51.000}
\emmoveto{125.083}{50.990}
\emlineto{125.280}{50.952}
\emmoveto{125.280}{50.942}
\emlineto{125.477}{50.906}
\emmoveto{125.477}{50.896}
\emlineto{125.673}{50.860}
\emmoveto{125.673}{50.850}
\emlineto{125.870}{50.815}
\emmoveto{125.870}{50.805}
\emlineto{126.067}{50.767}
\emmoveto{126.067}{50.757}
\emlineto{126.263}{50.724}
\emmoveto{126.263}{50.714}
\emlineto{126.460}{50.674}
\emmoveto{126.460}{50.664}
\emlineto{126.657}{50.631}
\emmoveto{126.657}{50.621}
\emlineto{126.853}{50.583}
\emmoveto{126.853}{50.573}
\emlineto{127.050}{50.536}
\emmoveto{127.050}{50.526}
\emlineto{127.247}{50.490}
\emmoveto{127.247}{50.480}
\emlineto{127.443}{50.445}
\emmoveto{127.443}{50.435}
\emlineto{127.640}{50.397}
\emmoveto{127.640}{50.387}
\emlineto{127.837}{50.353}
\emmoveto{127.837}{50.343}
\emlineto{128.033}{50.306}
\emmoveto{128.033}{50.296}
\emlineto{128.230}{50.259}
\emmoveto{128.230}{50.249}
\emlineto{128.427}{50.216}
\emmoveto{128.427}{50.206}
\emlineto{128.623}{50.167}
\emmoveto{128.623}{50.157}
\emlineto{128.820}{50.124}
\emmoveto{128.820}{50.114}
\emlineto{129.017}{50.076}
\emmoveto{129.017}{50.066}
\emlineto{129.213}{50.030}
\emmoveto{129.213}{50.020}
\emlineto{129.410}{49.984}
\emmoveto{129.410}{49.974}
\emlineto{129.607}{49.939}
\emmoveto{129.607}{49.929}
\emlineto{129.803}{49.891}
\emshow{86.340}{52.700}{vz}
\emshow{1.000}{10.000}{-7.00e-2}
\emshow{1.000}{17.000}{-6.30e-2}
\emshow{1.000}{24.000}{-5.60e-2}
\emshow{1.000}{31.000}{-4.90e-2}
\emshow{1.000}{38.000}{-4.20e-2}
\emshow{1.000}{45.000}{-3.50e-2}
\emshow{1.000}{52.000}{-2.80e-2}
\emshow{1.000}{59.000}{-2.10e-2}
\emshow{1.000}{66.000}{-1.40e-2}
\emshow{1.000}{73.000}{-7.00e-3}
\emshow{1.000}{80.000}{0.00e0}
\emshow{12.000}{5.000}{0.00e0}
\emshow{23.800}{5.000}{1.80e0}
\emshow{35.600}{5.000}{3.60e0}
\emshow{47.400}{5.000}{5.40e0}
\emshow{59.200}{5.000}{7.20e0}
\emshow{71.000}{5.000}{9.00e0}
\emshow{82.800}{5.000}{1.08e1}
\emshow{94.600}{5.000}{1.26e1}
\emshow{106.400}{5.000}{1.44e1}
\emshow{118.200}{5.000}{1.62e1}
\emshow{130.000}{5.000}{1.80e1}
\centerline{\bf {Fig. A. 2}}

\eject
\newcount\numpoint
\newcount\numpointo
\numpoint=1 \numpointo=1
\def\emmoveto#1#2{\offinterlineskip
\hbox to 0 true cm{\vbox to 0
true cm{\vskip - #2 true mm
\hskip #1 true mm \special{em:point
\the\numpoint}\vss}\hss}
\numpointo=\numpoint
\global\advance \numpoint by 1}
\def\emlineto#1#2{\offinterlineskip
\hbox to 0 true cm{\vbox to 0
true cm{\vskip - #2 true mm
\hskip #1 true mm \special{em:point
\the\numpoint}\vss}\hss}
\special{em:line
\the\numpointo,\the\numpoint}
\numpointo=\numpoint
\global\advance \numpoint by 1}
\def\emshow#1#2#3{\offinterlineskip
\hbox to 0 true cm{\vbox to 0
true cm{\vskip - #2 true mm
\hskip #1 true mm \vbox to 0
true cm{\vss\hbox{#3\hss
}}\vss}\hss}}
\special{em:linewidth 0.8pt}

\vrule width 0 mm height                0 mm depth 90.000 true mm

\special{em:linewidth 0.8pt}
\emmoveto{130.000}{10.000}
\emlineto{12.000}{10.000}
\emlineto{12.000}{80.000}
\emmoveto{71.000}{10.000}
\emlineto{71.000}{80.000}
\emmoveto{12.000}{45.000}
\emlineto{130.000}{45.000}
\emmoveto{130.000}{10.000}
\emlineto{130.000}{80.000}
\emlineto{12.000}{80.000}
\emlineto{12.000}{10.000}
\emlineto{130.000}{10.000}
\special{em:linewidth 0.4pt}
\emmoveto{12.000}{17.000}
\emlineto{130.000}{17.000}
\emmoveto{12.000}{24.000}
\emlineto{130.000}{24.000}
\emmoveto{12.000}{31.000}
\emlineto{130.000}{31.000}
\emmoveto{12.000}{38.000}
\emlineto{130.000}{38.000}
\emmoveto{12.000}{45.000}
\emlineto{130.000}{45.000}
\emmoveto{12.000}{52.000}
\emlineto{130.000}{52.000}
\emmoveto{12.000}{59.000}
\emlineto{130.000}{59.000}
\emmoveto{12.000}{66.000}
\emlineto{130.000}{66.000}
\emmoveto{12.000}{73.000}
\emlineto{130.000}{73.000}
\emmoveto{23.800}{10.000}
\emlineto{23.800}{80.000}
\emmoveto{35.600}{10.000}
\emlineto{35.600}{80.000}
\emmoveto{47.400}{10.000}
\emlineto{47.400}{80.000}
\emmoveto{59.200}{10.000}
\emlineto{59.200}{80.000}
\emmoveto{71.000}{10.000}
\emlineto{71.000}{80.000}
\emmoveto{82.800}{10.000}
\emlineto{82.800}{80.000}
\emmoveto{94.600}{10.000}
\emlineto{94.600}{80.000}
\emmoveto{106.400}{10.000}
\emlineto{106.400}{80.000}
\emmoveto{118.200}{10.000}
\emlineto{118.200}{80.000}
\special{em:linewidth 0.8pt}
\emmoveto{12.000}{10.000}
\emlineto{12.234}{10.011}
\emmoveto{12.234}{10.001}
\emlineto{12.468}{10.013}
\emmoveto{12.468}{10.003}
\emlineto{12.702}{10.016}
\emmoveto{12.702}{10.006}
\emlineto{12.937}{10.020}
\emmoveto{12.937}{10.010}
\emlineto{13.171}{10.026}
\emmoveto{13.171}{10.016}
\emlineto{13.405}{10.034}
\emmoveto{13.405}{10.024}
\emlineto{13.639}{10.042}
\emmoveto{13.639}{10.032}
\emlineto{13.873}{10.052}
\emmoveto{13.873}{10.042}
\emlineto{14.107}{10.063}
\emmoveto{14.107}{10.053}
\emlineto{14.341}{10.075}
\emmoveto{14.341}{10.065}
\emlineto{14.575}{10.089}
\emmoveto{14.575}{10.079}
\emlineto{14.810}{10.104}
\emmoveto{14.810}{10.094}
\emlineto{15.044}{10.121}
\emmoveto{15.044}{10.111}
\emlineto{15.278}{10.138}
\emmoveto{15.278}{10.128}
\emlineto{15.512}{10.157}
\emmoveto{15.512}{10.147}
\emlineto{15.746}{10.177}
\emmoveto{15.746}{10.167}
\emlineto{15.980}{10.199}
\emmoveto{15.980}{10.189}
\emlineto{16.214}{10.222}
\emmoveto{16.214}{10.212}
\emlineto{16.448}{10.246}
\emmoveto{16.448}{10.236}
\emlineto{16.683}{10.271}
\emmoveto{16.683}{10.261}
\emlineto{16.917}{10.298}
\emmoveto{16.917}{10.288}
\emlineto{17.151}{10.326}
\emmoveto{17.151}{10.316}
\emlineto{17.385}{10.355}
\emmoveto{17.385}{10.345}
\emlineto{17.619}{10.386}
\emmoveto{17.619}{10.376}
\emlineto{17.853}{10.417}
\emmoveto{17.853}{10.407}
\emlineto{18.087}{10.450}
\emmoveto{18.087}{10.440}
\emlineto{18.321}{10.485}
\emmoveto{18.321}{10.475}
\emlineto{18.556}{10.520}
\emmoveto{18.556}{10.510}
\emlineto{18.790}{10.557}
\emmoveto{18.790}{10.547}
\emlineto{19.024}{10.595}
\emmoveto{19.024}{10.585}
\emlineto{19.258}{10.635}
\emmoveto{19.258}{10.625}
\emlineto{19.492}{10.675}
\emmoveto{19.492}{10.665}
\emlineto{19.726}{10.717}
\emmoveto{19.726}{10.707}
\emlineto{19.960}{10.760}
\emmoveto{19.960}{10.750}
\emlineto{20.194}{10.804}
\emmoveto{20.194}{10.794}
\emlineto{20.429}{10.850}
\emmoveto{20.429}{10.840}
\emlineto{20.663}{10.897}
\emmoveto{20.663}{10.887}
\emlineto{20.897}{10.945}
\emmoveto{20.897}{10.935}
\emlineto{21.131}{10.994}
\emmoveto{21.131}{10.984}
\emlineto{21.365}{11.044}
\emmoveto{21.365}{11.034}
\emlineto{21.599}{11.096}
\emmoveto{21.599}{11.086}
\emlineto{21.833}{11.149}
\emmoveto{21.833}{11.139}
\emlineto{22.067}{11.203}
\emmoveto{22.067}{11.193}
\emlineto{22.302}{11.258}
\emmoveto{22.302}{11.248}
\emlineto{22.536}{11.314}
\emmoveto{22.536}{11.304}
\emlineto{22.770}{11.372}
\emmoveto{22.770}{11.362}
\emlineto{23.004}{11.431}
\emmoveto{23.004}{11.421}
\emlineto{23.238}{11.490}
\emmoveto{23.238}{11.480}
\emlineto{23.472}{11.552}
\emmoveto{23.472}{11.542}
\emlineto{23.706}{11.614}
\emmoveto{23.706}{11.604}
\emlineto{23.940}{11.677}
\emmoveto{23.940}{11.667}
\emlineto{24.175}{11.742}
\emmoveto{24.175}{11.732}
\emlineto{24.409}{11.807}
\emmoveto{24.409}{11.797}
\emlineto{24.643}{11.874}
\emmoveto{24.643}{11.864}
\emlineto{24.877}{11.942}
\emmoveto{24.877}{11.932}
\emlineto{25.111}{12.011}
\emmoveto{25.111}{12.001}
\emlineto{25.345}{12.081}
\emmoveto{25.345}{12.071}
\emlineto{25.579}{12.152}
\emmoveto{25.579}{12.142}
\emlineto{25.813}{12.225}
\emmoveto{25.813}{12.215}
\emlineto{26.048}{12.298}
\emmoveto{26.048}{12.288}
\emlineto{26.282}{12.373}
\emmoveto{26.282}{12.363}
\emlineto{26.516}{12.448}
\emmoveto{26.516}{12.438}
\emlineto{26.750}{12.525}
\emmoveto{26.750}{12.515}
\emlineto{26.984}{12.603}
\emmoveto{26.984}{12.593}
\emlineto{27.218}{12.682}
\emmoveto{27.218}{12.672}
\emlineto{27.452}{12.762}
\emmoveto{27.452}{12.752}
\emlineto{27.687}{12.842}
\emmoveto{27.687}{12.832}
\emlineto{27.921}{12.924}
\emmoveto{27.921}{12.914}
\emlineto{28.155}{13.007}
\emmoveto{28.155}{12.997}
\emlineto{28.389}{13.091}
\emmoveto{28.389}{13.081}
\emlineto{28.623}{13.176}
\emmoveto{28.623}{13.166}
\emlineto{28.857}{13.262}
\emmoveto{28.857}{13.252}
\emlineto{29.091}{13.349}
\emmoveto{29.091}{13.339}
\emlineto{29.325}{13.437}
\emmoveto{29.325}{13.427}
\emlineto{29.560}{13.526}
\emmoveto{29.560}{13.516}
\emlineto{29.794}{13.616}
\emmoveto{29.794}{13.606}
\emlineto{30.028}{13.707}
\emmoveto{30.028}{13.697}
\emlineto{30.262}{13.799}
\emmoveto{30.262}{13.789}
\emlineto{30.496}{13.892}
\emmoveto{30.496}{13.882}
\emlineto{30.730}{13.986}
\emmoveto{30.730}{13.976}
\emlineto{30.964}{14.080}
\emmoveto{30.964}{14.070}
\emlineto{31.198}{14.176}
\emmoveto{31.198}{14.166}
\emlineto{31.433}{14.272}
\emmoveto{31.433}{14.262}
\emlineto{31.667}{14.370}
\emmoveto{31.667}{14.360}
\emlineto{31.901}{14.468}
\emmoveto{31.901}{14.458}
\emlineto{32.135}{14.567}
\emmoveto{32.135}{14.557}
\emlineto{32.369}{14.668}
\emmoveto{32.369}{14.658}
\emlineto{32.603}{14.768}
\emmoveto{32.603}{14.758}
\emlineto{32.837}{14.870}
\emmoveto{32.837}{14.860}
\emlineto{33.071}{14.973}
\emmoveto{33.071}{14.963}
\emlineto{33.306}{15.077}
\emmoveto{33.306}{15.067}
\emlineto{33.540}{15.181}
\emmoveto{33.540}{15.171}
\emlineto{33.774}{15.286}
\emmoveto{33.774}{15.276}
\emlineto{34.008}{15.392}
\emmoveto{34.008}{15.382}
\emlineto{34.242}{15.499}
\emmoveto{34.242}{15.489}
\emlineto{34.476}{15.607}
\emmoveto{34.476}{15.597}
\emlineto{34.710}{15.715}
\emmoveto{34.710}{15.705}
\emlineto{34.944}{15.824}
\emmoveto{34.944}{15.814}
\emlineto{35.179}{15.934}
\emmoveto{35.179}{15.924}
\emlineto{35.413}{16.045}
\emmoveto{35.413}{16.035}
\emlineto{35.647}{16.157}
\emmoveto{35.647}{16.147}
\emlineto{35.881}{16.269}
\emmoveto{35.881}{16.259}
\emlineto{36.115}{16.382}
\emmoveto{36.115}{16.372}
\emlineto{36.349}{16.496}
\emmoveto{36.349}{16.486}
\emlineto{36.583}{16.610}
\emmoveto{36.583}{16.600}
\emlineto{36.817}{16.725}
\emmoveto{36.817}{16.715}
\emlineto{37.052}{16.841}
\emmoveto{37.052}{16.831}
\emlineto{37.286}{16.958}
\emmoveto{37.286}{16.948}
\emlineto{37.520}{17.075}
\emmoveto{37.520}{17.065}
\emlineto{37.754}{17.193}
\emmoveto{37.754}{17.183}
\emlineto{37.988}{17.312}
\emmoveto{37.988}{17.302}
\emlineto{38.222}{17.431}
\emmoveto{38.222}{17.421}
\emlineto{38.456}{17.551}
\emmoveto{38.456}{17.541}
\emlineto{38.690}{17.671}
\emmoveto{38.690}{17.661}
\emlineto{38.925}{17.793}
\emmoveto{38.925}{17.783}
\emlineto{39.159}{17.914}
\emmoveto{39.159}{17.904}
\emlineto{39.393}{18.037}
\emmoveto{39.393}{18.027}
\emlineto{39.627}{18.160}
\emmoveto{39.627}{18.150}
\emlineto{39.861}{18.284}
\emmoveto{39.861}{18.274}
\emlineto{40.095}{18.408}
\emmoveto{40.095}{18.398}
\emlineto{40.329}{18.533}
\emmoveto{40.329}{18.523}
\emlineto{40.563}{18.658}
\emmoveto{40.563}{18.648}
\emlineto{40.798}{18.784}
\emmoveto{40.798}{18.774}
\emlineto{41.032}{18.910}
\emmoveto{41.032}{18.900}
\emlineto{41.266}{19.037}
\emmoveto{41.266}{19.027}
\emlineto{41.500}{19.165}
\emmoveto{41.500}{19.155}
\emlineto{41.734}{19.293}
\emmoveto{41.734}{19.283}
\emlineto{41.968}{19.422}
\emmoveto{41.968}{19.412}
\emlineto{42.202}{19.551}
\emmoveto{42.202}{19.541}
\emlineto{42.437}{19.680}
\emmoveto{42.437}{19.670}
\emlineto{42.671}{19.810}
\emmoveto{42.671}{19.800}
\emlineto{42.905}{19.941}
\emmoveto{42.905}{19.931}
\emlineto{43.139}{20.072}
\emmoveto{43.139}{20.062}
\emlineto{43.373}{20.204}
\emmoveto{43.373}{20.194}
\emlineto{43.607}{20.335}
\emmoveto{43.607}{20.325}
\emlineto{43.841}{20.468}
\emmoveto{43.841}{20.458}
\emlineto{44.075}{20.601}
\emmoveto{44.075}{20.591}
\emlineto{44.310}{20.734}
\emmoveto{44.310}{20.724}
\emlineto{44.544}{20.868}
\emmoveto{44.544}{20.858}
\emlineto{44.778}{21.002}
\emmoveto{44.778}{20.992}
\emlineto{45.012}{21.136}
\emmoveto{45.012}{21.126}
\emlineto{45.246}{21.271}
\emmoveto{45.246}{21.261}
\emlineto{45.480}{21.406}
\emmoveto{45.480}{21.396}
\emlineto{45.714}{21.542}
\emmoveto{45.714}{21.532}
\emlineto{45.948}{21.678}
\emmoveto{45.948}{21.668}
\emlineto{46.183}{21.814}
\emmoveto{46.183}{21.804}
\emlineto{46.417}{21.951}
\emmoveto{46.417}{21.941}
\emlineto{46.651}{22.088}
\emmoveto{46.651}{22.078}
\emlineto{46.885}{22.225}
\emmoveto{46.885}{22.215}
\emlineto{47.119}{22.363}
\emmoveto{47.119}{22.353}
\emlineto{47.353}{22.501}
\emmoveto{47.353}{22.491}
\emlineto{47.587}{22.639}
\emmoveto{47.587}{22.629}
\emlineto{47.821}{22.778}
\emmoveto{47.821}{22.768}
\emlineto{48.056}{22.917}
\emmoveto{48.056}{22.907}
\emlineto{48.290}{23.056}
\emmoveto{48.290}{23.046}
\emlineto{48.524}{23.196}
\emmoveto{48.524}{23.186}
\emlineto{48.758}{23.335}
\emmoveto{48.758}{23.325}
\emlineto{48.992}{23.475}
\emmoveto{48.992}{23.465}
\emlineto{49.226}{23.616}
\emmoveto{49.226}{23.606}
\emlineto{49.460}{23.756}
\emmoveto{49.460}{23.746}
\emlineto{49.694}{23.897}
\emmoveto{49.694}{23.887}
\emlineto{49.929}{24.038}
\emmoveto{49.929}{24.028}
\emlineto{50.163}{24.179}
\emmoveto{50.163}{24.169}
\emlineto{50.397}{24.320}
\emmoveto{50.397}{24.310}
\emlineto{50.631}{24.462}
\emmoveto{50.631}{24.452}
\emlineto{50.865}{24.604}
\emmoveto{50.865}{24.594}
\emlineto{51.099}{24.746}
\emmoveto{51.099}{24.736}
\emlineto{51.333}{24.888}
\emmoveto{51.333}{24.878}
\emlineto{51.567}{25.030}
\emmoveto{51.567}{25.020}
\emlineto{51.802}{25.173}
\emmoveto{51.802}{25.163}
\emlineto{52.036}{25.315}
\emmoveto{52.036}{25.305}
\emlineto{52.270}{25.458}
\emmoveto{52.270}{25.448}
\emlineto{52.504}{25.601}
\emmoveto{52.504}{25.591}
\emlineto{52.738}{25.744}
\emmoveto{52.738}{25.734}
\emlineto{52.972}{25.887}
\emmoveto{52.972}{25.877}
\emlineto{53.206}{26.031}
\emmoveto{53.206}{26.021}
\emlineto{53.440}{26.174}
\emmoveto{53.440}{26.164}
\emlineto{53.675}{26.318}
\emmoveto{53.675}{26.308}
\emlineto{53.909}{26.462}
\emmoveto{53.909}{26.452}
\emlineto{54.143}{26.605}
\emmoveto{54.143}{26.595}
\emlineto{54.377}{26.749}
\emmoveto{54.377}{26.739}
\emlineto{54.611}{26.893}
\emmoveto{54.611}{26.883}
\emlineto{54.845}{27.037}
\emmoveto{54.845}{27.027}
\emlineto{55.079}{27.181}
\emmoveto{55.079}{27.171}
\emlineto{55.313}{27.325}
\emmoveto{55.313}{27.315}
\emlineto{55.548}{27.470}
\emmoveto{55.548}{27.460}
\emlineto{55.782}{27.614}
\emmoveto{55.782}{27.604}
\emlineto{56.016}{27.758}
\emmoveto{56.016}{27.748}
\emlineto{56.250}{27.903}
\emmoveto{56.250}{27.893}
\emlineto{56.484}{28.047}
\emmoveto{56.484}{28.037}
\emlineto{56.718}{28.191}
\emmoveto{56.718}{28.181}
\emlineto{56.952}{28.336}
\emmoveto{56.952}{28.326}
\emlineto{57.187}{28.480}
\emmoveto{57.187}{28.470}
\emlineto{57.421}{28.625}
\emmoveto{57.421}{28.615}
\emlineto{57.655}{28.769}
\emmoveto{57.655}{28.759}
\emlineto{57.889}{28.914}
\emmoveto{57.889}{28.904}
\emlineto{58.123}{29.058}
\emmoveto{58.123}{29.048}
\emlineto{58.357}{29.202}
\emmoveto{58.357}{29.192}
\emlineto{58.591}{29.347}
\emmoveto{58.591}{29.337}
\emlineto{58.825}{29.491}
\emmoveto{58.825}{29.481}
\emlineto{59.060}{29.635}
\emmoveto{59.060}{29.625}
\emlineto{59.294}{29.780}
\emmoveto{59.294}{29.770}
\emlineto{59.528}{29.924}
\emmoveto{59.528}{29.914}
\emlineto{59.762}{30.068}
\emmoveto{59.762}{30.058}
\emlineto{59.996}{30.212}
\emmoveto{59.996}{30.202}
\emlineto{60.230}{30.356}
\emmoveto{60.230}{30.346}
\emlineto{60.464}{30.500}
\emmoveto{60.464}{30.490}
\emlineto{60.698}{30.644}
\emmoveto{60.698}{30.634}
\emlineto{60.933}{30.788}
\emmoveto{60.933}{30.778}
\emlineto{61.167}{30.932}
\emmoveto{61.167}{30.922}
\emlineto{61.401}{31.076}
\emmoveto{61.401}{31.066}
\emlineto{61.635}{31.219}
\emmoveto{61.635}{31.209}
\emlineto{61.869}{31.363}
\emmoveto{61.869}{31.353}
\emlineto{62.103}{31.506}
\emmoveto{62.103}{31.496}
\emlineto{62.337}{31.649}
\emmoveto{62.337}{31.639}
\emlineto{62.571}{31.792}
\emmoveto{62.571}{31.782}
\emlineto{62.806}{31.935}
\emmoveto{62.806}{31.925}
\emlineto{63.040}{32.078}
\emmoveto{63.040}{32.068}
\emlineto{63.274}{32.221}
\emmoveto{63.274}{32.211}
\emlineto{63.508}{32.364}
\emmoveto{63.508}{32.354}
\emlineto{63.742}{32.506}
\emmoveto{63.742}{32.496}
\emlineto{63.976}{32.649}
\emmoveto{63.976}{32.639}
\emlineto{64.210}{32.791}
\emmoveto{64.210}{32.781}
\emlineto{64.444}{32.933}
\emmoveto{64.444}{32.923}
\emlineto{64.679}{33.075}
\emmoveto{64.679}{33.065}
\emlineto{64.913}{33.217}
\emmoveto{64.913}{33.207}
\emlineto{65.147}{33.359}
\emmoveto{65.147}{33.349}
\emlineto{65.381}{33.500}
\emmoveto{65.381}{33.490}
\emlineto{65.615}{33.641}
\emmoveto{65.615}{33.631}
\emlineto{65.849}{33.782}
\emmoveto{65.849}{33.772}
\emlineto{66.083}{33.923}
\emmoveto{66.083}{33.913}
\emlineto{66.317}{34.064}
\emmoveto{66.317}{34.054}
\emlineto{66.552}{34.205}
\emmoveto{66.552}{34.195}
\emlineto{66.786}{34.345}
\emmoveto{66.786}{34.335}
\emlineto{67.020}{34.485}
\emmoveto{67.020}{34.475}
\emlineto{67.254}{34.625}
\emmoveto{67.254}{34.615}
\emlineto{67.488}{34.765}
\emmoveto{67.488}{34.755}
\emlineto{67.722}{34.905}
\emmoveto{67.722}{34.895}
\emlineto{67.956}{35.044}
\emmoveto{67.956}{35.034}
\emlineto{68.190}{35.183}
\emmoveto{68.190}{35.173}
\emlineto{68.425}{35.322}
\emmoveto{68.425}{35.312}
\emlineto{68.659}{35.461}
\emmoveto{68.659}{35.451}
\emlineto{68.893}{35.600}
\emmoveto{68.893}{35.590}
\emlineto{69.127}{35.738}
\emmoveto{69.127}{35.728}
\emlineto{69.361}{35.876}
\emmoveto{69.361}{35.866}
\emlineto{69.595}{36.014}
\emmoveto{69.595}{36.004}
\emlineto{69.829}{36.152}
\emmoveto{69.829}{36.142}
\emlineto{70.063}{36.289}
\emmoveto{70.063}{36.279}
\emlineto{70.298}{36.426}
\emmoveto{70.298}{36.416}
\emlineto{70.532}{36.563}
\emmoveto{70.532}{36.553}
\emlineto{70.766}{36.700}
\emmoveto{70.766}{36.690}
\emlineto{71.000}{36.836}
\emmoveto{71.000}{36.826}
\emlineto{71.234}{36.972}
\emmoveto{71.234}{36.962}
\emlineto{71.468}{37.108}
\emmoveto{71.468}{37.098}
\emlineto{71.702}{37.244}
\emmoveto{71.702}{37.234}
\emlineto{71.937}{37.379}
\emmoveto{71.937}{37.369}
\emlineto{72.171}{37.515}
\emmoveto{72.171}{37.505}
\emlineto{72.405}{37.649}
\emmoveto{72.405}{37.639}
\emlineto{72.639}{37.784}
\emmoveto{72.639}{37.774}
\emlineto{72.873}{37.918}
\emmoveto{72.873}{37.908}
\emlineto{73.107}{38.052}
\emmoveto{73.107}{38.042}
\emlineto{73.341}{38.186}
\emmoveto{73.341}{38.176}
\emlineto{73.575}{38.320}
\emmoveto{73.575}{38.310}
\emlineto{73.810}{38.453}
\emmoveto{73.810}{38.443}
\emlineto{74.044}{38.586}
\emmoveto{74.044}{38.576}
\emlineto{74.278}{38.719}
\emmoveto{74.278}{38.709}
\emlineto{74.512}{38.851}
\emmoveto{74.512}{38.841}
\emlineto{74.746}{38.983}
\emmoveto{74.746}{38.973}
\emlineto{74.980}{39.115}
\emmoveto{74.980}{39.105}
\emlineto{75.214}{39.247}
\emmoveto{75.214}{39.237}
\emlineto{75.448}{39.378}
\emmoveto{75.448}{39.368}
\emlineto{75.683}{39.509}
\emmoveto{75.683}{39.499}
\emlineto{75.917}{39.640}
\emmoveto{75.917}{39.630}
\emlineto{76.151}{39.770}
\emmoveto{76.151}{39.760}
\emlineto{76.385}{39.900}
\emmoveto{76.385}{39.890}
\emlineto{76.619}{40.030}
\emmoveto{76.619}{40.020}
\emlineto{76.853}{40.159}
\emmoveto{76.853}{40.149}
\emlineto{77.087}{40.288}
\emmoveto{77.087}{40.278}
\emlineto{77.321}{40.417}
\emmoveto{77.321}{40.407}
\emlineto{77.556}{40.546}
\emmoveto{77.556}{40.536}
\emlineto{77.790}{40.674}
\emmoveto{77.790}{40.664}
\emlineto{78.024}{40.802}
\emmoveto{78.024}{40.792}
\emlineto{78.258}{40.929}
\emmoveto{78.258}{40.919}
\emlineto{78.492}{41.057}
\emmoveto{78.492}{41.047}
\emlineto{78.726}{41.184}
\emmoveto{78.726}{41.174}
\emlineto{78.960}{41.310}
\emmoveto{78.960}{41.300}
\emlineto{79.194}{41.437}
\emmoveto{79.194}{41.427}
\emlineto{79.429}{41.563}
\emmoveto{79.429}{41.553}
\emlineto{79.663}{41.688}
\emmoveto{79.663}{41.678}
\emlineto{79.897}{41.814}
\emmoveto{79.897}{41.804}
\emlineto{80.131}{41.939}
\emmoveto{80.131}{41.929}
\emlineto{80.365}{42.064}
\emmoveto{80.365}{42.054}
\emlineto{80.599}{42.188}
\emmoveto{80.599}{42.178}
\emlineto{80.833}{42.312}
\emmoveto{80.833}{42.302}
\emlineto{81.067}{42.436}
\emmoveto{81.067}{42.426}
\emlineto{81.302}{42.559}
\emmoveto{81.302}{42.549}
\emlineto{81.536}{42.682}
\emmoveto{81.536}{42.672}
\emlineto{81.770}{42.805}
\emmoveto{81.770}{42.795}
\emlineto{82.004}{42.928}
\emmoveto{82.004}{42.918}
\emlineto{82.238}{43.050}
\emmoveto{82.238}{43.040}
\emlineto{82.472}{43.171}
\emmoveto{82.472}{43.161}
\emlineto{82.706}{43.293}
\emmoveto{82.706}{43.283}
\emlineto{82.940}{43.414}
\emmoveto{82.940}{43.404}
\emlineto{83.175}{43.535}
\emmoveto{83.175}{43.525}
\emlineto{83.409}{43.655}
\emmoveto{83.409}{43.645}
\emlineto{83.643}{43.775}
\emmoveto{83.643}{43.765}
\emlineto{83.877}{43.895}
\emmoveto{83.877}{43.885}
\emlineto{84.111}{44.014}
\emmoveto{84.111}{44.004}
\emlineto{84.345}{44.133}
\emmoveto{84.345}{44.123}
\emlineto{84.579}{44.252}
\emmoveto{84.579}{44.242}
\emlineto{84.813}{44.370}
\emmoveto{84.813}{44.360}
\emlineto{85.048}{44.488}
\emmoveto{85.048}{44.478}
\emlineto{85.282}{44.606}
\emmoveto{85.282}{44.596}
\emlineto{85.516}{44.723}
\emmoveto{85.516}{44.713}
\emlineto{85.750}{44.840}
\emmoveto{85.750}{44.830}
\emlineto{85.984}{44.957}
\emmoveto{85.984}{44.947}
\emlineto{86.218}{45.073}
\emmoveto{86.218}{45.063}
\emlineto{86.452}{45.189}
\emmoveto{86.452}{45.179}
\emlineto{86.687}{45.305}
\emmoveto{86.687}{45.295}
\emlineto{86.921}{45.420}
\emmoveto{86.921}{45.410}
\emlineto{87.155}{45.535}
\emmoveto{87.155}{45.525}
\emlineto{87.389}{45.650}
\emmoveto{87.389}{45.640}
\emlineto{87.623}{45.764}
\emmoveto{87.623}{45.754}
\emlineto{87.857}{45.878}
\emmoveto{87.857}{45.868}
\emlineto{88.091}{45.991}
\emmoveto{88.091}{45.981}
\emlineto{88.325}{46.104}
\emmoveto{88.325}{46.094}
\emlineto{88.560}{46.217}
\emmoveto{88.560}{46.207}
\emlineto{88.794}{46.330}
\emmoveto{88.794}{46.320}
\emlineto{89.028}{46.442}
\emmoveto{89.028}{46.432}
\emlineto{89.262}{46.554}
\emmoveto{89.262}{46.544}
\emlineto{89.496}{46.665}
\emmoveto{89.496}{46.655}
\emlineto{89.730}{46.776}
\emmoveto{89.730}{46.766}
\emlineto{89.964}{46.887}
\emmoveto{89.964}{46.877}
\emlineto{90.198}{46.997}
\emmoveto{90.198}{46.987}
\emlineto{90.433}{47.107}
\emmoveto{90.433}{47.097}
\emlineto{90.667}{47.217}
\emmoveto{90.667}{47.207}
\emlineto{90.901}{47.326}
\emmoveto{90.901}{47.316}
\emlineto{91.135}{47.435}
\emmoveto{91.135}{47.425}
\emlineto{91.369}{47.544}
\emmoveto{91.369}{47.534}
\emlineto{91.603}{47.652}
\emmoveto{91.603}{47.642}
\emlineto{91.837}{47.760}
\emmoveto{91.837}{47.750}
\emlineto{92.071}{47.868}
\emmoveto{92.071}{47.858}
\emlineto{92.306}{47.975}
\emmoveto{92.306}{47.965}
\emlineto{92.540}{48.082}
\emmoveto{92.540}{48.072}
\emlineto{92.774}{48.189}
\emmoveto{92.774}{48.179}
\emlineto{93.008}{48.295}
\emmoveto{93.008}{48.285}
\emlineto{93.242}{48.401}
\emmoveto{93.242}{48.391}
\emlineto{93.476}{48.506}
\emmoveto{93.476}{48.496}
\emlineto{93.710}{48.611}
\emmoveto{93.710}{48.601}
\emlineto{93.944}{48.716}
\emmoveto{93.944}{48.706}
\emlineto{94.179}{48.821}
\emmoveto{94.179}{48.811}
\emlineto{94.413}{48.925}
\emmoveto{94.413}{48.915}
\emlineto{94.647}{49.029}
\emmoveto{94.647}{49.019}
\emlineto{94.881}{49.132}
\emmoveto{94.881}{49.122}
\emlineto{95.115}{49.235}
\emmoveto{95.115}{49.225}
\emlineto{95.349}{49.338}
\emmoveto{95.349}{49.328}
\emlineto{95.583}{49.440}
\emmoveto{95.583}{49.430}
\emlineto{95.817}{49.542}
\emmoveto{95.817}{49.532}
\emlineto{96.052}{49.644}
\emmoveto{96.052}{49.634}
\emlineto{96.286}{49.745}
\emmoveto{96.286}{49.735}
\emlineto{96.520}{49.846}
\emmoveto{96.520}{49.836}
\emlineto{96.754}{49.947}
\emmoveto{96.754}{49.937}
\emlineto{96.988}{50.047}
\emmoveto{96.988}{50.037}
\emlineto{97.222}{50.147}
\emmoveto{97.222}{50.137}
\emlineto{97.456}{50.247}
\emmoveto{97.456}{50.237}
\emlineto{97.690}{50.346}
\emmoveto{97.690}{50.336}
\emlineto{97.925}{50.445}
\emmoveto{97.925}{50.435}
\emlineto{98.159}{50.544}
\emmoveto{98.159}{50.534}
\emlineto{98.393}{50.642}
\emmoveto{98.393}{50.632}
\emlineto{98.627}{50.740}
\emmoveto{98.627}{50.730}
\emlineto{98.861}{50.838}
\emmoveto{98.861}{50.828}
\emlineto{99.095}{50.935}
\emmoveto{99.095}{50.925}
\emlineto{99.329}{51.032}
\emmoveto{99.329}{51.022}
\emlineto{99.563}{51.129}
\emmoveto{99.563}{51.119}
\emlineto{99.798}{51.225}
\emmoveto{99.798}{51.215}
\emlineto{100.032}{51.321}
\emmoveto{100.032}{51.311}
\emlineto{100.266}{51.416}
\emmoveto{100.266}{51.406}
\emlineto{100.500}{51.512}
\emmoveto{100.500}{51.502}
\emlineto{100.734}{51.606}
\emmoveto{100.734}{51.596}
\emlineto{100.968}{51.701}
\emmoveto{100.968}{51.691}
\emlineto{101.202}{51.795}
\emmoveto{101.202}{51.785}
\emlineto{101.437}{51.889}
\emmoveto{101.437}{51.879}
\emlineto{101.671}{51.983}
\emmoveto{101.671}{51.973}
\emlineto{101.905}{52.076}
\emmoveto{101.905}{52.066}
\emlineto{102.139}{52.169}
\emmoveto{102.139}{52.159}
\emlineto{102.373}{52.261}
\emmoveto{102.373}{52.251}
\emlineto{102.607}{52.354}
\emmoveto{102.607}{52.344}
\emlineto{102.841}{52.446}
\emmoveto{102.841}{52.436}
\emlineto{103.075}{52.537}
\emmoveto{103.075}{52.527}
\emlineto{103.310}{52.628}
\emmoveto{103.310}{52.618}
\emlineto{103.544}{52.719}
\emmoveto{103.544}{52.709}
\emlineto{103.778}{52.810}
\emmoveto{103.778}{52.800}
\emlineto{104.012}{52.900}
\emmoveto{104.012}{52.890}
\emlineto{104.246}{52.990}
\emmoveto{104.246}{52.980}
\emlineto{104.480}{53.080}
\emmoveto{104.480}{53.070}
\emlineto{104.714}{53.169}
\emmoveto{104.714}{53.159}
\emlineto{104.948}{53.258}
\emmoveto{104.948}{53.248}
\emlineto{105.183}{53.347}
\emmoveto{105.183}{53.337}
\emlineto{105.417}{53.435}
\emmoveto{105.417}{53.425}
\emlineto{105.651}{53.523}
\emmoveto{105.651}{53.513}
\emlineto{105.885}{53.611}
\emmoveto{105.885}{53.601}
\emlineto{106.119}{53.698}
\emmoveto{106.119}{53.688}
\emlineto{106.353}{53.785}
\emmoveto{106.353}{53.775}
\emlineto{106.587}{53.872}
\emmoveto{106.587}{53.862}
\emlineto{106.821}{53.959}
\emmoveto{106.821}{53.949}
\emlineto{107.056}{54.045}
\emmoveto{107.056}{54.035}
\emlineto{107.290}{54.130}
\emmoveto{107.290}{54.120}
\emlineto{107.524}{54.216}
\emmoveto{107.524}{54.206}
\emlineto{107.758}{54.301}
\emmoveto{107.758}{54.291}
\emlineto{107.992}{54.386}
\emmoveto{107.992}{54.376}
\emlineto{108.226}{54.471}
\emmoveto{108.226}{54.461}
\emlineto{108.460}{54.555}
\emmoveto{108.460}{54.545}
\emlineto{108.694}{54.639}
\emmoveto{108.694}{54.629}
\emlineto{108.929}{54.722}
\emmoveto{108.929}{54.712}
\emlineto{109.163}{54.806}
\emmoveto{109.163}{54.796}
\emlineto{109.397}{54.889}
\emmoveto{109.397}{54.879}
\emlineto{109.631}{54.971}
\emmoveto{109.631}{54.961}
\emlineto{109.865}{55.054}
\emmoveto{109.865}{55.044}
\emlineto{110.099}{55.136}
\emmoveto{110.099}{55.126}
\emlineto{110.333}{55.218}
\emmoveto{110.333}{55.208}
\emlineto{110.567}{55.299}
\emmoveto{110.567}{55.289}
\emlineto{110.802}{55.380}
\emmoveto{110.802}{55.370}
\emlineto{111.036}{55.461}
\emmoveto{111.036}{55.451}
\emlineto{111.270}{55.542}
\emmoveto{111.270}{55.532}
\emlineto{111.504}{55.622}
\emmoveto{111.504}{55.612}
\emlineto{111.738}{55.702}
\emmoveto{111.738}{55.692}
\emlineto{111.972}{55.782}
\emmoveto{111.972}{55.772}
\emlineto{112.206}{55.861}
\emmoveto{112.206}{55.851}
\emlineto{112.440}{55.940}
\emmoveto{112.440}{55.930}
\emlineto{112.675}{56.019}
\emmoveto{112.675}{56.009}
\emlineto{112.909}{56.097}
\emmoveto{112.909}{56.087}
\emlineto{113.143}{56.176}
\emmoveto{113.143}{56.166}
\emlineto{113.377}{56.253}
\emmoveto{113.377}{56.243}
\emlineto{113.611}{56.331}
\emmoveto{113.611}{56.321}
\emlineto{113.845}{56.408}
\emmoveto{113.845}{56.398}
\emlineto{114.079}{56.485}
\emmoveto{114.079}{56.475}
\emlineto{114.313}{56.562}
\emmoveto{114.313}{56.552}
\emlineto{114.548}{56.639}
\emmoveto{114.548}{56.629}
\emlineto{114.782}{56.715}
\emmoveto{114.782}{56.705}
\emlineto{115.016}{56.791}
\emmoveto{115.016}{56.781}
\emlineto{115.250}{56.866}
\emmoveto{115.250}{56.856}
\emlineto{115.484}{56.941}
\emmoveto{115.484}{56.931}
\emlineto{115.718}{57.017}
\emmoveto{115.718}{57.007}
\emlineto{115.952}{57.091}
\emmoveto{115.952}{57.081}
\emlineto{116.187}{57.166}
\emmoveto{116.187}{57.156}
\emlineto{116.421}{57.240}
\emmoveto{116.421}{57.230}
\emlineto{116.655}{57.314}
\emmoveto{116.655}{57.304}
\emlineto{116.889}{57.387}
\emmoveto{116.889}{57.377}
\emlineto{117.123}{57.461}
\emmoveto{117.123}{57.451}
\emlineto{117.357}{57.534}
\emmoveto{117.357}{57.524}
\emlineto{117.591}{57.607}
\emmoveto{117.591}{57.597}
\emlineto{117.825}{57.679}
\emmoveto{117.825}{57.669}
\emlineto{118.060}{57.751}
\emmoveto{118.060}{57.741}
\emlineto{118.294}{57.823}
\emmoveto{118.294}{57.813}
\emlineto{118.528}{57.895}
\emmoveto{118.528}{57.885}
\emlineto{118.762}{57.966}
\emmoveto{118.762}{57.956}
\emlineto{118.996}{58.037}
\emmoveto{118.996}{58.027}
\emlineto{119.230}{58.108}
\emmoveto{119.230}{58.098}
\emlineto{119.464}{58.179}
\emmoveto{119.464}{58.169}
\emlineto{119.698}{58.249}
\emmoveto{119.698}{58.239}
\emlineto{119.933}{58.319}
\emmoveto{119.933}{58.309}
\emlineto{120.167}{58.389}
\emmoveto{120.167}{58.379}
\emlineto{120.401}{58.459}
\emmoveto{120.401}{58.449}
\emlineto{120.635}{58.528}
\emmoveto{120.635}{58.518}
\emlineto{120.869}{58.597}
\emmoveto{120.869}{58.587}
\emlineto{121.103}{58.666}
\emmoveto{121.103}{58.656}
\emlineto{121.337}{58.734}
\emmoveto{121.337}{58.724}
\emlineto{121.571}{58.802}
\emmoveto{121.571}{58.792}
\emlineto{121.806}{58.870}
\emmoveto{121.806}{58.860}
\emlineto{122.040}{58.938}
\emmoveto{122.040}{58.928}
\emlineto{122.274}{59.005}
\emmoveto{122.274}{58.995}
\emlineto{122.508}{59.073}
\emmoveto{122.508}{59.063}
\emlineto{122.742}{59.139}
\emmoveto{122.742}{59.129}
\emlineto{122.976}{59.206}
\emmoveto{122.976}{59.196}
\emlineto{123.210}{59.273}
\emmoveto{123.210}{59.263}
\emlineto{123.444}{59.339}
\emmoveto{123.444}{59.329}
\emlineto{123.679}{59.405}
\emmoveto{123.679}{59.395}
\emlineto{123.913}{59.470}
\emmoveto{123.913}{59.460}
\emlineto{124.147}{59.536}
\emmoveto{124.147}{59.526}
\emlineto{124.381}{59.601}
\emmoveto{124.381}{59.591}
\emlineto{124.615}{59.666}
\emmoveto{124.615}{59.656}
\emlineto{124.849}{59.730}
\emmoveto{124.849}{59.720}
\emlineto{125.083}{59.795}
\emmoveto{125.083}{59.785}
\emlineto{125.317}{59.859}
\emmoveto{125.317}{59.849}
\emlineto{125.552}{59.923}
\emmoveto{125.552}{59.913}
\emlineto{125.786}{59.986}
\emmoveto{125.786}{59.976}
\emlineto{126.020}{60.050}
\emmoveto{126.020}{60.040}
\emlineto{126.254}{60.113}
\emmoveto{126.254}{60.103}
\emlineto{126.488}{60.176}
\emmoveto{126.488}{60.166}
\emlineto{126.722}{60.239}
\emmoveto{126.722}{60.229}
\emlineto{126.956}{60.301}
\emmoveto{126.956}{60.291}
\emlineto{127.190}{60.363}
\emmoveto{127.190}{60.353}
\emlineto{127.425}{60.425}
\emmoveto{127.425}{60.415}
\emlineto{127.659}{60.487}
\emmoveto{127.659}{60.477}
\emlineto{127.893}{60.549}
\emmoveto{127.893}{60.539}
\emlineto{128.127}{60.610}
\emmoveto{128.127}{60.600}
\emlineto{128.361}{60.671}
\emmoveto{128.361}{60.661}
\emlineto{128.595}{60.732}
\emmoveto{128.595}{60.722}
\emlineto{128.829}{60.792}
\emmoveto{128.829}{60.782}
\emlineto{129.063}{60.853}
\emmoveto{129.063}{60.843}
\emlineto{129.298}{60.913}
\emmoveto{129.298}{60.903}
\emlineto{129.532}{60.972}
\emmoveto{129.532}{60.962}
\emlineto{129.766}{61.032}
\emshow{36.780}{24.700}{wq}
\emmoveto{12.000}{10.000}
\emlineto{12.234}{12.755}
\emmoveto{12.234}{12.745}
\emlineto{12.468}{15.394}
\emmoveto{12.468}{15.384}
\emlineto{12.702}{17.931}
\emmoveto{12.702}{17.921}
\emlineto{12.937}{20.369}
\emmoveto{12.937}{20.359}
\emlineto{13.171}{22.712}
\emmoveto{13.171}{22.702}
\emlineto{13.405}{24.965}
\emmoveto{13.405}{24.955}
\emlineto{13.639}{27.130}
\emmoveto{13.639}{27.120}
\emlineto{13.873}{29.210}
\emmoveto{13.873}{29.200}
\emlineto{14.107}{31.210}
\emmoveto{14.107}{31.200}
\emlineto{14.341}{33.133}
\emmoveto{14.341}{33.123}
\emlineto{14.575}{34.980}
\emmoveto{14.575}{34.970}
\emlineto{14.810}{36.756}
\emmoveto{14.810}{36.746}
\emlineto{15.044}{38.462}
\emmoveto{15.044}{38.452}
\emlineto{15.278}{40.103}
\emmoveto{15.278}{40.093}
\emlineto{15.512}{41.679}
\emmoveto{15.512}{41.669}
\emlineto{15.746}{43.194}
\emmoveto{15.746}{43.184}
\emlineto{15.980}{44.650}
\emmoveto{15.980}{44.640}
\emlineto{16.214}{46.050}
\emmoveto{16.214}{46.040}
\emlineto{16.448}{47.395}
\emmoveto{16.448}{47.385}
\emlineto{16.683}{48.687}
\emmoveto{16.683}{48.677}
\emlineto{16.917}{49.930}
\emmoveto{16.917}{49.920}
\emlineto{17.151}{51.124}
\emmoveto{17.151}{51.114}
\emlineto{17.385}{52.272}
\emmoveto{17.385}{52.262}
\emlineto{17.619}{53.375}
\emmoveto{17.619}{53.365}
\emlineto{17.853}{54.435}
\emmoveto{17.853}{54.425}
\emlineto{18.087}{55.453}
\emmoveto{18.087}{55.443}
\emlineto{18.321}{56.432}
\emmoveto{18.321}{56.422}
\emlineto{18.556}{57.374}
\emmoveto{18.556}{57.364}
\emlineto{18.790}{58.278}
\emmoveto{18.790}{58.268}
\emlineto{19.024}{59.147}
\emmoveto{19.024}{59.137}
\emlineto{19.258}{59.983}
\emmoveto{19.258}{59.973}
\emlineto{19.492}{60.775}
\emlineto{19.492}{60.785}
\emmoveto{19.492}{60.775}
\emlineto{19.726}{62.870}
\emlineto{19.726}{62.880}
\emmoveto{19.726}{62.870}
\emlineto{19.960}{59.031}
\emlineto{19.960}{59.041}
\emmoveto{19.960}{59.031}
\emlineto{20.194}{60.386}
\emlineto{20.194}{60.396}
\emmoveto{20.194}{60.386}
\emlineto{20.429}{58.021}
\emmoveto{20.429}{58.011}
\emlineto{20.663}{55.737}
\emlineto{20.663}{55.747}
\emmoveto{20.663}{55.737}
\emlineto{20.897}{56.795}
\emlineto{20.897}{56.805}
\emmoveto{20.897}{56.795}
\emlineto{21.131}{54.633}
\emlineto{21.131}{54.643}
\emmoveto{21.131}{54.633}
\emlineto{21.365}{49.581}
\emlineto{21.365}{49.591}
\emmoveto{21.365}{49.581}
\emlineto{21.599}{53.655}
\emlineto{21.599}{53.665}
\emmoveto{21.599}{53.655}
\emlineto{21.833}{48.981}
\emlineto{21.833}{48.991}
\emmoveto{21.833}{48.981}
\emlineto{22.067}{50.117}
\emlineto{22.067}{50.127}
\emmoveto{22.067}{50.117}
\emlineto{22.302}{48.565}
\emmoveto{22.302}{48.555}
\emlineto{22.536}{47.091}
\emlineto{22.536}{47.101}
\emmoveto{22.536}{47.091}
\emlineto{22.770}{48.365}
\emlineto{22.770}{48.375}
\emmoveto{22.770}{48.365}
\emlineto{23.004}{47.160}
\emlineto{23.004}{47.170}
\emmoveto{23.004}{47.160}
\emlineto{23.238}{48.452}
\emlineto{23.238}{48.462}
\emmoveto{23.238}{48.452}
\emlineto{23.472}{47.397}
\emmoveto{23.472}{47.387}
\emlineto{23.706}{46.416}
\emlineto{23.706}{46.426}
\emmoveto{23.706}{46.416}
\emlineto{23.940}{47.777}
\emlineto{23.940}{47.787}
\emmoveto{23.940}{47.777}
\emlineto{24.175}{46.945}
\emmoveto{24.175}{46.935}
\emlineto{24.409}{46.181}
\emlineto{24.409}{46.191}
\emmoveto{24.409}{46.181}
\emlineto{24.643}{47.760}
\emlineto{24.643}{47.770}
\emmoveto{24.643}{47.760}
\emlineto{24.877}{47.130}
\emlineto{24.877}{47.140}
\emmoveto{24.877}{47.130}
\emlineto{25.111}{48.599}
\emlineto{25.111}{48.609}
\emmoveto{25.111}{48.599}
\emlineto{25.345}{48.085}
\emlineto{25.345}{48.095}
\emmoveto{25.345}{48.085}
\emlineto{25.579}{49.590}
\emlineto{25.579}{49.600}
\emmoveto{25.579}{49.590}
\emlineto{25.813}{47.129}
\emlineto{25.813}{47.139}
\emmoveto{25.813}{47.129}
\emlineto{26.048}{48.714}
\emmoveto{26.048}{48.704}
\emlineto{26.282}{50.278}
\emlineto{26.282}{50.288}
\emmoveto{26.282}{50.278}
\emlineto{26.516}{49.906}
\emmoveto{26.516}{49.896}
\emlineto{26.750}{49.587}
\emmoveto{26.750}{49.577}
\emlineto{26.984}{49.332}
\emmoveto{26.984}{49.322}
\emlineto{27.218}{49.128}
\emlineto{27.218}{49.138}
\emmoveto{27.218}{49.128}
\emlineto{27.452}{50.666}
\emmoveto{27.452}{50.656}
\emlineto{27.687}{52.085}
\emlineto{27.687}{52.095}
\emmoveto{27.687}{52.085}
\emlineto{27.921}{51.585}
\emlineto{27.921}{51.595}
\emmoveto{27.921}{51.585}
\emlineto{28.155}{52.680}
\emlineto{28.155}{52.690}
\emmoveto{28.155}{52.680}
\emlineto{28.389}{51.830}
\emlineto{28.389}{51.840}
\emmoveto{28.389}{51.830}
\emlineto{28.623}{52.367}
\emmoveto{28.623}{52.357}
\emlineto{28.857}{52.844}
\emlineto{28.857}{52.854}
\emmoveto{28.857}{52.844}
\emlineto{29.091}{53.001}
\emlineto{29.091}{53.011}
\emmoveto{29.091}{53.001}
\emlineto{29.325}{51.535}
\emlineto{29.325}{51.545}
\emmoveto{29.325}{51.535}
\emlineto{29.560}{53.001}
\emlineto{29.560}{53.011}
\emmoveto{29.560}{53.001}
\emlineto{29.794}{51.306}
\emlineto{29.794}{51.316}
\emmoveto{29.794}{51.306}
\emlineto{30.028}{52.487}
\emmoveto{30.028}{52.477}
\emlineto{30.262}{53.499}
\emlineto{30.262}{53.509}
\emmoveto{30.262}{53.499}
\emlineto{30.496}{51.532}
\emlineto{30.496}{51.542}
\emmoveto{30.496}{51.532}
\emlineto{30.730}{52.337}
\emlineto{30.730}{52.347}
\emmoveto{30.730}{52.337}
\emlineto{30.964}{51.731}
\emmoveto{30.964}{51.721}
\emlineto{31.198}{51.043}
\emlineto{31.198}{51.053}
\emmoveto{31.198}{51.043}
\emlineto{31.433}{51.592}
\emmoveto{31.433}{51.582}
\emlineto{31.667}{52.199}
\emlineto{31.667}{52.209}
\emmoveto{31.667}{52.199}
\emlineto{31.901}{51.357}
\emmoveto{31.901}{51.347}
\emlineto{32.135}{50.462}
\emlineto{32.135}{50.472}
\emmoveto{32.135}{50.462}
\emlineto{32.369}{50.919}
\emlineto{32.369}{50.929}
\emmoveto{32.369}{50.919}
\emlineto{32.603}{49.974}
\emlineto{32.603}{49.984}
\emmoveto{32.603}{49.974}
\emlineto{32.837}{50.363}
\emmoveto{32.837}{50.353}
\emlineto{33.071}{50.703}
\emmoveto{33.071}{50.693}
\emlineto{33.306}{51.000}
\emlineto{33.306}{51.010}
\emmoveto{33.306}{51.000}
\emlineto{33.540}{49.985}
\emlineto{33.540}{49.995}
\emmoveto{33.540}{49.985}
\emlineto{33.774}{50.266}
\emmoveto{33.774}{50.256}
\emlineto{34.008}{50.518}
\emmoveto{34.008}{50.508}
\emlineto{34.242}{50.747}
\emlineto{34.242}{50.757}
\emmoveto{34.242}{50.747}
\emlineto{34.476}{49.902}
\emlineto{34.476}{49.912}
\emmoveto{34.476}{49.902}
\emlineto{34.710}{50.145}
\emmoveto{34.710}{50.135}
\emlineto{34.944}{50.374}
\emmoveto{34.944}{50.364}
\emlineto{35.179}{50.591}
\emlineto{35.179}{50.601}
\emmoveto{35.179}{50.591}
\emlineto{35.413}{49.795}
\emlineto{35.413}{49.805}
\emmoveto{35.413}{49.795}
\emlineto{35.647}{50.042}
\emmoveto{35.647}{50.032}
\emlineto{35.881}{50.274}
\emmoveto{35.881}{50.264}
\emlineto{36.115}{50.495}
\emmoveto{36.115}{50.485}
\emlineto{36.349}{50.700}
\emmoveto{36.349}{50.690}
\emlineto{36.583}{50.882}
\emmoveto{36.583}{50.872}
\emlineto{36.817}{51.038}
\emmoveto{36.817}{51.028}
\emlineto{37.052}{51.164}
\emmoveto{37.052}{51.154}
\emlineto{37.286}{51.257}
\emmoveto{37.286}{51.247}
\emlineto{37.520}{51.314}
\emmoveto{37.520}{51.304}
\emlineto{37.754}{51.333}
\emmoveto{37.754}{51.323}
\emlineto{37.988}{51.312}
\emmoveto{37.988}{51.302}
\emlineto{38.222}{51.251}
\emmoveto{38.222}{51.241}
\emlineto{38.456}{51.149}
\emmoveto{38.456}{51.139}
\emlineto{38.690}{50.996}
\emlineto{38.690}{51.006}
\emmoveto{38.690}{50.996}
\emlineto{38.925}{52.729}
\emlineto{38.925}{52.739}
\emmoveto{38.925}{52.729}
\emlineto{39.159}{52.496}
\emmoveto{39.159}{52.486}
\emlineto{39.393}{52.203}
\emlineto{39.393}{52.213}
\emmoveto{39.393}{52.203}
\emlineto{39.627}{52.726}
\emlineto{39.627}{52.736}
\emmoveto{39.627}{52.726}
\emlineto{39.861}{52.368}
\emmoveto{39.861}{52.358}
\emlineto{40.095}{51.962}
\emmoveto{40.095}{51.952}
\emlineto{40.329}{51.511}
\emlineto{40.329}{51.521}
\emmoveto{40.329}{51.511}
\emlineto{40.563}{52.839}
\emlineto{40.563}{52.849}
\emmoveto{40.563}{52.839}
\emlineto{40.798}{52.327}
\emmoveto{40.798}{52.317}
\emlineto{41.032}{51.764}
\emlineto{41.032}{51.774}
\emmoveto{41.032}{51.764}
\emlineto{41.266}{51.971}
\emlineto{41.266}{51.981}
\emmoveto{41.266}{51.971}
\emlineto{41.500}{52.317}
\emlineto{41.500}{52.327}
\emmoveto{41.500}{52.317}
\emlineto{41.734}{51.669}
\emlineto{41.734}{51.679}
\emmoveto{41.734}{51.669}
\emlineto{41.968}{52.715}
\emlineto{41.968}{52.725}
\emmoveto{41.968}{52.715}
\emlineto{42.202}{52.009}
\emlineto{42.202}{52.019}
\emmoveto{42.202}{52.009}
\emlineto{42.437}{52.220}
\emlineto{42.437}{52.230}
\emmoveto{42.437}{52.220}
\emlineto{42.671}{51.472}
\emlineto{42.671}{51.482}
\emmoveto{42.671}{51.472}
\emlineto{42.905}{52.372}
\emlineto{42.905}{52.382}
\emmoveto{42.905}{52.372}
\emlineto{43.139}{52.500}
\emlineto{43.139}{52.510}
\emmoveto{43.139}{52.500}
\emlineto{43.373}{51.693}
\emlineto{43.373}{51.703}
\emmoveto{43.373}{51.693}
\emlineto{43.607}{51.782}
\emlineto{43.607}{51.792}
\emmoveto{43.607}{51.782}
\emlineto{43.841}{52.566}
\emlineto{43.841}{52.576}
\emmoveto{43.841}{52.566}
\emlineto{44.075}{52.609}
\emlineto{44.075}{52.619}
\emmoveto{44.075}{52.609}
\emlineto{44.310}{51.748}
\emlineto{44.310}{51.758}
\emmoveto{44.310}{51.748}
\emlineto{44.544}{51.773}
\emmoveto{44.544}{51.763}
\emlineto{44.778}{51.762}
\emlineto{44.778}{51.772}
\emmoveto{44.778}{51.762}
\emlineto{45.012}{52.430}
\emlineto{45.012}{52.440}
\emmoveto{45.012}{52.430}
\emlineto{45.246}{52.401}
\emmoveto{45.246}{52.391}
\emlineto{45.480}{52.346}
\emmoveto{45.480}{52.336}
\emlineto{45.714}{52.275}
\emmoveto{45.714}{52.265}
\emlineto{45.948}{52.179}
\emlineto{45.948}{52.189}
\emmoveto{45.948}{52.179}
\emlineto{46.183}{52.735}
\emlineto{46.183}{52.745}
\emmoveto{46.183}{52.735}
\emlineto{46.417}{52.620}
\emmoveto{46.417}{52.610}
\emlineto{46.651}{52.479}
\emmoveto{46.651}{52.469}
\emlineto{46.885}{52.311}
\emlineto{46.885}{52.321}
\emmoveto{46.885}{52.311}
\emlineto{47.119}{52.774}
\emlineto{47.119}{52.784}
\emmoveto{47.119}{52.774}
\emlineto{47.353}{52.585}
\emmoveto{47.353}{52.575}
\emlineto{47.587}{52.359}
\emlineto{47.587}{52.369}
\emmoveto{47.587}{52.359}
\emlineto{47.821}{52.934}
\emlineto{47.821}{52.944}
\emmoveto{47.821}{52.934}
\emlineto{48.056}{52.678}
\emlineto{48.056}{52.688}
\emmoveto{48.056}{52.678}
\emlineto{48.290}{53.017}
\emlineto{48.290}{53.027}
\emmoveto{48.290}{53.017}
\emlineto{48.524}{52.729}
\emmoveto{48.524}{52.719}
\emlineto{48.758}{52.403}
\emlineto{48.758}{52.413}
\emmoveto{48.758}{52.403}
\emlineto{48.992}{53.455}
\emlineto{48.992}{53.465}
\emmoveto{48.992}{53.455}
\emlineto{49.226}{53.103}
\emmoveto{49.226}{53.093}
\emlineto{49.460}{52.713}
\emlineto{49.460}{52.723}
\emmoveto{49.460}{52.713}
\emlineto{49.694}{53.089}
\emlineto{49.694}{53.099}
\emmoveto{49.694}{53.089}
\emlineto{49.929}{53.253}
\emlineto{49.929}{53.263}
\emmoveto{49.929}{53.253}
\emlineto{50.163}{52.811}
\emlineto{50.163}{52.821}
\emmoveto{50.163}{52.811}
\emlineto{50.397}{53.114}
\emlineto{50.397}{53.124}
\emmoveto{50.397}{53.114}
\emlineto{50.631}{53.205}
\emlineto{50.631}{53.215}
\emmoveto{50.631}{53.205}
\emlineto{50.865}{53.459}
\emlineto{50.865}{53.469}
\emmoveto{50.865}{53.459}
\emlineto{51.099}{52.939}
\emlineto{51.099}{52.949}
\emmoveto{51.099}{52.939}
\emlineto{51.333}{53.150}
\emlineto{51.333}{53.160}
\emmoveto{51.333}{53.150}
\emlineto{51.567}{53.151}
\emlineto{51.567}{53.161}
\emmoveto{51.567}{53.151}
\emlineto{51.802}{53.328}
\emmoveto{51.802}{53.318}
\emlineto{52.036}{53.464}
\emlineto{52.036}{53.474}
\emmoveto{52.036}{53.464}
\emlineto{52.270}{53.402}
\emlineto{52.270}{53.412}
\emmoveto{52.270}{53.402}
\emlineto{52.504}{53.518}
\emmoveto{52.504}{53.508}
\emlineto{52.738}{53.595}
\emlineto{52.738}{53.605}
\emmoveto{52.738}{53.595}
\emlineto{52.972}{52.942}
\emlineto{52.972}{52.952}
\emmoveto{52.972}{52.942}
\emlineto{53.206}{53.526}
\emlineto{53.206}{53.536}
\emmoveto{53.206}{53.526}
\emlineto{53.440}{53.569}
\emmoveto{53.440}{53.559}
\emlineto{53.675}{53.586}
\emmoveto{53.675}{53.576}
\emlineto{53.909}{53.576}
\emlineto{53.909}{53.586}
\emmoveto{53.909}{53.576}
\emlineto{54.143}{53.372}
\emlineto{54.143}{53.382}
\emmoveto{54.143}{53.372}
\emlineto{54.377}{53.350}
\emmoveto{54.377}{53.340}
\emlineto{54.611}{53.304}
\emmoveto{54.611}{53.294}
\emlineto{54.845}{53.233}
\emlineto{54.845}{53.243}
\emmoveto{54.845}{53.233}
\emlineto{55.079}{53.660}
\emlineto{55.079}{53.670}
\emmoveto{55.079}{53.660}
\emlineto{55.313}{53.577}
\emmoveto{55.313}{53.567}
\emlineto{55.548}{53.470}
\emmoveto{55.548}{53.460}
\emlineto{55.782}{53.339}
\emlineto{55.782}{53.349}
\emmoveto{55.782}{53.339}
\emlineto{56.016}{53.695}
\emlineto{56.016}{53.705}
\emmoveto{56.016}{53.695}
\emlineto{56.250}{54.220}
\emlineto{56.250}{54.230}
\emmoveto{56.250}{54.220}
\emlineto{56.484}{54.063}
\emmoveto{56.484}{54.053}
\emlineto{56.718}{53.883}
\emmoveto{56.718}{53.873}
\emlineto{56.952}{53.680}
\emlineto{56.952}{53.690}
\emmoveto{56.952}{53.680}
\emlineto{57.187}{53.948}
\emlineto{57.187}{53.958}
\emmoveto{57.187}{53.948}
\emlineto{57.421}{54.388}
\emlineto{57.421}{54.398}
\emmoveto{57.421}{54.388}
\emlineto{57.655}{54.161}
\emmoveto{57.655}{54.151}
\emlineto{57.889}{53.910}
\emmoveto{57.889}{53.900}
\emlineto{58.123}{53.638}
\emlineto{58.123}{53.648}
\emmoveto{58.123}{53.638}
\emlineto{58.357}{54.474}
\emlineto{58.357}{54.484}
\emmoveto{58.357}{54.474}
\emlineto{58.591}{54.181}
\emlineto{58.591}{54.191}
\emmoveto{58.591}{54.181}
\emlineto{58.825}{54.521}
\emlineto{58.825}{54.531}
\emmoveto{58.825}{54.521}
\emlineto{59.060}{54.211}
\emmoveto{59.060}{54.201}
\emlineto{59.294}{53.869}
\emlineto{59.294}{53.879}
\emmoveto{59.294}{53.869}
\emlineto{59.528}{54.610}
\emlineto{59.528}{54.620}
\emmoveto{59.528}{54.610}
\emlineto{59.762}{54.248}
\emlineto{59.762}{54.258}
\emmoveto{59.762}{54.248}
\emlineto{59.996}{54.508}
\emlineto{59.996}{54.518}
\emmoveto{59.996}{54.508}
\emlineto{60.230}{54.121}
\emlineto{60.230}{54.131}
\emmoveto{60.230}{54.121}
\emlineto{60.464}{54.361}
\emmoveto{60.464}{54.351}
\emlineto{60.698}{54.565}
\emlineto{60.698}{54.575}
\emmoveto{60.698}{54.565}
\emlineto{60.933}{54.572}
\emlineto{60.933}{54.582}
\emmoveto{60.933}{54.572}
\emlineto{61.167}{54.764}
\emmoveto{61.167}{54.754}
\emlineto{61.401}{54.920}
\emlineto{61.401}{54.930}
\emmoveto{61.401}{54.920}
\emlineto{61.635}{54.456}
\emlineto{61.635}{54.466}
\emmoveto{61.635}{54.456}
\emlineto{61.869}{54.605}
\emmoveto{61.869}{54.595}
\emlineto{62.103}{54.730}
\emmoveto{62.103}{54.720}
\emlineto{62.337}{54.830}
\emlineto{62.337}{54.840}
\emmoveto{62.337}{54.830}
\emlineto{62.571}{54.735}
\emlineto{62.571}{54.745}
\emmoveto{62.571}{54.735}
\emlineto{62.806}{54.828}
\emmoveto{62.806}{54.818}
\emlineto{63.040}{54.897}
\emmoveto{63.040}{54.887}
\emlineto{63.274}{54.952}
\emmoveto{63.274}{54.942}
\emlineto{63.508}{54.994}
\emmoveto{63.508}{54.984}
\emlineto{63.742}{55.024}
\emmoveto{63.742}{55.014}
\emlineto{63.976}{55.040}
\emmoveto{63.976}{55.030}
\emlineto{64.210}{55.044}
\emmoveto{64.210}{55.034}
\emlineto{64.444}{55.036}
\emmoveto{64.444}{55.026}
\emlineto{64.679}{55.016}
\emmoveto{64.679}{55.006}
\emlineto{64.913}{54.984}
\emmoveto{64.913}{54.974}
\emlineto{65.147}{54.940}
\emmoveto{65.147}{54.930}
\emlineto{65.381}{54.884}
\emmoveto{65.381}{54.874}
\emlineto{65.615}{54.807}
\emlineto{65.615}{54.817}
\emmoveto{65.615}{54.807}
\emlineto{65.849}{55.693}
\emlineto{65.849}{55.703}
\emmoveto{65.849}{55.693}
\emlineto{66.083}{55.610}
\emmoveto{66.083}{55.600}
\emlineto{66.317}{55.505}
\emmoveto{66.317}{55.495}
\emlineto{66.552}{55.380}
\emlineto{66.552}{55.390}
\emmoveto{66.552}{55.380}
\emlineto{66.786}{55.824}
\emlineto{66.786}{55.834}
\emmoveto{66.786}{55.824}
\emlineto{67.020}{55.695}
\emmoveto{67.020}{55.685}
\emlineto{67.254}{55.536}
\emlineto{67.254}{55.546}
\emmoveto{67.254}{55.536}
\emlineto{67.488}{55.569}
\emlineto{67.488}{55.579}
\emmoveto{67.488}{55.569}
\emlineto{67.722}{55.409}
\emmoveto{67.722}{55.399}
\emlineto{67.956}{55.219}
\emlineto{67.956}{55.229}
\emmoveto{67.956}{55.219}
\emlineto{68.190}{55.589}
\emlineto{68.190}{55.599}
\emmoveto{68.190}{55.589}
\emlineto{68.425}{55.387}
\emlineto{68.425}{55.397}
\emmoveto{68.425}{55.387}
\emlineto{68.659}{55.732}
\emlineto{68.659}{55.742}
\emmoveto{68.659}{55.732}
\emlineto{68.893}{55.509}
\emlineto{68.893}{55.519}
\emmoveto{68.893}{55.509}
\emlineto{69.127}{55.829}
\emlineto{69.127}{55.839}
\emmoveto{69.127}{55.829}
\emlineto{69.361}{55.585}
\emlineto{69.361}{55.595}
\emmoveto{69.361}{55.585}
\emlineto{69.595}{55.891}
\emmoveto{69.595}{55.881}
\emlineto{69.829}{56.163}
\emlineto{69.829}{56.173}
\emmoveto{69.829}{56.163}
\emlineto{70.063}{55.888}
\emlineto{70.063}{55.898}
\emmoveto{70.063}{55.888}
\emlineto{70.298}{56.148}
\emlineto{70.298}{56.158}
\emmoveto{70.298}{56.148}
\emlineto{70.532}{56.045}
\emlineto{70.532}{56.055}
\emmoveto{70.532}{56.045}
\emlineto{70.766}{55.741}
\emlineto{70.766}{55.751}
\emmoveto{70.766}{55.741}
\emlineto{71.000}{55.978}
\emmoveto{71.000}{55.968}
\emlineto{71.234}{56.193}
\emmoveto{71.234}{56.183}
\emlineto{71.468}{56.385}
\emlineto{71.468}{56.395}
\emmoveto{71.468}{56.385}
\emlineto{71.702}{56.042}
\emlineto{71.702}{56.052}
\emmoveto{71.702}{56.042}
\emlineto{71.937}{56.223}
\emlineto{71.937}{56.233}
\emmoveto{71.937}{56.223}
\emlineto{72.171}{56.054}
\emlineto{72.171}{56.064}
\emmoveto{72.171}{56.054}
\emlineto{72.405}{56.225}
\emmoveto{72.405}{56.215}
\emlineto{72.639}{56.374}
\emmoveto{72.639}{56.364}
\emlineto{72.873}{56.502}
\emlineto{72.873}{56.512}
\emmoveto{72.873}{56.502}
\emlineto{73.107}{56.296}
\emlineto{73.107}{56.306}
\emmoveto{73.107}{56.296}
\emlineto{73.341}{56.424}
\emmoveto{73.341}{56.414}
\emlineto{73.575}{56.532}
\emmoveto{73.575}{56.522}
\emlineto{73.810}{56.618}
\emlineto{73.810}{56.628}
\emmoveto{73.810}{56.618}
\emlineto{74.044}{56.378}
\emlineto{74.044}{56.388}
\emmoveto{74.044}{56.378}
\emlineto{74.278}{56.466}
\emmoveto{74.278}{56.456}
\emlineto{74.512}{56.534}
\emmoveto{74.512}{56.524}
\emlineto{74.746}{56.581}
\emlineto{74.746}{56.591}
\emmoveto{74.746}{56.581}
\emlineto{74.980}{56.822}
\emlineto{74.980}{56.832}
\emmoveto{74.980}{56.822}
\emlineto{75.214}{56.870}
\emmoveto{75.214}{56.860}
\emlineto{75.448}{56.888}
\emlineto{75.448}{56.898}
\emmoveto{75.448}{56.888}
\emlineto{75.683}{56.590}
\emlineto{75.683}{56.600}
\emmoveto{75.683}{56.590}
\emlineto{75.917}{56.600}
\emlineto{75.917}{56.610}
\emmoveto{75.917}{56.600}
\emlineto{76.151}{57.109}
\emlineto{76.151}{57.119}
\emmoveto{76.151}{57.109}
\emlineto{76.385}{56.787}
\emlineto{76.385}{56.797}
\emmoveto{76.385}{56.787}
\emlineto{76.619}{56.769}
\emlineto{76.619}{56.779}
\emmoveto{76.619}{56.769}
\emlineto{76.853}{57.246}
\emlineto{76.853}{57.256}
\emmoveto{76.853}{57.246}
\emlineto{77.087}{56.901}
\emlineto{77.087}{56.911}
\emmoveto{77.087}{56.901}
\emlineto{77.321}{57.357}
\emlineto{77.321}{57.367}
\emmoveto{77.321}{57.357}
\emlineto{77.556}{56.997}
\emlineto{77.556}{57.007}
\emmoveto{77.556}{56.997}
\emlineto{77.790}{57.433}
\emlineto{77.790}{57.443}
\emmoveto{77.790}{57.433}
\emlineto{78.024}{57.057}
\emlineto{78.024}{57.067}
\emmoveto{78.024}{57.057}
\emlineto{78.258}{57.474}
\emlineto{78.258}{57.484}
\emmoveto{78.258}{57.474}
\emlineto{78.492}{57.385}
\emlineto{78.492}{57.395}
\emmoveto{78.492}{57.385}
\emlineto{78.726}{57.481}
\emlineto{78.726}{57.491}
\emmoveto{78.726}{57.481}
\emlineto{78.960}{57.376}
\emlineto{78.960}{57.386}
\emmoveto{78.960}{57.376}
\emlineto{79.194}{57.455}
\emlineto{79.194}{57.465}
\emmoveto{79.194}{57.455}
\emlineto{79.429}{57.333}
\emlineto{79.429}{57.343}
\emmoveto{79.429}{57.333}
\emlineto{79.663}{57.407}
\emmoveto{79.663}{57.397}
\emlineto{79.897}{57.452}
\emlineto{79.897}{57.462}
\emmoveto{79.897}{57.452}
\emlineto{80.131}{57.306}
\emlineto{80.131}{57.316}
\emmoveto{80.131}{57.306}
\emlineto{80.365}{57.346}
\emlineto{80.365}{57.356}
\emmoveto{80.365}{57.346}
\emlineto{80.599}{57.668}
\emlineto{80.599}{57.678}
\emmoveto{80.599}{57.668}
\emlineto{80.833}{57.691}
\emlineto{80.833}{57.701}
\emmoveto{80.833}{57.691}
\emlineto{81.067}{57.994}
\emlineto{81.067}{58.004}
\emmoveto{81.067}{57.994}
\emlineto{81.302}{57.520}
\emlineto{81.302}{57.530}
\emmoveto{81.302}{57.520}
\emlineto{81.536}{57.521}
\emlineto{81.536}{57.531}
\emmoveto{81.536}{57.521}
\emlineto{81.770}{57.798}
\emlineto{81.770}{57.808}
\emmoveto{81.770}{57.798}
\emlineto{82.004}{57.783}
\emlineto{82.004}{57.793}
\emmoveto{82.004}{57.783}
\emlineto{82.238}{58.042}
\emlineto{82.238}{58.052}
\emmoveto{82.238}{58.042}
\emlineto{82.472}{58.021}
\emmoveto{82.472}{58.011}
\emlineto{82.706}{57.983}
\emmoveto{82.706}{57.973}
\emlineto{82.940}{57.927}
\emlineto{82.940}{57.937}
\emmoveto{82.940}{57.927}
\emlineto{83.175}{58.153}
\emlineto{83.175}{58.163}
\emmoveto{83.175}{58.153}
\emlineto{83.409}{58.102}
\emmoveto{83.409}{58.092}
\emlineto{83.643}{58.025}
\emlineto{83.643}{58.035}
\emmoveto{83.643}{58.025}
\emlineto{83.877}{58.226}
\emlineto{83.877}{58.236}
\emmoveto{83.877}{58.226}
\emlineto{84.111}{58.154}
\emmoveto{84.111}{58.144}
\emlineto{84.345}{58.066}
\emmoveto{84.345}{58.056}
\emlineto{84.579}{57.960}
\emlineto{84.579}{57.970}
\emmoveto{84.579}{57.960}
\emlineto{84.813}{58.594}
\emlineto{84.813}{58.604}
\emmoveto{84.813}{58.594}
\emlineto{85.048}{58.494}
\emmoveto{85.048}{58.484}
\emlineto{85.282}{58.377}
\emmoveto{85.282}{58.367}
\emlineto{85.516}{58.244}
\emlineto{85.516}{58.254}
\emmoveto{85.516}{58.244}
\emlineto{85.750}{58.575}
\emlineto{85.750}{58.585}
\emmoveto{85.750}{58.575}
\emlineto{85.984}{58.439}
\emlineto{85.984}{58.449}
\emmoveto{85.984}{58.439}
\emlineto{86.218}{58.561}
\emlineto{86.218}{58.571}
\emmoveto{86.218}{58.561}
\emlineto{86.452}{58.411}
\emlineto{86.452}{58.421}
\emmoveto{86.452}{58.411}
\emlineto{86.687}{58.711}
\emlineto{86.687}{58.721}
\emmoveto{86.687}{58.711}
\emlineto{86.921}{58.549}
\emlineto{86.921}{58.559}
\emmoveto{86.921}{58.549}
\emlineto{87.155}{58.834}
\emlineto{87.155}{58.844}
\emmoveto{87.155}{58.834}
\emlineto{87.389}{58.659}
\emlineto{87.389}{58.669}
\emmoveto{87.389}{58.659}
\emlineto{87.623}{58.929}
\emlineto{87.623}{58.939}
\emmoveto{87.623}{58.929}
\emlineto{87.857}{58.751}
\emmoveto{87.857}{58.741}
\emlineto{88.091}{58.547}
\emlineto{88.091}{58.557}
\emmoveto{88.091}{58.547}
\emlineto{88.325}{58.807}
\emmoveto{88.325}{58.797}
\emlineto{88.560}{59.040}
\emlineto{88.560}{59.050}
\emmoveto{88.560}{59.040}
\emlineto{88.794}{58.827}
\emlineto{88.794}{58.837}
\emmoveto{88.794}{58.827}
\emlineto{89.028}{59.056}
\emlineto{89.028}{59.066}
\emmoveto{89.028}{59.056}
\emlineto{89.262}{58.831}
\emlineto{89.262}{58.841}
\emmoveto{89.262}{58.831}
\emlineto{89.496}{59.056}
\emmoveto{89.496}{59.046}
\emlineto{89.730}{59.253}
\emlineto{89.730}{59.263}
\emmoveto{89.730}{59.253}
\emlineto{89.964}{59.010}
\emlineto{89.964}{59.020}
\emmoveto{89.964}{59.010}
\emlineto{90.198}{59.215}
\emmoveto{90.198}{59.205}
\emlineto{90.433}{59.392}
\emlineto{90.433}{59.402}
\emmoveto{90.433}{59.392}
\emlineto{90.667}{58.884}
\emlineto{90.667}{58.894}
\emmoveto{90.667}{58.884}
\emlineto{90.901}{59.070}
\emmoveto{90.901}{59.060}
\emlineto{91.135}{59.238}
\emmoveto{91.135}{59.228}
\emlineto{91.369}{59.399}
\emmoveto{91.369}{59.389}
\emlineto{91.603}{59.554}
\emmoveto{91.603}{59.544}
\emlineto{91.837}{59.692}
\emlineto{91.837}{59.702}
\emmoveto{91.837}{59.692}
\emlineto{92.071}{59.155}
\emlineto{92.071}{59.165}
\emmoveto{92.071}{59.155}
\emlineto{92.306}{59.301}
\emmoveto{92.306}{59.291}
\emlineto{92.540}{59.431}
\emmoveto{92.540}{59.421}
\emlineto{92.774}{59.545}
\emlineto{92.774}{59.555}
\emmoveto{92.774}{59.545}
\emlineto{93.008}{59.423}
\emlineto{93.008}{59.433}
\emmoveto{93.008}{59.423}
\emlineto{93.242}{59.544}
\emmoveto{93.242}{59.534}
\emlineto{93.476}{59.639}
\emlineto{93.476}{59.649}
\emmoveto{93.476}{59.639}
\emlineto{93.710}{59.501}
\emlineto{93.710}{59.511}
\emmoveto{93.710}{59.501}
\emlineto{93.944}{59.604}
\emmoveto{93.944}{59.594}
\emlineto{94.179}{59.681}
\emlineto{94.179}{59.691}
\emmoveto{94.179}{59.681}
\emlineto{94.413}{59.527}
\emlineto{94.413}{59.537}
\emmoveto{94.413}{59.527}
\emlineto{94.647}{60.030}
\emlineto{94.647}{60.040}
\emmoveto{94.647}{60.030}
\emlineto{94.881}{60.098}
\emlineto{94.881}{60.108}
\emmoveto{94.881}{60.098}
\emlineto{95.115}{59.929}
\emlineto{95.115}{59.939}
\emmoveto{95.115}{59.929}
\emlineto{95.349}{59.986}
\emlineto{95.349}{59.996}
\emmoveto{95.349}{59.986}
\emlineto{95.583}{59.808}
\emlineto{95.583}{59.818}
\emmoveto{95.583}{59.808}
\emlineto{95.817}{59.854}
\emlineto{95.817}{59.864}
\emmoveto{95.817}{59.854}
\emlineto{96.052}{60.088}
\emlineto{96.052}{60.098}
\emmoveto{96.052}{60.088}
\emlineto{96.286}{60.122}
\emlineto{96.286}{60.132}
\emmoveto{96.286}{60.122}
\emlineto{96.520}{59.924}
\emlineto{96.520}{59.934}
\emmoveto{96.520}{59.924}
\emlineto{96.754}{60.368}
\emlineto{96.754}{60.378}
\emmoveto{96.754}{60.368}
\emlineto{96.988}{60.160}
\emlineto{96.988}{60.170}
\emmoveto{96.988}{60.160}
\emlineto{97.222}{60.172}
\emlineto{97.222}{60.182}
\emmoveto{97.222}{60.172}
\emlineto{97.456}{60.373}
\emlineto{97.456}{60.383}
\emmoveto{97.456}{60.373}
\emlineto{97.690}{60.374}
\emlineto{97.690}{60.384}
\emmoveto{97.690}{60.374}
\emlineto{97.925}{60.147}
\emlineto{97.925}{60.157}
\emmoveto{97.925}{60.147}
\emlineto{98.159}{60.332}
\emlineto{98.159}{60.342}
\emmoveto{98.159}{60.332}
\emlineto{98.393}{60.317}
\emlineto{98.393}{60.327}
\emmoveto{98.393}{60.317}
\emlineto{98.627}{60.491}
\emlineto{98.627}{60.501}
\emmoveto{98.627}{60.491}
\emlineto{98.861}{60.466}
\emlineto{98.861}{60.476}
\emmoveto{98.861}{60.466}
\emlineto{99.095}{60.629}
\emlineto{99.095}{60.639}
\emmoveto{99.095}{60.629}
\emlineto{99.329}{60.375}
\emlineto{99.329}{60.385}
\emmoveto{99.329}{60.375}
\emlineto{99.563}{60.747}
\emlineto{99.563}{60.757}
\emmoveto{99.563}{60.747}
\emlineto{99.798}{60.483}
\emlineto{99.798}{60.493}
\emmoveto{99.798}{60.483}
\emlineto{100.032}{60.636}
\emmoveto{100.032}{60.626}
\emlineto{100.266}{60.764}
\emlineto{100.266}{60.774}
\emmoveto{100.266}{60.764}
\emlineto{100.500}{60.703}
\emlineto{100.500}{60.713}
\emmoveto{100.500}{60.703}
\emlineto{100.734}{60.840}
\emmoveto{100.734}{60.830}
\emlineto{100.968}{60.953}
\emlineto{100.968}{60.963}
\emmoveto{100.968}{60.953}
\emlineto{101.202}{60.663}
\emlineto{101.202}{60.673}
\emmoveto{101.202}{60.663}
\emlineto{101.437}{60.787}
\emmoveto{101.437}{60.777}
\emlineto{101.671}{60.885}
\emlineto{101.671}{60.895}
\emmoveto{101.671}{60.885}
\emlineto{101.905}{61.200}
\emlineto{101.905}{61.210}
\emmoveto{101.905}{61.200}
\emlineto{102.139}{60.893}
\emlineto{102.139}{60.903}
\emmoveto{102.139}{60.893}
\emlineto{102.373}{60.997}
\emmoveto{102.373}{60.987}
\emlineto{102.607}{61.086}
\emmoveto{102.607}{61.076}
\emlineto{102.841}{61.170}
\emmoveto{102.841}{61.160}
\emlineto{103.075}{61.250}
\emmoveto{103.075}{61.240}
\emlineto{103.310}{61.314}
\emlineto{103.310}{61.324}
\emmoveto{103.310}{61.314}
\emlineto{103.544}{60.983}
\emlineto{103.544}{60.993}
\emmoveto{103.544}{60.983}
\emlineto{103.778}{61.059}
\emmoveto{103.778}{61.049}
\emlineto{104.012}{61.121}
\emmoveto{104.012}{61.111}
\emlineto{104.246}{61.178}
\emmoveto{104.246}{61.168}
\emlineto{104.480}{61.230}
\emmoveto{104.480}{61.220}
\emlineto{104.714}{61.277}
\emmoveto{104.714}{61.267}
\emlineto{104.948}{61.321}
\emmoveto{104.948}{61.311}
\emlineto{105.183}{61.359}
\emmoveto{105.183}{61.349}
\emlineto{105.417}{61.383}
\emlineto{105.417}{61.393}
\emmoveto{105.417}{61.383}
\emlineto{105.651}{61.809}
\emlineto{105.651}{61.819}
\emmoveto{105.651}{61.809}
\emlineto{105.885}{61.632}
\emlineto{105.885}{61.642}
\emmoveto{105.885}{61.632}
\emlineto{106.119}{61.664}
\emmoveto{106.119}{61.654}
\emlineto{106.353}{61.680}
\emmoveto{106.353}{61.670}
\emlineto{106.587}{61.693}
\emmoveto{106.587}{61.683}
\emlineto{106.821}{61.691}
\emlineto{106.821}{61.701}
\emmoveto{106.821}{61.691}
\emlineto{107.056}{61.497}
\emlineto{107.056}{61.507}
\emmoveto{107.056}{61.497}
\emlineto{107.290}{61.888}
\emlineto{107.290}{61.898}
\emmoveto{107.290}{61.888}
\emlineto{107.524}{61.894}
\emmoveto{107.524}{61.884}
\emlineto{107.758}{61.875}
\emlineto{107.758}{61.885}
\emmoveto{107.758}{61.875}
\emlineto{107.992}{61.666}
\emlineto{107.992}{61.676}
\emmoveto{107.992}{61.666}
\emlineto{108.226}{62.039}
\emlineto{108.226}{62.049}
\emmoveto{108.226}{62.039}
\emlineto{108.460}{62.018}
\emlineto{108.460}{62.028}
\emmoveto{108.460}{62.018}
\emlineto{108.694}{61.799}
\emlineto{108.694}{61.809}
\emmoveto{108.694}{61.799}
\emlineto{108.929}{62.157}
\emlineto{108.929}{62.167}
\emmoveto{108.929}{62.157}
\emlineto{109.163}{62.124}
\emlineto{109.163}{62.134}
\emmoveto{109.163}{62.124}
\emlineto{109.397}{61.894}
\emlineto{109.397}{61.904}
\emmoveto{109.397}{61.894}
\emlineto{109.631}{62.240}
\emlineto{109.631}{62.250}
\emmoveto{109.631}{62.240}
\emlineto{109.865}{62.003}
\emlineto{109.865}{62.013}
\emmoveto{109.865}{62.003}
\emlineto{110.099}{62.339}
\emlineto{110.099}{62.349}
\emmoveto{110.099}{62.339}
\emlineto{110.333}{62.096}
\emlineto{110.333}{62.106}
\emmoveto{110.333}{62.096}
\emlineto{110.567}{62.423}
\emlineto{110.567}{62.433}
\emmoveto{110.567}{62.423}
\emlineto{110.802}{62.173}
\emlineto{110.802}{62.183}
\emmoveto{110.802}{62.173}
\emlineto{111.036}{62.491}
\emlineto{111.036}{62.501}
\emmoveto{111.036}{62.491}
\emlineto{111.270}{62.235}
\emlineto{111.270}{62.245}
\emmoveto{111.270}{62.235}
\emlineto{111.504}{62.545}
\emlineto{111.504}{62.555}
\emmoveto{111.504}{62.545}
\emlineto{111.738}{62.281}
\emlineto{111.738}{62.291}
\emmoveto{111.738}{62.281}
\emlineto{111.972}{62.583}
\emlineto{111.972}{62.593}
\emmoveto{111.972}{62.583}
\emlineto{112.206}{62.313}
\emlineto{112.206}{62.323}
\emmoveto{112.206}{62.313}
\emlineto{112.440}{62.420}
\emlineto{112.440}{62.430}
\emmoveto{112.440}{62.420}
\emlineto{112.675}{62.709}
\emlineto{112.675}{62.719}
\emmoveto{112.675}{62.709}
\emlineto{112.909}{62.429}
\emlineto{112.909}{62.439}
\emmoveto{112.909}{62.429}
\emlineto{113.143}{62.525}
\emlineto{113.143}{62.535}
\emmoveto{113.143}{62.525}
\emlineto{113.377}{62.802}
\emlineto{113.377}{62.812}
\emmoveto{113.377}{62.802}
\emlineto{113.611}{62.513}
\emlineto{113.611}{62.523}
\emmoveto{113.611}{62.513}
\emlineto{113.845}{62.598}
\emlineto{113.845}{62.608}
\emmoveto{113.845}{62.598}
\emlineto{114.079}{62.862}
\emlineto{114.079}{62.872}
\emmoveto{114.079}{62.862}
\emlineto{114.313}{62.940}
\emlineto{114.313}{62.950}
\emmoveto{114.313}{62.940}
\emlineto{114.548}{62.639}
\emlineto{114.548}{62.649}
\emmoveto{114.548}{62.639}
\emlineto{114.782}{62.720}
\emmoveto{114.782}{62.710}
\emlineto{115.016}{62.777}
\emlineto{115.016}{62.787}
\emmoveto{115.016}{62.777}
\emlineto{115.250}{63.022}
\emlineto{115.250}{63.032}
\emmoveto{115.250}{63.022}
\emlineto{115.484}{63.092}
\emmoveto{115.484}{63.082}
\emlineto{115.718}{63.148}
\emmoveto{115.718}{63.138}
\emlineto{115.952}{63.192}
\emlineto{115.952}{63.202}
\emmoveto{115.952}{63.192}
\emlineto{116.187}{62.870}
\emlineto{116.187}{62.880}
\emmoveto{116.187}{62.870}
\emlineto{116.421}{62.927}
\emmoveto{116.421}{62.917}
\emlineto{116.655}{62.970}
\emmoveto{116.655}{62.960}
\emlineto{116.889}{63.011}
\emmoveto{116.889}{63.001}
\emlineto{117.123}{63.048}
\emmoveto{117.123}{63.038}
\emlineto{117.357}{63.081}
\emmoveto{117.357}{63.071}
\emlineto{117.591}{63.112}
\emmoveto{117.591}{63.102}
\emlineto{117.825}{63.139}
\emmoveto{117.825}{63.129}
\emlineto{118.060}{63.163}
\emmoveto{118.060}{63.153}
\emlineto{118.294}{63.174}
\emlineto{118.294}{63.184}
\emmoveto{118.294}{63.174}
\emlineto{118.528}{63.558}
\emlineto{118.528}{63.568}
\emmoveto{118.528}{63.558}
\emlineto{118.762}{63.582}
\emmoveto{118.762}{63.572}
\emlineto{118.996}{63.593}
\emmoveto{118.996}{63.583}
\emlineto{119.230}{63.591}
\emlineto{119.230}{63.601}
\emmoveto{119.230}{63.591}
\emlineto{119.464}{63.423}
\emlineto{119.464}{63.433}
\emmoveto{119.464}{63.423}
\emlineto{119.698}{63.434}
\emmoveto{119.698}{63.424}
\emlineto{119.933}{63.423}
\emlineto{119.933}{63.433}
\emmoveto{119.933}{63.423}
\emlineto{120.167}{63.782}
\emlineto{120.167}{63.792}
\emmoveto{120.167}{63.782}
\emlineto{120.401}{63.774}
\emlineto{120.401}{63.784}
\emmoveto{120.401}{63.774}
\emlineto{120.635}{63.593}
\emlineto{120.635}{63.603}
\emmoveto{120.635}{63.593}
\emlineto{120.869}{63.579}
\emlineto{120.869}{63.589}
\emmoveto{120.869}{63.579}
\emlineto{121.103}{63.925}
\emlineto{121.103}{63.935}
\emmoveto{121.103}{63.925}
\emlineto{121.337}{63.905}
\emlineto{121.337}{63.915}
\emmoveto{121.337}{63.905}
\emlineto{121.571}{63.713}
\emlineto{121.571}{63.723}
\emmoveto{121.571}{63.713}
\emlineto{121.806}{63.687}
\emlineto{121.806}{63.697}
\emmoveto{121.806}{63.687}
\emlineto{122.040}{64.020}
\emlineto{122.040}{64.030}
\emmoveto{122.040}{64.020}
\emlineto{122.274}{63.820}
\emlineto{122.274}{63.830}
\emmoveto{122.274}{63.820}
\emlineto{122.508}{63.785}
\emlineto{122.508}{63.795}
\emmoveto{122.508}{63.785}
\emlineto{122.742}{63.941}
\emlineto{122.742}{63.951}
\emmoveto{122.742}{63.941}
\emlineto{122.976}{63.900}
\emlineto{122.976}{63.910}
\emmoveto{122.976}{63.900}
\emlineto{123.210}{64.216}
\emlineto{123.210}{64.226}
\emmoveto{123.210}{64.216}
\emlineto{123.444}{64.004}
\emlineto{123.444}{64.014}
\emmoveto{123.444}{64.004}
\emlineto{123.679}{64.312}
\emlineto{123.679}{64.322}
\emmoveto{123.679}{64.312}
\emlineto{123.913}{64.095}
\emlineto{123.913}{64.105}
\emmoveto{123.913}{64.095}
\emlineto{124.147}{64.041}
\emlineto{124.147}{64.051}
\emmoveto{124.147}{64.041}
\emlineto{124.381}{64.176}
\emlineto{124.381}{64.186}
\emmoveto{124.381}{64.176}
\emlineto{124.615}{64.115}
\emlineto{124.615}{64.125}
\emmoveto{124.615}{64.115}
\emlineto{124.849}{64.255}
\emmoveto{124.849}{64.245}
\emlineto{125.083}{64.371}
\emlineto{125.083}{64.381}
\emmoveto{125.083}{64.371}
\emlineto{125.317}{64.303}
\emlineto{125.317}{64.313}
\emmoveto{125.317}{64.303}
\emlineto{125.552}{64.424}
\emlineto{125.552}{64.434}
\emmoveto{125.552}{64.424}
\emlineto{125.786}{64.350}
\emlineto{125.786}{64.360}
\emmoveto{125.786}{64.350}
\emlineto{126.020}{64.465}
\emlineto{126.020}{64.475}
\emmoveto{126.020}{64.465}
\emlineto{126.254}{64.225}
\emlineto{126.254}{64.235}
\emmoveto{126.254}{64.225}
\emlineto{126.488}{64.496}
\emlineto{126.488}{64.506}
\emmoveto{126.488}{64.496}
\emlineto{126.722}{64.603}
\emlineto{126.722}{64.613}
\emmoveto{126.722}{64.603}
\emlineto{126.956}{64.356}
\emlineto{126.956}{64.366}
\emmoveto{126.956}{64.356}
\emlineto{127.190}{64.458}
\emlineto{127.190}{64.468}
\emmoveto{127.190}{64.458}
\emlineto{127.425}{64.717}
\emlineto{127.425}{64.727}
\emmoveto{127.425}{64.717}
\emlineto{127.659}{64.814}
\emlineto{127.659}{64.824}
\emmoveto{127.659}{64.814}
\emlineto{127.893}{64.557}
\emlineto{127.893}{64.567}
\emmoveto{127.893}{64.557}
\emlineto{128.127}{64.659}
\emmoveto{128.127}{64.649}
\emlineto{128.361}{64.748}
\emmoveto{128.361}{64.738}
\emlineto{128.595}{64.825}
\emlineto{128.595}{64.835}
\emmoveto{128.595}{64.825}
\emlineto{128.829}{64.717}
\emlineto{128.829}{64.727}
\emmoveto{128.829}{64.717}
\emlineto{129.063}{64.808}
\emmoveto{129.063}{64.798}
\emlineto{129.298}{64.887}
\emmoveto{129.298}{64.877}
\emlineto{129.532}{64.964}
\emmoveto{129.532}{64.954}
\emlineto{129.766}{65.037}
\emshow{36.780}{54.100}{wz}
\emshow{1.000}{10.000}{-4.00e-2}
\emshow{1.000}{17.000}{-3.60e-2}
\emshow{1.000}{24.000}{-3.20e-2}
\emshow{1.000}{31.000}{-2.80e-2}
\emshow{1.000}{38.000}{-2.40e-2}
\emshow{1.000}{45.000}{-2.00e-2}
\emshow{1.000}{52.000}{-1.60e-2}
\emshow{1.000}{59.000}{-1.20e-2}
\emshow{1.000}{66.000}{-8.00e-3}
\emshow{1.000}{73.000}{-4.00e-3}
\emshow{1.000}{80.000}{0.00e0}
\emshow{12.000}{5.000}{0.00e0}
\emshow{23.800}{5.000}{1.51e0}
\emshow{35.600}{5.000}{3.02e0}
\emshow{47.400}{5.000}{4.54e0}
\emshow{59.200}{5.000}{6.05e0}
\emshow{71.000}{5.000}{7.56e0}
\emshow{82.800}{5.000}{9.07e0}
\emshow{94.600}{5.000}{1.06e1}
\emshow{106.400}{5.000}{1.21e1}
\emshow{118.200}{5.000}{1.36e1}
\emshow{130.000}{5.000}{1.51e1}

\centerline{\bf {Fig. A. 3}}

\eject
\newcount\numpoint
\newcount\numpointo
\numpoint=1 \numpointo=1
\def\emmoveto#1#2{\offinterlineskip
\hbox to 0 true cm{\vbox to 0
true cm{\vskip - #2 true mm
\hskip #1 true mm \special{em:point
\the\numpoint}\vss}\hss}
\numpointo=\numpoint
\global\advance \numpoint by 1}
\def\emlineto#1#2{\offinterlineskip
\hbox to 0 true cm{\vbox to 0
true cm{\vskip - #2 true mm
\hskip #1 true mm \special{em:point
\the\numpoint}\vss}\hss}
\special{em:line
\the\numpointo,\the\numpoint}
\numpointo=\numpoint
\global\advance \numpoint by 1}
\def\emshow#1#2#3{\offinterlineskip
\hbox to 0 true cm{\vbox to 0
true cm{\vskip - #2 true mm
\hskip #1 true mm \vbox to 0
true cm{\vss\hbox{#3\hss
}}\vss}\hss}}
\special{em:linewidth 0.8pt}

\vrule width 0 mm height                0 mm depth 90.000 true mm

\special{em:linewidth 0.8pt}
\emmoveto{130.000}{10.000}
\emlineto{12.000}{10.000}
\emlineto{12.000}{80.000}
\emmoveto{71.000}{10.000}
\emlineto{71.000}{80.000}
\emmoveto{12.000}{45.000}
\emlineto{130.000}{45.000}
\emmoveto{130.000}{10.000}
\emlineto{130.000}{80.000}
\emlineto{12.000}{80.000}
\emlineto{12.000}{10.000}
\emlineto{130.000}{10.000}
\special{em:linewidth 0.4pt}
\emmoveto{12.000}{17.000}
\emlineto{130.000}{17.000}
\emmoveto{12.000}{24.000}
\emlineto{130.000}{24.000}
\emmoveto{12.000}{31.000}
\emlineto{130.000}{31.000}
\emmoveto{12.000}{38.000}
\emlineto{130.000}{38.000}
\emmoveto{12.000}{45.000}
\emlineto{130.000}{45.000}
\emmoveto{12.000}{52.000}
\emlineto{130.000}{52.000}
\emmoveto{12.000}{59.000}
\emlineto{130.000}{59.000}
\emmoveto{12.000}{66.000}
\emlineto{130.000}{66.000}
\emmoveto{12.000}{73.000}
\emlineto{130.000}{73.000}
\emmoveto{23.800}{10.000}
\emlineto{23.800}{80.000}
\emmoveto{35.600}{10.000}
\emlineto{35.600}{80.000}
\emmoveto{47.400}{10.000}
\emlineto{47.400}{80.000}
\emmoveto{59.200}{10.000}
\emlineto{59.200}{80.000}
\emmoveto{71.000}{10.000}
\emlineto{71.000}{80.000}
\emmoveto{82.800}{10.000}
\emlineto{82.800}{80.000}
\emmoveto{94.600}{10.000}
\emlineto{94.600}{80.000}
\emmoveto{106.400}{10.000}
\emlineto{106.400}{80.000}
\emmoveto{118.200}{10.000}
\emlineto{118.200}{80.000}
\special{em:linewidth 0.8pt}
\emmoveto{12.000}{10.000}
\emlineto{12.262}{10.092}
\emmoveto{12.262}{10.082}
\emlineto{12.524}{10.171}
\emmoveto{12.524}{10.161}
\emlineto{12.787}{10.247}
\emmoveto{12.787}{10.237}
\emlineto{13.049}{10.321}
\emmoveto{13.049}{10.311}
\emlineto{13.311}{10.391}
\emmoveto{13.311}{10.381}
\emlineto{13.573}{10.458}
\emmoveto{13.573}{10.448}
\emlineto{13.836}{10.523}
\emmoveto{13.836}{10.513}
\emlineto{14.098}{10.585}
\emmoveto{14.098}{10.575}
\emlineto{14.360}{10.645}
\emmoveto{14.360}{10.635}
\emlineto{14.622}{10.702}
\emmoveto{14.622}{10.692}
\emlineto{14.884}{10.758}
\emmoveto{14.884}{10.748}
\emlineto{15.147}{10.811}
\emmoveto{15.147}{10.801}
\emlineto{15.409}{10.862}
\emmoveto{15.409}{10.852}
\emlineto{15.671}{10.911}
\emmoveto{15.671}{10.901}
\emlineto{15.933}{10.958}
\emmoveto{15.933}{10.948}
\emlineto{16.196}{11.003}
\emmoveto{16.196}{10.993}
\emlineto{16.458}{11.047}
\emmoveto{16.458}{11.037}
\emlineto{16.720}{11.089}
\emmoveto{16.720}{11.079}
\emlineto{16.982}{11.129}
\emmoveto{16.982}{11.119}
\emlineto{17.244}{11.167}
\emmoveto{17.244}{11.157}
\emlineto{17.507}{11.205}
\emmoveto{17.507}{11.195}
\emlineto{17.769}{11.240}
\emmoveto{17.769}{11.230}
\emlineto{18.031}{11.275}
\emmoveto{18.031}{11.265}
\emlineto{18.293}{11.308}
\emmoveto{18.293}{11.298}
\emlineto{18.556}{11.339}
\emmoveto{18.556}{11.329}
\emlineto{18.818}{11.370}
\emmoveto{18.818}{11.360}
\emlineto{19.080}{11.399}
\emmoveto{19.080}{11.389}
\emlineto{19.342}{11.427}
\emmoveto{19.342}{11.417}
\emlineto{19.604}{11.454}
\emmoveto{19.604}{11.444}
\emlineto{19.867}{11.480}
\emmoveto{19.867}{11.470}
\emlineto{20.129}{11.505}
\emmoveto{20.129}{11.495}
\emlineto{20.391}{11.529}
\emmoveto{20.391}{11.519}
\emlineto{20.653}{11.553}
\emmoveto{20.653}{11.543}
\emlineto{20.916}{11.575}
\emmoveto{20.916}{11.565}
\emlineto{21.178}{11.600}
\emmoveto{21.178}{11.590}
\emlineto{21.440}{11.627}
\emmoveto{21.440}{11.617}
\emlineto{21.702}{11.657}
\emmoveto{21.702}{11.647}
\emlineto{21.964}{11.688}
\emmoveto{21.964}{11.678}
\emlineto{22.227}{11.720}
\emmoveto{22.227}{11.710}
\emlineto{22.489}{11.754}
\emmoveto{22.489}{11.744}
\emlineto{22.751}{11.789}
\emmoveto{22.751}{11.779}
\emlineto{23.013}{11.825}
\emmoveto{23.013}{11.815}
\emlineto{23.276}{11.862}
\emmoveto{23.276}{11.852}
\emlineto{23.538}{11.900}
\emmoveto{23.538}{11.890}
\emlineto{23.800}{11.939}
\emmoveto{23.800}{11.929}
\emlineto{24.062}{11.978}
\emmoveto{24.062}{11.968}
\emlineto{24.324}{12.017}
\emmoveto{24.324}{12.007}
\emlineto{24.587}{12.057}
\emmoveto{24.587}{12.047}
\emlineto{24.849}{12.098}
\emmoveto{24.849}{12.088}
\emlineto{25.111}{12.138}
\emmoveto{25.111}{12.128}
\emlineto{25.373}{12.179}
\emmoveto{25.373}{12.169}
\emlineto{25.636}{12.219}
\emmoveto{25.636}{12.209}
\emlineto{25.898}{12.260}
\emmoveto{25.898}{12.250}
\emlineto{26.160}{12.301}
\emmoveto{26.160}{12.291}
\emlineto{26.422}{12.341}
\emmoveto{26.422}{12.331}
\emlineto{26.684}{12.382}
\emmoveto{26.684}{12.372}
\emlineto{26.947}{12.422}
\emmoveto{26.947}{12.412}
\emlineto{27.209}{12.462}
\emmoveto{27.209}{12.452}
\emlineto{27.471}{12.501}
\emmoveto{27.471}{12.491}
\emlineto{27.733}{12.541}
\emmoveto{27.733}{12.531}
\emlineto{27.996}{12.579}
\emmoveto{27.996}{12.569}
\emlineto{28.258}{12.618}
\emmoveto{28.258}{12.608}
\emlineto{28.520}{12.656}
\emmoveto{28.520}{12.646}
\emlineto{28.782}{12.694}
\emmoveto{28.782}{12.684}
\emlineto{29.044}{12.731}
\emmoveto{29.044}{12.721}
\emlineto{29.307}{12.767}
\emmoveto{29.307}{12.757}
\emlineto{29.569}{12.804}
\emmoveto{29.569}{12.794}
\emlineto{29.831}{12.839}
\emmoveto{29.831}{12.829}
\emlineto{30.093}{12.875}
\emmoveto{30.093}{12.865}
\emlineto{30.356}{12.910}
\emmoveto{30.356}{12.900}
\emlineto{30.618}{12.945}
\emmoveto{30.618}{12.935}
\emlineto{30.880}{12.980}
\emmoveto{30.880}{12.970}
\emlineto{31.142}{13.014}
\emmoveto{31.142}{13.004}
\emlineto{31.404}{13.049}
\emmoveto{31.404}{13.039}
\emlineto{31.667}{13.084}
\emmoveto{31.667}{13.074}
\emlineto{31.929}{13.119}
\emmoveto{31.929}{13.109}
\emlineto{32.191}{13.154}
\emmoveto{32.191}{13.144}
\emlineto{32.453}{13.189}
\emmoveto{32.453}{13.179}
\emlineto{32.716}{13.224}
\emmoveto{32.716}{13.214}
\emlineto{32.978}{13.260}
\emmoveto{32.978}{13.250}
\emlineto{33.240}{13.296}
\emmoveto{33.240}{13.286}
\emlineto{33.502}{13.332}
\emmoveto{33.502}{13.322}
\emlineto{33.764}{13.368}
\emmoveto{33.764}{13.358}
\emlineto{34.027}{13.404}
\emmoveto{34.027}{13.394}
\emlineto{34.289}{13.441}
\emmoveto{34.289}{13.431}
\emlineto{34.551}{13.478}
\emmoveto{34.551}{13.468}
\emlineto{34.813}{13.515}
\emmoveto{34.813}{13.505}
\emlineto{35.076}{13.552}
\emmoveto{35.076}{13.542}
\emlineto{35.338}{13.589}
\emmoveto{35.338}{13.579}
\emlineto{35.600}{13.627}
\emmoveto{35.600}{13.617}
\emlineto{35.862}{13.665}
\emmoveto{35.862}{13.655}
\emlineto{36.124}{13.702}
\emmoveto{36.124}{13.692}
\emlineto{36.387}{13.740}
\emmoveto{36.387}{13.730}
\emlineto{36.649}{13.778}
\emmoveto{36.649}{13.768}
\emlineto{36.911}{13.816}
\emmoveto{36.911}{13.806}
\emlineto{37.173}{13.855}
\emmoveto{37.173}{13.845}
\emlineto{37.436}{13.893}
\emmoveto{37.436}{13.883}
\emlineto{37.698}{13.931}
\emmoveto{37.698}{13.921}
\emlineto{37.960}{13.969}
\emmoveto{37.960}{13.959}
\emlineto{38.222}{14.008}
\emmoveto{38.222}{13.998}
\emlineto{38.484}{14.046}
\emmoveto{38.484}{14.036}
\emlineto{38.747}{14.084}
\emmoveto{38.747}{14.074}
\emlineto{39.009}{14.122}
\emmoveto{39.009}{14.112}
\emlineto{39.271}{14.160}
\emmoveto{39.271}{14.150}
\emlineto{39.533}{14.198}
\emmoveto{39.533}{14.188}
\emlineto{39.796}{14.236}
\emmoveto{39.796}{14.226}
\emlineto{40.058}{14.274}
\emmoveto{40.058}{14.264}
\emlineto{40.320}{14.312}
\emmoveto{40.320}{14.302}
\emlineto{40.582}{14.349}
\emmoveto{40.582}{14.339}
\emlineto{40.844}{14.387}
\emmoveto{40.844}{14.377}
\emlineto{41.107}{14.425}
\emmoveto{41.107}{14.415}
\emlineto{41.369}{14.462}
\emmoveto{41.369}{14.452}
\emlineto{41.631}{14.500}
\emmoveto{41.631}{14.490}
\emlineto{41.893}{14.537}
\emmoveto{41.893}{14.527}
\emlineto{42.156}{14.575}
\emmoveto{42.156}{14.565}
\emlineto{42.418}{14.613}
\emmoveto{42.418}{14.603}
\emlineto{42.680}{14.650}
\emmoveto{42.680}{14.640}
\emlineto{42.942}{14.688}
\emmoveto{42.942}{14.678}
\emlineto{43.204}{14.725}
\emmoveto{43.204}{14.715}
\emlineto{43.467}{14.763}
\emmoveto{43.467}{14.753}
\emlineto{43.729}{14.801}
\emmoveto{43.729}{14.791}
\emlineto{43.991}{14.838}
\emmoveto{43.991}{14.828}
\emlineto{44.253}{14.876}
\emmoveto{44.253}{14.866}
\emlineto{44.516}{14.914}
\emmoveto{44.516}{14.904}
\emlineto{44.778}{14.952}
\emmoveto{44.778}{14.942}
\emlineto{45.040}{14.990}
\emmoveto{45.040}{14.980}
\emlineto{45.302}{15.028}
\emmoveto{45.302}{15.018}
\emlineto{45.564}{15.066}
\emmoveto{45.564}{15.056}
\emlineto{45.827}{15.104}
\emmoveto{45.827}{15.094}
\emlineto{46.089}{15.142}
\emmoveto{46.089}{15.132}
\emlineto{46.351}{15.181}
\emmoveto{46.351}{15.171}
\emlineto{46.613}{15.219}
\emmoveto{46.613}{15.209}
\emlineto{46.876}{15.258}
\emmoveto{46.876}{15.248}
\emlineto{47.138}{15.296}
\emmoveto{47.138}{15.286}
\emlineto{47.400}{15.335}
\emmoveto{47.400}{15.325}
\emlineto{47.662}{15.373}
\emmoveto{47.662}{15.363}
\emlineto{47.924}{15.412}
\emmoveto{47.924}{15.402}
\emlineto{48.187}{15.451}
\emmoveto{48.187}{15.441}
\emlineto{48.449}{15.489}
\emmoveto{48.449}{15.479}
\emlineto{48.711}{15.528}
\emmoveto{48.711}{15.518}
\emlineto{48.973}{15.567}
\emmoveto{48.973}{15.557}
\emlineto{49.236}{15.606}
\emmoveto{49.236}{15.596}
\emlineto{49.498}{15.645}
\emmoveto{49.498}{15.635}
\emlineto{49.760}{15.684}
\emmoveto{49.760}{15.674}
\emlineto{50.022}{15.723}
\emmoveto{50.022}{15.713}
\emlineto{50.284}{15.762}
\emmoveto{50.284}{15.752}
\emlineto{50.547}{15.801}
\emmoveto{50.547}{15.791}
\emlineto{50.809}{15.840}
\emmoveto{50.809}{15.830}
\emlineto{51.071}{15.879}
\emmoveto{51.071}{15.869}
\emlineto{51.333}{15.918}
\emmoveto{51.333}{15.908}
\emlineto{51.596}{15.957}
\emmoveto{51.596}{15.947}
\emlineto{51.858}{15.996}
\emmoveto{51.858}{15.986}
\emlineto{52.120}{16.036}
\emmoveto{52.120}{16.026}
\emlineto{52.382}{16.075}
\emmoveto{52.382}{16.065}
\emlineto{52.644}{16.114}
\emmoveto{52.644}{16.104}
\emlineto{52.907}{16.153}
\emmoveto{52.907}{16.143}
\emlineto{53.169}{16.193}
\emmoveto{53.169}{16.183}
\emlineto{53.431}{16.232}
\emmoveto{53.431}{16.222}
\emlineto{53.693}{16.271}
\emmoveto{53.693}{16.261}
\emlineto{53.956}{16.311}
\emmoveto{53.956}{16.301}
\emlineto{54.218}{16.350}
\emmoveto{54.218}{16.340}
\emlineto{54.480}{16.389}
\emmoveto{54.480}{16.379}
\emlineto{54.742}{16.429}
\emmoveto{54.742}{16.419}
\emlineto{55.004}{16.468}
\emmoveto{55.004}{16.458}
\emlineto{55.267}{16.508}
\emmoveto{55.267}{16.498}
\emlineto{55.529}{16.548}
\emmoveto{55.529}{16.538}
\emlineto{55.791}{16.587}
\emmoveto{55.791}{16.577}
\emlineto{56.053}{16.627}
\emmoveto{56.053}{16.617}
\emlineto{56.316}{16.667}
\emmoveto{56.316}{16.657}
\emlineto{56.578}{16.707}
\emmoveto{56.578}{16.697}
\emlineto{56.840}{16.747}
\emmoveto{56.840}{16.737}
\emlineto{57.102}{16.787}
\emmoveto{57.102}{16.777}
\emlineto{57.364}{16.827}
\emmoveto{57.364}{16.817}
\emlineto{57.627}{16.867}
\emmoveto{57.627}{16.857}
\emlineto{57.889}{16.907}
\emmoveto{57.889}{16.897}
\emlineto{58.151}{16.947}
\emmoveto{58.151}{16.937}
\emlineto{58.413}{16.987}
\emmoveto{58.413}{16.977}
\emlineto{58.676}{17.027}
\emmoveto{58.676}{17.017}
\emlineto{58.938}{17.068}
\emmoveto{58.938}{17.058}
\emlineto{59.200}{17.108}
\emmoveto{59.200}{17.098}
\emlineto{59.462}{17.148}
\emmoveto{59.462}{17.138}
\emlineto{59.724}{17.189}
\emmoveto{59.724}{17.179}
\emlineto{59.987}{17.229}
\emmoveto{59.987}{17.219}
\emlineto{60.249}{17.270}
\emmoveto{60.249}{17.260}
\emlineto{60.511}{17.311}
\emmoveto{60.511}{17.301}
\emlineto{60.773}{17.351}
\emmoveto{60.773}{17.341}
\emlineto{61.036}{17.392}
\emmoveto{61.036}{17.382}
\emlineto{61.298}{17.433}
\emmoveto{61.298}{17.423}
\emlineto{61.560}{17.474}
\emmoveto{61.560}{17.464}
\emlineto{61.822}{17.514}
\emmoveto{61.822}{17.504}
\emlineto{62.084}{17.555}
\emmoveto{62.084}{17.545}
\emlineto{62.347}{17.596}
\emmoveto{62.347}{17.586}
\emlineto{62.609}{17.637}
\emmoveto{62.609}{17.627}
\emlineto{62.871}{17.678}
\emmoveto{62.871}{17.668}
\emlineto{63.133}{17.720}
\emmoveto{63.133}{17.710}
\emlineto{63.396}{17.761}
\emmoveto{63.396}{17.751}
\emlineto{63.658}{17.802}
\emmoveto{63.658}{17.792}
\emlineto{63.920}{17.843}
\emmoveto{63.920}{17.833}
\emlineto{64.182}{17.885}
\emmoveto{64.182}{17.875}
\emlineto{64.444}{17.926}
\emmoveto{64.444}{17.916}
\emlineto{64.707}{17.967}
\emmoveto{64.707}{17.957}
\emlineto{64.969}{18.009}
\emmoveto{64.969}{17.999}
\emlineto{65.231}{18.051}
\emmoveto{65.231}{18.041}
\emlineto{65.493}{18.092}
\emmoveto{65.493}{18.082}
\emlineto{65.756}{18.134}
\emmoveto{65.756}{18.124}
\emlineto{66.018}{18.176}
\emmoveto{66.018}{18.166}
\emlineto{66.280}{18.217}
\emmoveto{66.280}{18.207}
\emlineto{66.542}{18.259}
\emmoveto{66.542}{18.249}
\emlineto{66.804}{18.301}
\emmoveto{66.804}{18.291}
\emlineto{67.067}{18.343}
\emmoveto{67.067}{18.333}
\emlineto{67.329}{18.385}
\emmoveto{67.329}{18.375}
\emlineto{67.591}{18.427}
\emmoveto{67.591}{18.417}
\emlineto{67.853}{18.469}
\emmoveto{67.853}{18.459}
\emlineto{68.116}{18.512}
\emmoveto{68.116}{18.502}
\emlineto{68.378}{18.554}
\emmoveto{68.378}{18.544}
\emlineto{68.640}{18.596}
\emmoveto{68.640}{18.586}
\emlineto{68.902}{18.639}
\emmoveto{68.902}{18.629}
\emlineto{69.164}{18.681}
\emmoveto{69.164}{18.671}
\emlineto{69.427}{18.724}
\emmoveto{69.427}{18.714}
\emlineto{69.689}{18.766}
\emmoveto{69.689}{18.756}
\emlineto{69.951}{18.809}
\emmoveto{69.951}{18.799}
\emlineto{70.213}{18.852}
\emmoveto{70.213}{18.842}
\emlineto{70.476}{18.894}
\emmoveto{70.476}{18.884}
\emlineto{70.738}{18.937}
\emmoveto{70.738}{18.927}
\emlineto{71.000}{18.980}
\emmoveto{71.000}{18.970}
\emlineto{71.262}{19.023}
\emmoveto{71.262}{19.013}
\emlineto{71.524}{19.066}
\emmoveto{71.524}{19.056}
\emlineto{71.787}{19.109}
\emmoveto{71.787}{19.099}
\emlineto{72.049}{19.153}
\emmoveto{72.049}{19.143}
\emlineto{72.311}{19.196}
\emmoveto{72.311}{19.186}
\emlineto{72.573}{19.239}
\emmoveto{72.573}{19.229}
\emlineto{72.836}{19.283}
\emmoveto{72.836}{19.273}
\emlineto{73.098}{19.326}
\emmoveto{73.098}{19.316}
\emlineto{73.360}{19.370}
\emmoveto{73.360}{19.360}
\emlineto{73.622}{19.413}
\emmoveto{73.622}{19.403}
\emlineto{73.884}{19.457}
\emmoveto{73.884}{19.447}
\emlineto{74.147}{19.501}
\emmoveto{74.147}{19.491}
\emlineto{74.409}{19.545}
\emmoveto{74.409}{19.535}
\emlineto{74.671}{19.588}
\emmoveto{74.671}{19.578}
\emlineto{74.933}{19.632}
\emmoveto{74.933}{19.622}
\emlineto{75.196}{19.677}
\emmoveto{75.196}{19.667}
\emlineto{75.458}{19.721}
\emmoveto{75.458}{19.711}
\emlineto{75.720}{19.765}
\emmoveto{75.720}{19.755}
\emlineto{75.982}{19.809}
\emmoveto{75.982}{19.799}
\emlineto{76.244}{19.854}
\emmoveto{76.244}{19.844}
\emlineto{76.507}{19.898}
\emmoveto{76.507}{19.888}
\emlineto{76.769}{19.943}
\emmoveto{76.769}{19.933}
\emlineto{77.031}{19.987}
\emmoveto{77.031}{19.977}
\emlineto{77.293}{20.032}
\emmoveto{77.293}{20.022}
\emlineto{77.556}{20.077}
\emmoveto{77.556}{20.067}
\emlineto{77.818}{20.121}
\emmoveto{77.818}{20.111}
\emlineto{78.080}{20.166}
\emmoveto{78.080}{20.156}
\emlineto{78.342}{20.211}
\emmoveto{78.342}{20.201}
\emlineto{78.604}{20.256}
\emmoveto{78.604}{20.246}
\emlineto{78.867}{20.302}
\emmoveto{78.867}{20.292}
\emlineto{79.129}{20.347}
\emmoveto{79.129}{20.337}
\emlineto{79.391}{20.392}
\emmoveto{79.391}{20.382}
\emlineto{79.653}{20.438}
\emmoveto{79.653}{20.428}
\emlineto{79.916}{20.483}
\emmoveto{79.916}{20.473}
\emlineto{80.178}{20.529}
\emmoveto{80.178}{20.519}
\emlineto{80.440}{20.574}
\emmoveto{80.440}{20.564}
\emlineto{80.702}{20.620}
\emmoveto{80.702}{20.610}
\emlineto{80.964}{20.666}
\emmoveto{80.964}{20.656}
\emlineto{81.227}{20.712}
\emmoveto{81.227}{20.702}
\emlineto{81.489}{20.758}
\emmoveto{81.489}{20.748}
\emlineto{81.751}{20.804}
\emmoveto{81.751}{20.794}
\emlineto{82.013}{20.850}
\emmoveto{82.013}{20.840}
\emlineto{82.276}{20.896}
\emmoveto{82.276}{20.886}
\emlineto{82.538}{20.943}
\emmoveto{82.538}{20.933}
\emlineto{82.800}{20.989}
\emmoveto{82.800}{20.979}
\emlineto{83.062}{21.036}
\emmoveto{83.062}{21.026}
\emlineto{83.324}{21.083}
\emmoveto{83.324}{21.073}
\emlineto{83.587}{21.129}
\emmoveto{83.587}{21.119}
\emlineto{83.849}{21.176}
\emmoveto{83.849}{21.166}
\emlineto{84.111}{21.223}
\emmoveto{84.111}{21.213}
\emlineto{84.373}{21.270}
\emmoveto{84.373}{21.260}
\emlineto{84.636}{21.317}
\emmoveto{84.636}{21.307}
\emlineto{84.898}{21.364}
\emmoveto{84.898}{21.354}
\emlineto{85.160}{21.412}
\emmoveto{85.160}{21.402}
\emlineto{85.422}{21.459}
\emmoveto{85.422}{21.449}
\emlineto{85.684}{21.507}
\emmoveto{85.684}{21.497}
\emlineto{85.947}{21.554}
\emmoveto{85.947}{21.544}
\emlineto{86.209}{21.602}
\emmoveto{86.209}{21.592}
\emlineto{86.471}{21.650}
\emmoveto{86.471}{21.640}
\emlineto{86.733}{21.698}
\emmoveto{86.733}{21.688}
\emlineto{86.996}{21.746}
\emmoveto{86.996}{21.736}
\emlineto{87.258}{21.794}
\emmoveto{87.258}{21.784}
\emlineto{87.520}{21.842}
\emmoveto{87.520}{21.832}
\emlineto{87.782}{21.891}
\emmoveto{87.782}{21.881}
\emlineto{88.044}{21.939}
\emmoveto{88.044}{21.929}
\emlineto{88.307}{21.988}
\emmoveto{88.307}{21.978}
\emlineto{88.569}{22.036}
\emmoveto{88.569}{22.026}
\emlineto{88.831}{22.085}
\emmoveto{88.831}{22.075}
\emlineto{89.093}{22.134}
\emmoveto{89.093}{22.124}
\emlineto{89.356}{22.183}
\emmoveto{89.356}{22.173}
\emlineto{89.618}{22.232}
\emmoveto{89.618}{22.222}
\emlineto{89.880}{22.281}
\emmoveto{89.880}{22.271}
\emlineto{90.142}{22.330}
\emmoveto{90.142}{22.320}
\emlineto{90.404}{22.380}
\emmoveto{90.404}{22.370}
\emlineto{90.667}{22.429}
\emmoveto{90.667}{22.419}
\emlineto{90.929}{22.479}
\emmoveto{90.929}{22.469}
\emlineto{91.191}{22.529}
\emmoveto{91.191}{22.519}
\emlineto{91.453}{22.579}
\emmoveto{91.453}{22.569}
\emlineto{91.716}{22.629}
\emmoveto{91.716}{22.619}
\emlineto{91.978}{22.679}
\emmoveto{91.978}{22.669}
\emlineto{92.240}{22.729}
\emmoveto{92.240}{22.719}
\emlineto{92.502}{22.779}
\emmoveto{92.502}{22.769}
\emlineto{92.764}{22.830}
\emmoveto{92.764}{22.820}
\emlineto{93.027}{22.881}
\emmoveto{93.027}{22.871}
\emlineto{93.289}{22.931}
\emmoveto{93.289}{22.921}
\emlineto{93.551}{22.982}
\emmoveto{93.551}{22.972}
\emlineto{93.813}{23.033}
\emmoveto{93.813}{23.023}
\emlineto{94.076}{23.084}
\emmoveto{94.076}{23.074}
\emlineto{94.338}{23.135}
\emmoveto{94.338}{23.125}
\emlineto{94.600}{23.187}
\emmoveto{94.600}{23.177}
\emlineto{94.862}{23.238}
\emmoveto{94.862}{23.228}
\emlineto{95.124}{23.290}
\emmoveto{95.124}{23.280}
\emlineto{95.387}{23.342}
\emmoveto{95.387}{23.332}
\emlineto{95.649}{23.393}
\emmoveto{95.649}{23.383}
\emlineto{95.911}{23.445}
\emmoveto{95.911}{23.435}
\emlineto{96.173}{23.498}
\emmoveto{96.173}{23.488}
\emlineto{96.436}{23.550}
\emmoveto{96.436}{23.540}
\emlineto{96.698}{23.602}
\emmoveto{96.698}{23.592}
\emlineto{96.960}{23.655}
\emmoveto{96.960}{23.645}
\emlineto{97.222}{23.707}
\emmoveto{97.222}{23.697}
\emlineto{97.484}{23.760}
\emmoveto{97.484}{23.750}
\emlineto{97.747}{23.813}
\emmoveto{97.747}{23.803}
\emlineto{98.009}{23.866}
\emmoveto{98.009}{23.856}
\emlineto{98.271}{23.920}
\emmoveto{98.271}{23.910}
\emlineto{98.533}{23.973}
\emmoveto{98.533}{23.963}
\emlineto{98.796}{24.026}
\emmoveto{98.796}{24.016}
\emlineto{99.058}{24.080}
\emmoveto{99.058}{24.070}
\emlineto{99.320}{24.134}
\emmoveto{99.320}{24.124}
\emlineto{99.582}{24.188}
\emmoveto{99.582}{24.178}
\emlineto{99.844}{24.242}
\emmoveto{99.844}{24.232}
\emlineto{100.107}{24.296}
\emmoveto{100.107}{24.286}
\emlineto{100.369}{24.351}
\emmoveto{100.369}{24.341}
\emlineto{100.631}{24.405}
\emmoveto{100.631}{24.395}
\emlineto{100.893}{24.460}
\emmoveto{100.893}{24.450}
\emlineto{101.156}{24.515}
\emmoveto{101.156}{24.505}
\emlineto{101.418}{24.570}
\emmoveto{101.418}{24.560}
\emlineto{101.680}{24.625}
\emmoveto{101.680}{24.615}
\emlineto{101.942}{24.680}
\emmoveto{101.942}{24.670}
\emlineto{102.204}{24.736}
\emmoveto{102.204}{24.726}
\emlineto{102.467}{24.791}
\emmoveto{102.467}{24.781}
\emlineto{102.729}{24.847}
\emmoveto{102.729}{24.837}
\emlineto{102.991}{24.903}
\emmoveto{102.991}{24.893}
\emlineto{103.253}{24.959}
\emmoveto{103.253}{24.949}
\emlineto{103.516}{25.016}
\emmoveto{103.516}{25.006}
\emlineto{103.778}{25.072}
\emmoveto{103.778}{25.062}
\emlineto{104.040}{25.129}
\emmoveto{104.040}{25.119}
\emlineto{104.302}{25.186}
\emmoveto{104.302}{25.176}
\emlineto{104.564}{25.243}
\emmoveto{104.564}{25.233}
\emlineto{104.827}{25.300}
\emmoveto{104.827}{25.290}
\emlineto{105.089}{25.357}
\emmoveto{105.089}{25.347}
\emlineto{105.351}{25.415}
\emmoveto{105.351}{25.405}
\emlineto{105.613}{25.472}
\emmoveto{105.613}{25.462}
\emlineto{105.876}{25.530}
\emmoveto{105.876}{25.520}
\emlineto{106.138}{25.588}
\emmoveto{106.138}{25.578}
\emlineto{106.400}{25.646}
\emmoveto{106.400}{25.636}
\emlineto{106.662}{25.705}
\emmoveto{106.662}{25.695}
\emlineto{106.924}{25.763}
\emmoveto{106.924}{25.753}
\emlineto{107.187}{25.822}
\emmoveto{107.187}{25.812}
\emlineto{107.449}{25.881}
\emmoveto{107.449}{25.871}
\emlineto{107.711}{25.940}
\emmoveto{107.711}{25.930}
\emlineto{107.973}{26.000}
\emmoveto{107.973}{25.990}
\emlineto{108.236}{26.059}
\emmoveto{108.236}{26.049}
\emlineto{108.498}{26.119}
\emmoveto{108.498}{26.109}
\emlineto{108.760}{26.179}
\emmoveto{108.760}{26.169}
\emlineto{109.022}{26.239}
\emmoveto{109.022}{26.229}
\emlineto{109.284}{26.299}
\emmoveto{109.284}{26.289}
\emlineto{109.547}{26.360}
\emmoveto{109.547}{26.350}
\emlineto{109.809}{26.421}
\emmoveto{109.809}{26.411}
\emlineto{110.071}{26.481}
\emmoveto{110.071}{26.471}
\emlineto{110.333}{26.543}
\emmoveto{110.333}{26.533}
\emlineto{110.596}{26.604}
\emmoveto{110.596}{26.594}
\emlineto{110.858}{26.666}
\emmoveto{110.858}{26.656}
\emlineto{111.120}{26.727}
\emmoveto{111.120}{26.717}
\emlineto{111.382}{26.789}
\emmoveto{111.382}{26.779}
\emlineto{111.644}{26.851}
\emmoveto{111.644}{26.841}
\emlineto{111.907}{26.914}
\emmoveto{111.907}{26.904}
\emlineto{112.169}{26.976}
\emmoveto{112.169}{26.966}
\emlineto{112.431}{27.039}
\emmoveto{112.431}{27.029}
\emlineto{112.693}{27.102}
\emmoveto{112.693}{27.092}
\emlineto{112.956}{27.166}
\emmoveto{112.956}{27.156}
\emlineto{113.218}{27.229}
\emmoveto{113.218}{27.219}
\emlineto{113.480}{27.293}
\emmoveto{113.480}{27.283}
\emlineto{113.742}{27.357}
\emmoveto{113.742}{27.347}
\emlineto{114.004}{27.421}
\emmoveto{114.004}{27.411}
\emlineto{114.267}{27.486}
\emmoveto{114.267}{27.476}
\emlineto{114.529}{27.550}
\emmoveto{114.529}{27.540}
\emlineto{114.791}{27.615}
\emmoveto{114.791}{27.605}
\emlineto{115.053}{27.681}
\emmoveto{115.053}{27.671}
\emlineto{115.316}{27.746}
\emmoveto{115.316}{27.736}
\emlineto{115.578}{27.812}
\emmoveto{115.578}{27.802}
\emlineto{115.840}{27.878}
\emmoveto{115.840}{27.868}
\emlineto{116.102}{27.944}
\emmoveto{116.102}{27.934}
\emlineto{116.364}{28.010}
\emmoveto{116.364}{28.000}
\emlineto{116.627}{28.077}
\emmoveto{116.627}{28.067}
\emlineto{116.889}{28.144}
\emmoveto{116.889}{28.134}
\emlineto{117.151}{28.211}
\emmoveto{117.151}{28.201}
\emlineto{117.413}{28.278}
\emmoveto{117.413}{28.268}
\emlineto{117.676}{28.346}
\emmoveto{117.676}{28.336}
\emlineto{117.938}{28.414}
\emmoveto{117.938}{28.404}
\emlineto{118.200}{28.482}
\emmoveto{118.200}{28.472}
\emlineto{118.462}{28.551}
\emmoveto{118.462}{28.541}
\emlineto{118.724}{28.620}
\emmoveto{118.724}{28.610}
\emlineto{118.987}{28.689}
\emmoveto{118.987}{28.679}
\emlineto{119.249}{28.758}
\emmoveto{119.249}{28.748}
\emlineto{119.511}{28.828}
\emmoveto{119.511}{28.818}
\emlineto{119.773}{28.898}
\emmoveto{119.773}{28.888}
\emlineto{120.036}{28.968}
\emmoveto{120.036}{28.958}
\emlineto{120.298}{29.039}
\emmoveto{120.298}{29.029}
\emlineto{120.560}{29.110}
\emmoveto{120.560}{29.100}
\emlineto{120.822}{29.181}
\emmoveto{120.822}{29.171}
\emlineto{121.084}{29.252}
\emmoveto{121.084}{29.242}
\emlineto{121.347}{29.324}
\emmoveto{121.347}{29.314}
\emlineto{121.609}{29.396}
\emmoveto{121.609}{29.386}
\emlineto{121.871}{29.468}
\emmoveto{121.871}{29.458}
\emlineto{122.133}{29.541}
\emmoveto{122.133}{29.531}
\emlineto{122.396}{29.614}
\emmoveto{122.396}{29.604}
\emlineto{122.658}{29.687}
\emmoveto{122.658}{29.677}
\emlineto{122.920}{29.761}
\emmoveto{122.920}{29.751}
\emlineto{123.182}{29.835}
\emmoveto{123.182}{29.825}
\emlineto{123.444}{29.909}
\emmoveto{123.444}{29.899}
\emlineto{123.707}{29.984}
\emmoveto{123.707}{29.974}
\emlineto{123.969}{30.059}
\emmoveto{123.969}{30.049}
\emlineto{124.231}{30.134}
\emmoveto{124.231}{30.124}
\emlineto{124.493}{30.210}
\emmoveto{124.493}{30.200}
\emlineto{124.756}{30.286}
\emmoveto{124.756}{30.276}
\emlineto{125.018}{30.362}
\emmoveto{125.018}{30.352}
\emlineto{125.280}{30.439}
\emmoveto{125.280}{30.429}
\emlineto{125.542}{30.516}
\emmoveto{125.542}{30.506}
\emlineto{125.804}{30.594}
\emmoveto{125.804}{30.584}
\emlineto{126.067}{30.671}
\emmoveto{126.067}{30.661}
\emlineto{126.329}{30.750}
\emmoveto{126.329}{30.740}
\emlineto{126.591}{30.828}
\emmoveto{126.591}{30.818}
\emlineto{126.853}{30.907}
\emmoveto{126.853}{30.897}
\emlineto{127.116}{30.986}
\emmoveto{127.116}{30.976}
\emlineto{127.378}{31.066}
\emmoveto{127.378}{31.056}
\emlineto{127.640}{31.146}
\emmoveto{127.640}{31.136}
\emlineto{127.902}{31.227}
\emmoveto{127.902}{31.217}
\emlineto{128.164}{31.308}
\emmoveto{128.164}{31.298}
\emlineto{128.427}{31.389}
\emmoveto{128.427}{31.379}
\emlineto{128.689}{31.471}
\emmoveto{128.689}{31.461}
\emlineto{128.951}{31.553}
\emmoveto{128.951}{31.543}
\emlineto{129.213}{31.636}
\emmoveto{129.213}{31.626}
\emlineto{129.476}{31.719}
\emmoveto{129.476}{31.709}
\emlineto{129.738}{31.802}
\emshow{95.780}{17.700}{vz}
\emmoveto{12.000}{10.000}
\emlineto{12.262}{10.094}
\emmoveto{12.262}{10.084}
\emlineto{12.524}{10.178}
\emmoveto{12.524}{10.168}
\emlineto{12.787}{10.262}
\emmoveto{12.787}{10.252}
\emlineto{13.049}{10.346}
\emmoveto{13.049}{10.336}
\emlineto{13.311}{10.430}
\emmoveto{13.311}{10.420}
\emlineto{13.573}{10.514}
\emmoveto{13.573}{10.504}
\emlineto{13.836}{10.598}
\emmoveto{13.836}{10.588}
\emlineto{14.098}{10.682}
\emmoveto{14.098}{10.672}
\emlineto{14.360}{10.766}
\emmoveto{14.360}{10.756}
\emlineto{14.622}{10.850}
\emmoveto{14.622}{10.840}
\emlineto{14.884}{10.934}
\emmoveto{14.884}{10.924}
\emlineto{15.147}{11.018}
\emmoveto{15.147}{11.008}
\emlineto{15.409}{11.102}
\emmoveto{15.409}{11.092}
\emlineto{15.671}{11.186}
\emmoveto{15.671}{11.176}
\emlineto{15.933}{11.270}
\emmoveto{15.933}{11.260}
\emlineto{16.196}{11.355}
\emmoveto{16.196}{11.345}
\emlineto{16.458}{11.439}
\emmoveto{16.458}{11.429}
\emlineto{16.720}{11.523}
\emmoveto{16.720}{11.513}
\emlineto{16.982}{11.607}
\emmoveto{16.982}{11.597}
\emlineto{17.244}{11.691}
\emmoveto{17.244}{11.681}
\emlineto{17.507}{11.775}
\emmoveto{17.507}{11.765}
\emlineto{17.769}{11.860}
\emmoveto{17.769}{11.850}
\emlineto{18.031}{11.944}
\emmoveto{18.031}{11.934}
\emlineto{18.293}{12.028}
\emmoveto{18.293}{12.018}
\emlineto{18.556}{12.112}
\emmoveto{18.556}{12.102}
\emlineto{18.818}{12.196}
\emmoveto{18.818}{12.186}
\emlineto{19.080}{12.281}
\emmoveto{19.080}{12.271}
\emlineto{19.342}{12.365}
\emmoveto{19.342}{12.355}
\emlineto{19.604}{12.449}
\emmoveto{19.604}{12.439}
\emlineto{19.867}{12.534}
\emmoveto{19.867}{12.524}
\emlineto{20.129}{12.618}
\emmoveto{20.129}{12.608}
\emlineto{20.391}{12.703}
\emmoveto{20.391}{12.693}
\emlineto{20.653}{12.787}
\emmoveto{20.653}{12.777}
\emlineto{20.916}{12.872}
\emmoveto{20.916}{12.862}
\emlineto{21.178}{12.956}
\emmoveto{21.178}{12.946}
\emlineto{21.440}{13.041}
\emmoveto{21.440}{13.031}
\emlineto{21.702}{13.125}
\emmoveto{21.702}{13.115}
\emlineto{21.964}{13.210}
\emmoveto{21.964}{13.200}
\emlineto{22.227}{13.294}
\emmoveto{22.227}{13.284}
\emlineto{22.489}{13.379}
\emmoveto{22.489}{13.369}
\emlineto{22.751}{13.464}
\emmoveto{22.751}{13.454}
\emlineto{23.013}{13.549}
\emmoveto{23.013}{13.539}
\emlineto{23.276}{13.633}
\emmoveto{23.276}{13.623}
\emlineto{23.538}{13.718}
\emmoveto{23.538}{13.708}
\emlineto{23.800}{13.803}
\emmoveto{23.800}{13.793}
\emlineto{24.062}{13.888}
\emmoveto{24.062}{13.878}
\emlineto{24.324}{13.973}
\emmoveto{24.324}{13.963}
\emlineto{24.587}{14.058}
\emmoveto{24.587}{14.048}
\emlineto{24.849}{14.143}
\emmoveto{24.849}{14.133}
\emlineto{25.111}{14.228}
\emmoveto{25.111}{14.218}
\emlineto{25.373}{14.313}
\emmoveto{25.373}{14.303}
\emlineto{25.636}{14.398}
\emmoveto{25.636}{14.388}
\emlineto{25.898}{14.483}
\emmoveto{25.898}{14.473}
\emlineto{26.160}{14.568}
\emmoveto{26.160}{14.558}
\emlineto{26.422}{14.654}
\emmoveto{26.422}{14.644}
\emlineto{26.684}{14.739}
\emmoveto{26.684}{14.729}
\emlineto{26.947}{14.824}
\emmoveto{26.947}{14.814}
\emlineto{27.209}{14.910}
\emmoveto{27.209}{14.900}
\emlineto{27.471}{14.995}
\emmoveto{27.471}{14.985}
\emlineto{27.733}{15.081}
\emmoveto{27.733}{15.071}
\emlineto{27.996}{15.166}
\emmoveto{27.996}{15.156}
\emlineto{28.258}{15.252}
\emmoveto{28.258}{15.242}
\emlineto{28.520}{15.338}
\emmoveto{28.520}{15.328}
\emlineto{28.782}{15.423}
\emmoveto{28.782}{15.413}
\emlineto{29.044}{15.509}
\emmoveto{29.044}{15.499}
\emlineto{29.307}{15.595}
\emmoveto{29.307}{15.585}
\emlineto{29.569}{15.681}
\emmoveto{29.569}{15.671}
\emlineto{29.831}{15.767}
\emmoveto{29.831}{15.757}
\emlineto{30.093}{15.853}
\emmoveto{30.093}{15.843}
\emlineto{30.356}{15.939}
\emmoveto{30.356}{15.929}
\emlineto{30.618}{16.025}
\emmoveto{30.618}{16.015}
\emlineto{30.880}{16.111}
\emmoveto{30.880}{16.101}
\emlineto{31.142}{16.198}
\emmoveto{31.142}{16.188}
\emlineto{31.404}{16.284}
\emmoveto{31.404}{16.274}
\emlineto{31.667}{16.370}
\emmoveto{31.667}{16.360}
\emlineto{31.929}{16.457}
\emmoveto{31.929}{16.447}
\emlineto{32.191}{16.543}
\emmoveto{32.191}{16.533}
\emlineto{32.453}{16.630}
\emmoveto{32.453}{16.620}
\emlineto{32.716}{16.717}
\emmoveto{32.716}{16.707}
\emlineto{32.978}{16.804}
\emmoveto{32.978}{16.794}
\emlineto{33.240}{16.890}
\emmoveto{33.240}{16.880}
\emlineto{33.502}{16.977}
\emmoveto{33.502}{16.967}
\emlineto{33.764}{17.064}
\emmoveto{33.764}{17.054}
\emlineto{34.027}{17.151}
\emmoveto{34.027}{17.141}
\emlineto{34.289}{17.238}
\emmoveto{34.289}{17.228}
\emlineto{34.551}{17.326}
\emmoveto{34.551}{17.316}
\emlineto{34.813}{17.413}
\emmoveto{34.813}{17.403}
\emlineto{35.076}{17.500}
\emmoveto{35.076}{17.490}
\emlineto{35.338}{17.588}
\emmoveto{35.338}{17.578}
\emlineto{35.600}{17.675}
\emmoveto{35.600}{17.665}
\emlineto{35.862}{17.763}
\emmoveto{35.862}{17.753}
\emlineto{36.124}{17.850}
\emmoveto{36.124}{17.840}
\emlineto{36.387}{17.938}
\emmoveto{36.387}{17.928}
\emlineto{36.649}{18.026}
\emmoveto{36.649}{18.016}
\emlineto{36.911}{18.114}
\emmoveto{36.911}{18.104}
\emlineto{37.173}{18.202}
\emmoveto{37.173}{18.192}
\emlineto{37.436}{18.290}
\emmoveto{37.436}{18.280}
\emlineto{37.698}{18.378}
\emmoveto{37.698}{18.368}
\emlineto{37.960}{18.467}
\emmoveto{37.960}{18.457}
\emlineto{38.222}{18.555}
\emmoveto{38.222}{18.545}
\emlineto{38.484}{18.644}
\emmoveto{38.484}{18.634}
\emlineto{38.747}{18.732}
\emmoveto{38.747}{18.722}
\emlineto{39.009}{18.821}
\emmoveto{39.009}{18.811}
\emlineto{39.271}{18.910}
\emmoveto{39.271}{18.900}
\emlineto{39.533}{18.998}
\emmoveto{39.533}{18.988}
\emlineto{39.796}{19.087}
\emmoveto{39.796}{19.077}
\emlineto{40.058}{19.176}
\emmoveto{40.058}{19.166}
\emlineto{40.320}{19.266}
\emmoveto{40.320}{19.256}
\emlineto{40.582}{19.355}
\emmoveto{40.582}{19.345}
\emlineto{40.844}{19.444}
\emmoveto{40.844}{19.434}
\emlineto{41.107}{19.534}
\emmoveto{41.107}{19.524}
\emlineto{41.369}{19.623}
\emmoveto{41.369}{19.613}
\emlineto{41.631}{19.713}
\emmoveto{41.631}{19.703}
\emlineto{41.893}{19.803}
\emmoveto{41.893}{19.793}
\emlineto{42.156}{19.893}
\emmoveto{42.156}{19.883}
\emlineto{42.418}{19.983}
\emmoveto{42.418}{19.973}
\emlineto{42.680}{20.073}
\emmoveto{42.680}{20.063}
\emlineto{42.942}{20.163}
\emmoveto{42.942}{20.153}
\emlineto{43.204}{20.253}
\emmoveto{43.204}{20.243}
\emlineto{43.467}{20.344}
\emmoveto{43.467}{20.334}
\emlineto{43.729}{20.434}
\emmoveto{43.729}{20.424}
\emlineto{43.991}{20.525}
\emmoveto{43.991}{20.515}
\emlineto{44.253}{20.616}
\emmoveto{44.253}{20.606}
\emlineto{44.516}{20.707}
\emmoveto{44.516}{20.697}
\emlineto{44.778}{20.798}
\emmoveto{44.778}{20.788}
\emlineto{45.040}{20.889}
\emmoveto{45.040}{20.879}
\emlineto{45.302}{20.980}
\emmoveto{45.302}{20.970}
\emlineto{45.564}{21.072}
\emmoveto{45.564}{21.062}
\emlineto{45.827}{21.163}
\emmoveto{45.827}{21.153}
\emlineto{46.089}{21.255}
\emmoveto{46.089}{21.245}
\emlineto{46.351}{21.347}
\emmoveto{46.351}{21.337}
\emlineto{46.613}{21.439}
\emmoveto{46.613}{21.429}
\emlineto{46.876}{21.531}
\emmoveto{46.876}{21.521}
\emlineto{47.138}{21.623}
\emmoveto{47.138}{21.613}
\emlineto{47.400}{21.715}
\emmoveto{47.400}{21.705}
\emlineto{47.662}{21.808}
\emmoveto{47.662}{21.798}
\emlineto{47.924}{21.901}
\emmoveto{47.924}{21.891}
\emlineto{48.187}{21.993}
\emmoveto{48.187}{21.983}
\emlineto{48.449}{22.086}
\emmoveto{48.449}{22.076}
\emlineto{48.711}{22.179}
\emmoveto{48.711}{22.169}
\emlineto{48.973}{22.272}
\emmoveto{48.973}{22.262}
\emlineto{49.236}{22.366}
\emmoveto{49.236}{22.356}
\emlineto{49.498}{22.459}
\emmoveto{49.498}{22.449}
\emlineto{49.760}{22.553}
\emmoveto{49.760}{22.543}
\emlineto{50.022}{22.647}
\emmoveto{50.022}{22.637}
\emlineto{50.284}{22.741}
\emmoveto{50.284}{22.731}
\emlineto{50.547}{22.835}
\emmoveto{50.547}{22.825}
\emlineto{50.809}{22.929}
\emmoveto{50.809}{22.919}
\emlineto{51.071}{23.023}
\emmoveto{51.071}{23.013}
\emlineto{51.333}{23.118}
\emmoveto{51.333}{23.108}
\emlineto{51.596}{23.212}
\emmoveto{51.596}{23.202}
\emlineto{51.858}{23.307}
\emmoveto{51.858}{23.297}
\emlineto{52.120}{23.402}
\emmoveto{52.120}{23.392}
\emlineto{52.382}{23.497}
\emmoveto{52.382}{23.487}
\emlineto{52.644}{23.593}
\emmoveto{52.644}{23.583}
\emlineto{52.907}{23.688}
\emmoveto{52.907}{23.678}
\emlineto{53.169}{23.784}
\emmoveto{53.169}{23.774}
\emlineto{53.431}{23.880}
\emmoveto{53.431}{23.870}
\emlineto{53.693}{23.976}
\emmoveto{53.693}{23.966}
\emlineto{53.956}{24.072}
\emmoveto{53.956}{24.062}
\emlineto{54.218}{24.168}
\emmoveto{54.218}{24.158}
\emlineto{54.480}{24.265}
\emmoveto{54.480}{24.255}
\emlineto{54.742}{24.361}
\emmoveto{54.742}{24.351}
\emlineto{55.004}{24.458}
\emmoveto{55.004}{24.448}
\emlineto{55.267}{24.555}
\emmoveto{55.267}{24.545}
\emlineto{55.529}{24.653}
\emmoveto{55.529}{24.643}
\emlineto{55.791}{24.750}
\emmoveto{55.791}{24.740}
\emlineto{56.053}{24.848}
\emmoveto{56.053}{24.838}
\emlineto{56.316}{24.945}
\emmoveto{56.316}{24.935}
\emlineto{56.578}{25.043}
\emmoveto{56.578}{25.033}
\emlineto{56.840}{25.141}
\emmoveto{56.840}{25.131}
\emlineto{57.102}{25.240}
\emmoveto{57.102}{25.230}
\emlineto{57.364}{25.338}
\emmoveto{57.364}{25.328}
\emlineto{57.627}{25.437}
\emmoveto{57.627}{25.427}
\emlineto{57.889}{25.536}
\emmoveto{57.889}{25.526}
\emlineto{58.151}{25.635}
\emmoveto{58.151}{25.625}
\emlineto{58.413}{25.735}
\emmoveto{58.413}{25.725}
\emlineto{58.676}{25.834}
\emmoveto{58.676}{25.824}
\emlineto{58.938}{25.934}
\emmoveto{58.938}{25.924}
\emlineto{59.200}{26.034}
\emmoveto{59.200}{26.024}
\emlineto{59.462}{26.134}
\emmoveto{59.462}{26.124}
\emlineto{59.724}{26.234}
\emmoveto{59.724}{26.224}
\emlineto{59.987}{26.335}
\emmoveto{59.987}{26.325}
\emlineto{60.249}{26.436}
\emmoveto{60.249}{26.426}
\emlineto{60.511}{26.537}
\emmoveto{60.511}{26.527}
\emlineto{60.773}{26.638}
\emmoveto{60.773}{26.628}
\emlineto{61.036}{26.739}
\emmoveto{61.036}{26.729}
\emlineto{61.298}{26.841}
\emmoveto{61.298}{26.831}
\emlineto{61.560}{26.943}
\emmoveto{61.560}{26.933}
\emlineto{61.822}{27.045}
\emmoveto{61.822}{27.035}
\emlineto{62.084}{27.148}
\emmoveto{62.084}{27.138}
\emlineto{62.347}{27.250}
\emmoveto{62.347}{27.240}
\emlineto{62.609}{27.353}
\emmoveto{62.609}{27.343}
\emlineto{62.871}{27.456}
\emmoveto{62.871}{27.446}
\emlineto{63.133}{27.559}
\emmoveto{63.133}{27.549}
\emlineto{63.396}{27.663}
\emmoveto{63.396}{27.653}
\emlineto{63.658}{27.767}
\emmoveto{63.658}{27.757}
\emlineto{63.920}{27.871}
\emmoveto{63.920}{27.861}
\emlineto{64.182}{27.975}
\emmoveto{64.182}{27.965}
\emlineto{64.444}{28.079}
\emmoveto{64.444}{28.069}
\emlineto{64.707}{28.184}
\emmoveto{64.707}{28.174}
\emlineto{64.969}{28.289}
\emmoveto{64.969}{28.279}
\emlineto{65.231}{28.395}
\emmoveto{65.231}{28.385}
\emlineto{65.493}{28.500}
\emmoveto{65.493}{28.490}
\emlineto{65.756}{28.606}
\emmoveto{65.756}{28.596}
\emlineto{66.018}{28.712}
\emmoveto{66.018}{28.702}
\emlineto{66.280}{28.818}
\emmoveto{66.280}{28.808}
\emlineto{66.542}{28.925}
\emmoveto{66.542}{28.915}
\emlineto{66.804}{29.032}
\emmoveto{66.804}{29.022}
\emlineto{67.067}{29.139}
\emmoveto{67.067}{29.129}
\emlineto{67.329}{29.246}
\emmoveto{67.329}{29.236}
\emlineto{67.591}{29.354}
\emmoveto{67.591}{29.344}
\emlineto{67.853}{29.462}
\emmoveto{67.853}{29.452}
\emlineto{68.116}{29.570}
\emmoveto{68.116}{29.560}
\emlineto{68.378}{29.679}
\emmoveto{68.378}{29.669}
\emlineto{68.640}{29.788}
\emmoveto{68.640}{29.778}
\emlineto{68.902}{29.897}
\emmoveto{68.902}{29.887}
\emlineto{69.164}{30.006}
\emmoveto{69.164}{29.996}
\emlineto{69.427}{30.116}
\emmoveto{69.427}{30.106}
\emlineto{69.689}{30.226}
\emmoveto{69.689}{30.216}
\emlineto{69.951}{30.336}
\emmoveto{69.951}{30.326}
\emlineto{70.213}{30.447}
\emmoveto{70.213}{30.437}
\emlineto{70.476}{30.558}
\emmoveto{70.476}{30.548}
\emlineto{70.738}{30.669}
\emmoveto{70.738}{30.659}
\emlineto{71.000}{30.781}
\emmoveto{71.000}{30.771}
\emlineto{71.262}{30.893}
\emmoveto{71.262}{30.883}
\emlineto{71.524}{31.005}
\emmoveto{71.524}{30.995}
\emlineto{71.787}{31.117}
\emmoveto{71.787}{31.107}
\emlineto{72.049}{31.230}
\emmoveto{72.049}{31.220}
\emlineto{72.311}{31.343}
\emmoveto{72.311}{31.333}
\emlineto{72.573}{31.457}
\emmoveto{72.573}{31.447}
\emlineto{72.836}{31.571}
\emmoveto{72.836}{31.561}
\emlineto{73.098}{31.685}
\emmoveto{73.098}{31.675}
\emlineto{73.360}{31.799}
\emmoveto{73.360}{31.789}
\emlineto{73.622}{31.914}
\emmoveto{73.622}{31.904}
\emlineto{73.884}{32.029}
\emmoveto{73.884}{32.019}
\emlineto{74.147}{32.145}
\emmoveto{74.147}{32.135}
\emlineto{74.409}{32.261}
\emmoveto{74.409}{32.251}
\emlineto{74.671}{32.377}
\emmoveto{74.671}{32.367}
\emlineto{74.933}{32.494}
\emmoveto{74.933}{32.484}
\emlineto{75.196}{32.611}
\emmoveto{75.196}{32.601}
\emlineto{75.458}{32.728}
\emmoveto{75.458}{32.718}
\emlineto{75.720}{32.846}
\emmoveto{75.720}{32.836}
\emlineto{75.982}{32.964}
\emmoveto{75.982}{32.954}
\emlineto{76.244}{33.083}
\emmoveto{76.244}{33.073}
\emlineto{76.507}{33.202}
\emmoveto{76.507}{33.192}
\emlineto{76.769}{33.321}
\emmoveto{76.769}{33.311}
\emlineto{77.031}{33.441}
\emmoveto{77.031}{33.431}
\emlineto{77.293}{33.561}
\emmoveto{77.293}{33.551}
\emlineto{77.556}{33.681}
\emmoveto{77.556}{33.671}
\emlineto{77.818}{33.802}
\emmoveto{77.818}{33.792}
\emlineto{78.080}{33.924}
\emmoveto{78.080}{33.914}
\emlineto{78.342}{34.045}
\emmoveto{78.342}{34.035}
\emlineto{78.604}{34.167}
\emmoveto{78.604}{34.157}
\emlineto{78.867}{34.290}
\emmoveto{78.867}{34.280}
\emlineto{79.129}{34.413}
\emmoveto{79.129}{34.403}
\emlineto{79.391}{34.536}
\emmoveto{79.391}{34.526}
\emlineto{79.653}{34.660}
\emmoveto{79.653}{34.650}
\emlineto{79.916}{34.785}
\emmoveto{79.916}{34.775}
\emlineto{80.178}{34.909}
\emmoveto{80.178}{34.899}
\emlineto{80.440}{35.034}
\emmoveto{80.440}{35.024}
\emlineto{80.702}{35.160}
\emmoveto{80.702}{35.150}
\emlineto{80.964}{35.286}
\emmoveto{80.964}{35.276}
\emlineto{81.227}{35.413}
\emmoveto{81.227}{35.403}
\emlineto{81.489}{35.540}
\emmoveto{81.489}{35.530}
\emlineto{81.751}{35.667}
\emmoveto{81.751}{35.657}
\emlineto{82.013}{35.795}
\emmoveto{82.013}{35.785}
\emlineto{82.276}{35.924}
\emmoveto{82.276}{35.914}
\emlineto{82.538}{36.053}
\emmoveto{82.538}{36.043}
\emlineto{82.800}{36.182}
\emmoveto{82.800}{36.172}
\emlineto{83.062}{36.312}
\emmoveto{83.062}{36.302}
\emlineto{83.324}{36.442}
\emmoveto{83.324}{36.432}
\emlineto{83.587}{36.573}
\emmoveto{83.587}{36.563}
\emlineto{83.849}{36.705}
\emmoveto{83.849}{36.695}
\emlineto{84.111}{36.837}
\emmoveto{84.111}{36.827}
\emlineto{84.373}{36.969}
\emmoveto{84.373}{36.959}
\emlineto{84.636}{37.102}
\emmoveto{84.636}{37.092}
\emlineto{84.898}{37.236}
\emmoveto{84.898}{37.226}
\emlineto{85.160}{37.370}
\emmoveto{85.160}{37.360}
\emlineto{85.422}{37.504}
\emmoveto{85.422}{37.494}
\emlineto{85.684}{37.639}
\emmoveto{85.684}{37.629}
\emlineto{85.947}{37.775}
\emmoveto{85.947}{37.765}
\emlineto{86.209}{37.912}
\emmoveto{86.209}{37.902}
\emlineto{86.471}{38.048}
\emmoveto{86.471}{38.038}
\emlineto{86.733}{38.186}
\emmoveto{86.733}{38.176}
\emlineto{86.996}{38.324}
\emmoveto{86.996}{38.314}
\emlineto{87.258}{38.462}
\emmoveto{87.258}{38.452}
\emlineto{87.520}{38.602}
\emmoveto{87.520}{38.592}
\emlineto{87.782}{38.741}
\emmoveto{87.782}{38.731}
\emlineto{88.044}{38.882}
\emmoveto{88.044}{38.872}
\emlineto{88.307}{39.023}
\emmoveto{88.307}{39.013}
\emlineto{88.569}{39.165}
\emmoveto{88.569}{39.155}
\emlineto{88.831}{39.307}
\emmoveto{88.831}{39.297}
\emlineto{89.093}{39.450}
\emmoveto{89.093}{39.440}
\emlineto{89.356}{39.593}
\emmoveto{89.356}{39.583}
\emlineto{89.618}{39.738}
\emmoveto{89.618}{39.728}
\emlineto{89.880}{39.882}
\emmoveto{89.880}{39.872}
\emlineto{90.142}{40.028}
\emmoveto{90.142}{40.018}
\emlineto{90.404}{40.174}
\emmoveto{90.404}{40.164}
\emlineto{90.667}{40.321}
\emmoveto{90.667}{40.311}
\emlineto{90.929}{40.469}
\emmoveto{90.929}{40.459}
\emlineto{91.191}{40.617}
\emmoveto{91.191}{40.607}
\emlineto{91.453}{40.766}
\emmoveto{91.453}{40.756}
\emlineto{91.716}{40.916}
\emmoveto{91.716}{40.906}
\emlineto{91.978}{41.066}
\emmoveto{91.978}{41.056}
\emlineto{92.240}{41.217}
\emmoveto{92.240}{41.207}
\emlineto{92.502}{41.369}
\emmoveto{92.502}{41.359}
\emlineto{92.764}{41.522}
\emmoveto{92.764}{41.512}
\emlineto{93.027}{41.675}
\emmoveto{93.027}{41.665}
\emlineto{93.289}{41.829}
\emmoveto{93.289}{41.819}
\emlineto{93.551}{41.984}
\emmoveto{93.551}{41.974}
\emlineto{93.813}{42.140}
\emmoveto{93.813}{42.130}
\emlineto{94.076}{42.296}
\emmoveto{94.076}{42.286}
\emlineto{94.338}{42.454}
\emmoveto{94.338}{42.444}
\emlineto{94.600}{42.612}
\emmoveto{94.600}{42.602}
\emlineto{94.862}{42.771}
\emmoveto{94.862}{42.761}
\emlineto{95.124}{42.930}
\emmoveto{95.124}{42.920}
\emlineto{95.387}{43.091}
\emmoveto{95.387}{43.081}
\emlineto{95.649}{43.252}
\emmoveto{95.649}{43.242}
\emlineto{95.911}{43.415}
\emmoveto{95.911}{43.405}
\emlineto{96.173}{43.578}
\emmoveto{96.173}{43.568}
\emlineto{96.436}{43.742}
\emmoveto{96.436}{43.732}
\emlineto{96.698}{43.907}
\emmoveto{96.698}{43.897}
\emlineto{96.960}{44.073}
\emmoveto{96.960}{44.063}
\emlineto{97.222}{44.240}
\emmoveto{97.222}{44.230}
\emlineto{97.484}{44.407}
\emmoveto{97.484}{44.397}
\emlineto{97.747}{44.576}
\emmoveto{97.747}{44.566}
\emlineto{98.009}{44.746}
\emmoveto{98.009}{44.736}
\emlineto{98.271}{44.916}
\emmoveto{98.271}{44.906}
\emlineto{98.533}{45.088}
\emmoveto{98.533}{45.078}
\emlineto{98.796}{45.260}
\emmoveto{98.796}{45.250}
\emlineto{99.058}{45.434}
\emmoveto{99.058}{45.424}
\emlineto{99.320}{45.608}
\emmoveto{99.320}{45.598}
\emlineto{99.582}{45.784}
\emmoveto{99.582}{45.774}
\emlineto{99.844}{45.961}
\emmoveto{99.844}{45.951}
\emlineto{100.107}{46.138}
\emmoveto{100.107}{46.128}
\emlineto{100.369}{46.317}
\emmoveto{100.369}{46.307}
\emlineto{100.631}{46.497}
\emmoveto{100.631}{46.487}
\emlineto{100.893}{46.678}
\emmoveto{100.893}{46.668}
\emlineto{101.156}{46.860}
\emmoveto{101.156}{46.850}
\emlineto{101.418}{47.043}
\emmoveto{101.418}{47.033}
\emlineto{101.680}{47.228}
\emmoveto{101.680}{47.218}
\emlineto{101.942}{47.413}
\emmoveto{101.942}{47.403}
\emlineto{102.204}{47.600}
\emmoveto{102.204}{47.590}
\emlineto{102.467}{47.788}
\emmoveto{102.467}{47.778}
\emlineto{102.729}{47.977}
\emmoveto{102.729}{47.967}
\emlineto{102.991}{48.167}
\emmoveto{102.991}{48.157}
\emlineto{103.253}{48.359}
\emmoveto{103.253}{48.349}
\emlineto{103.516}{48.551}
\emmoveto{103.516}{48.541}
\emlineto{103.778}{48.745}
\emmoveto{103.778}{48.735}
\emlineto{104.040}{48.941}
\emmoveto{104.040}{48.931}
\emlineto{104.302}{49.137}
\emmoveto{104.302}{49.127}
\emlineto{104.564}{49.335}
\emmoveto{104.564}{49.325}
\emlineto{104.827}{49.535}
\emmoveto{104.827}{49.525}
\emlineto{105.089}{49.736}
\emmoveto{105.089}{49.726}
\emlineto{105.351}{49.938}
\emmoveto{105.351}{49.928}
\emlineto{105.613}{50.141}
\emmoveto{105.613}{50.131}
\emlineto{105.876}{50.346}
\emmoveto{105.876}{50.336}
\emlineto{106.138}{50.553}
\emmoveto{106.138}{50.543}
\emlineto{106.400}{50.760}
\emmoveto{106.400}{50.750}
\emlineto{106.662}{50.970}
\emmoveto{106.662}{50.960}
\emlineto{106.924}{51.181}
\emmoveto{106.924}{51.171}
\emlineto{107.187}{51.393}
\emmoveto{107.187}{51.383}
\emlineto{107.449}{51.607}
\emmoveto{107.449}{51.597}
\emlineto{107.711}{51.823}
\emmoveto{107.711}{51.813}
\emlineto{107.973}{52.040}
\emmoveto{107.973}{52.030}
\emlineto{108.236}{52.258}
\emmoveto{108.236}{52.248}
\emlineto{108.498}{52.479}
\emmoveto{108.498}{52.469}
\emlineto{108.760}{52.701}
\emmoveto{108.760}{52.691}
\emlineto{109.022}{52.925}
\emmoveto{109.022}{52.915}
\emlineto{109.284}{53.150}
\emmoveto{109.284}{53.140}
\emlineto{109.547}{53.378}
\emmoveto{109.547}{53.368}
\emlineto{109.809}{53.607}
\emmoveto{109.809}{53.597}
\emlineto{110.071}{53.837}
\emmoveto{110.071}{53.827}
\emlineto{110.333}{54.070}
\emmoveto{110.333}{54.060}
\emlineto{110.596}{54.305}
\emmoveto{110.596}{54.295}
\emlineto{110.858}{54.541}
\emmoveto{110.858}{54.531}
\emlineto{111.120}{54.779}
\emmoveto{111.120}{54.769}
\emlineto{111.382}{55.020}
\emmoveto{111.382}{55.010}
\emlineto{111.644}{55.262}
\emmoveto{111.644}{55.252}
\emlineto{111.907}{55.506}
\emmoveto{111.907}{55.496}
\emlineto{112.169}{55.752}
\emmoveto{112.169}{55.742}
\emlineto{112.431}{56.001}
\emmoveto{112.431}{55.991}
\emlineto{112.693}{56.251}
\emmoveto{112.693}{56.241}
\emlineto{112.956}{56.504}
\emmoveto{112.956}{56.494}
\emlineto{113.218}{56.759}
\emmoveto{113.218}{56.749}
\emlineto{113.480}{57.015}
\emmoveto{113.480}{57.005}
\emlineto{113.742}{57.275}
\emmoveto{113.742}{57.265}
\emlineto{114.004}{57.536}
\emmoveto{114.004}{57.526}
\emlineto{114.267}{57.800}
\emmoveto{114.267}{57.790}
\emlineto{114.529}{58.066}
\emmoveto{114.529}{58.056}
\emlineto{114.791}{58.334}
\emmoveto{114.791}{58.324}
\emlineto{115.053}{58.605}
\emmoveto{115.053}{58.595}
\emlineto{115.316}{58.878}
\emmoveto{115.316}{58.868}
\emlineto{115.578}{59.154}
\emmoveto{115.578}{59.144}
\emlineto{115.840}{59.432}
\emmoveto{115.840}{59.422}
\emlineto{116.102}{59.712}
\emmoveto{116.102}{59.702}
\emlineto{116.364}{59.996}
\emmoveto{116.364}{59.986}
\emlineto{116.627}{60.282}
\emmoveto{116.627}{60.272}
\emlineto{116.889}{60.570}
\emmoveto{116.889}{60.560}
\emlineto{117.151}{60.861}
\emmoveto{117.151}{60.851}
\emlineto{117.413}{61.155}
\emmoveto{117.413}{61.145}
\emlineto{117.676}{61.452}
\emmoveto{117.676}{61.442}
\emlineto{117.938}{61.751}
\emmoveto{117.938}{61.741}
\emlineto{118.200}{62.053}
\emmoveto{118.200}{62.043}
\emlineto{118.462}{62.359}
\emmoveto{118.462}{62.349}
\emlineto{118.724}{62.666}
\emmoveto{118.724}{62.656}
\emlineto{118.987}{62.977}
\emmoveto{118.987}{62.967}
\emlineto{119.249}{63.291}
\emmoveto{119.249}{63.281}
\emlineto{119.511}{63.608}
\emmoveto{119.511}{63.598}
\emlineto{119.773}{63.928}
\emmoveto{119.773}{63.918}
\emlineto{120.036}{64.251}
\emmoveto{120.036}{64.241}
\emlineto{120.298}{64.576}
\emmoveto{120.298}{64.566}
\emlineto{120.560}{64.905}
\emmoveto{120.560}{64.895}
\emlineto{120.822}{65.237}
\emmoveto{120.822}{65.227}
\emlineto{121.084}{65.573}
\emmoveto{121.084}{65.563}
\emlineto{121.347}{65.911}
\emmoveto{121.347}{65.901}
\emlineto{121.609}{66.252}
\emmoveto{121.609}{66.242}
\emlineto{121.871}{66.597}
\emmoveto{121.871}{66.587}
\emlineto{122.133}{66.944}
\emmoveto{122.133}{66.934}
\emlineto{122.396}{67.295}
\emmoveto{122.396}{67.285}
\emlineto{122.658}{67.649}
\emmoveto{122.658}{67.639}
\emlineto{122.920}{68.006}
\emmoveto{122.920}{67.996}
\emlineto{123.182}{68.365}
\emmoveto{123.182}{68.355}
\emlineto{123.444}{68.728}
\emmoveto{123.444}{68.718}
\emlineto{123.707}{69.094}
\emmoveto{123.707}{69.084}
\emlineto{123.969}{69.462}
\emmoveto{123.969}{69.452}
\emlineto{124.231}{69.834}
\emmoveto{124.231}{69.824}
\emlineto{124.493}{70.208}
\emmoveto{124.493}{70.198}
\emlineto{124.756}{70.584}
\emmoveto{124.756}{70.574}
\emlineto{125.018}{70.963}
\emmoveto{125.018}{70.953}
\emlineto{125.280}{71.344}
\emmoveto{125.280}{71.334}
\emlineto{125.542}{71.727}
\emmoveto{125.542}{71.717}
\emlineto{125.804}{72.112}
\emmoveto{125.804}{72.102}
\emlineto{126.067}{72.498}
\emmoveto{126.067}{72.488}
\emlineto{126.329}{72.886}
\emmoveto{126.329}{72.876}
\emlineto{126.591}{73.274}
\emmoveto{126.591}{73.264}
\emlineto{126.853}{73.663}
\emmoveto{126.853}{73.653}
\emlineto{127.116}{74.052}
\emmoveto{127.116}{74.042}
\emlineto{127.378}{74.440}
\emmoveto{127.378}{74.430}
\emlineto{127.640}{74.827}
\emmoveto{127.640}{74.817}
\emlineto{127.902}{75.212}
\emmoveto{127.902}{75.202}
\emlineto{128.164}{75.595}
\emmoveto{128.164}{75.585}
\emlineto{128.427}{75.974}
\emmoveto{128.427}{75.964}
\emlineto{128.689}{76.348}
\emmoveto{128.689}{76.338}
\emlineto{128.951}{76.717}
\emmoveto{128.951}{76.707}
\emlineto{129.213}{77.078}
\emmoveto{129.213}{77.068}
\emlineto{129.476}{77.431}
\emmoveto{129.476}{77.421}
\emlineto{129.738}{77.773}
\emshow{95.780}{38.700}{vq}
\emshow{1.000}{10.000}{0.00e0}
\emshow{1.000}{17.000}{1.00e-1}
\emshow{1.000}{24.000}{2.00e-1}
\emshow{1.000}{31.000}{3.00e-1}
\emshow{1.000}{38.000}{4.00e-1}
\emshow{1.000}{45.000}{5.00e-1}
\emshow{1.000}{52.000}{6.00e-1}
\emshow{1.000}{59.000}{7.00e-1}
\emshow{1.000}{66.000}{8.00e-1}
\emshow{1.000}{73.000}{9.00e-1}
\emshow{1.000}{80.000}{1.00e0}
\emshow{12.000}{5.000}{0.00e0}
\emshow{23.800}{5.000}{1.35e0}
\emshow{35.600}{5.000}{2.70e0}
\emshow{47.400}{5.000}{4.05e0}
\emshow{59.200}{5.000}{5.40e0}
\emshow{71.000}{5.000}{6.75e0}
\emshow{82.800}{5.000}{8.10e0}
\emshow{94.600}{5.000}{9.45e0}
\emshow{106.400}{5.000}{1.08e1}
\emshow{118.200}{5.000}{1.22e1}
\emshow{130.000}{5.000}{1.35e1}

\centerline{\bf {Fig. B. 1}}

 
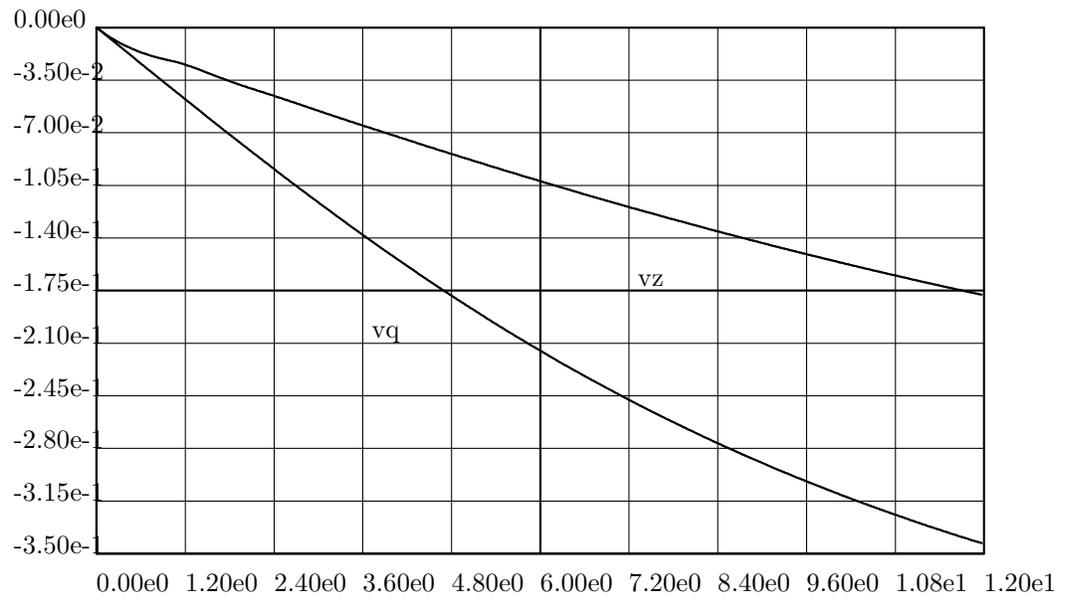
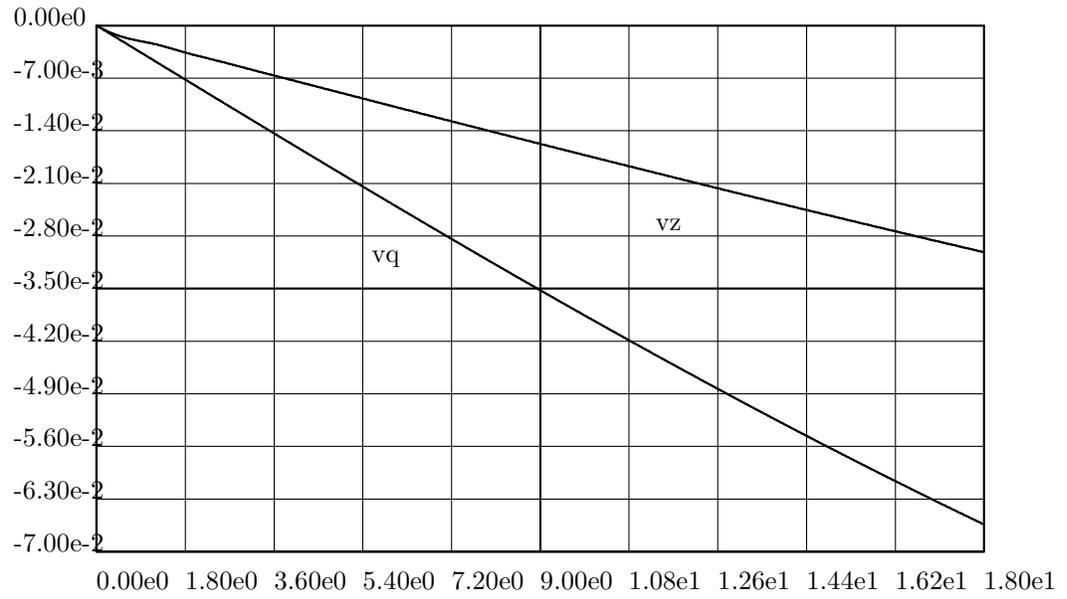
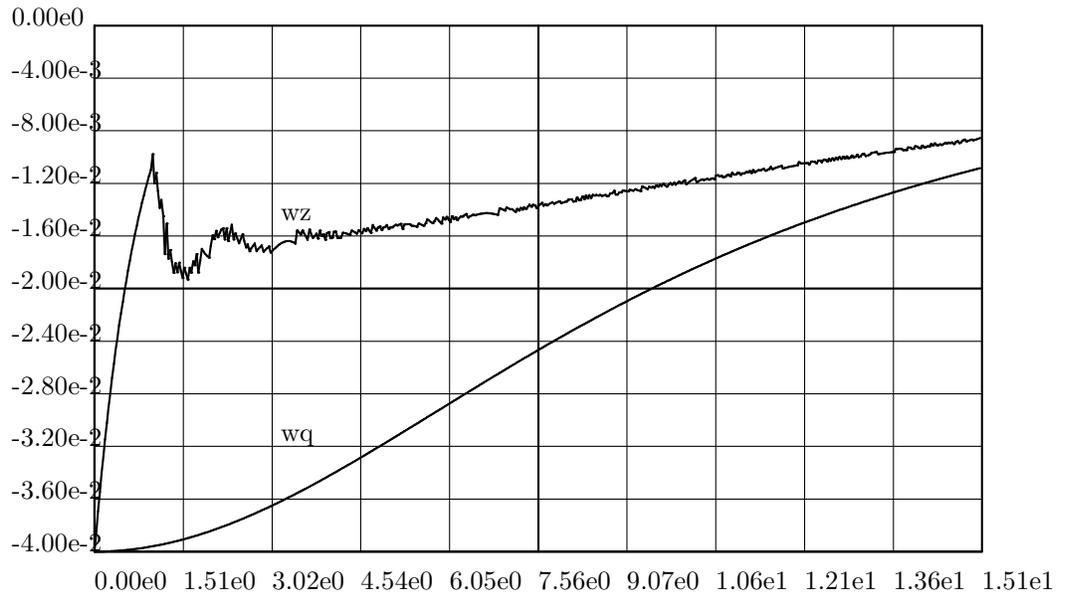
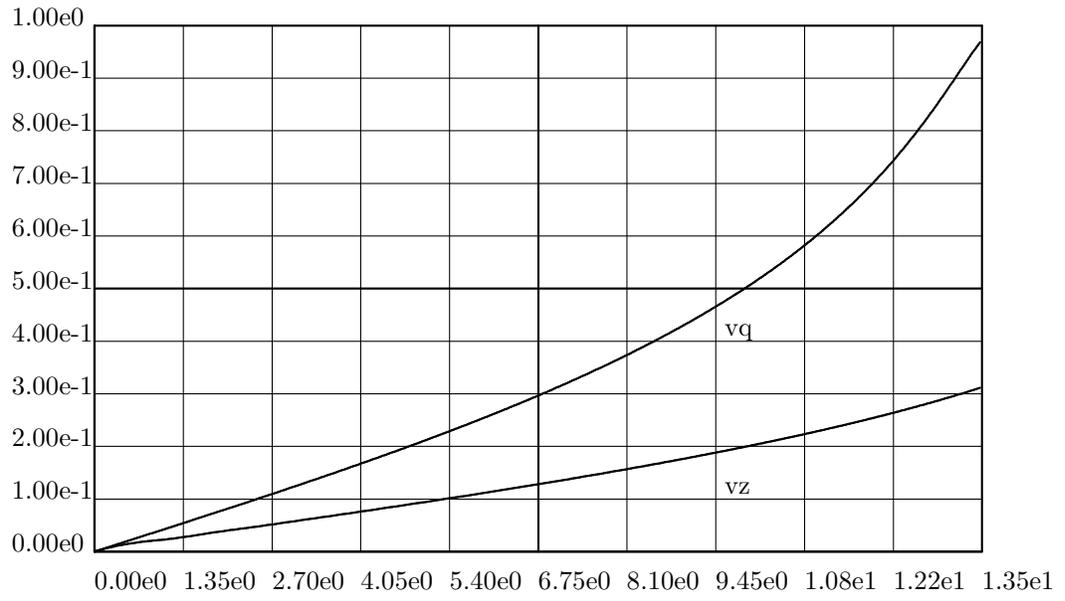
\end{document}